\pdfoutput=1
\documentclass[
    twocolumn,groupedaddress,superscriptaddress,
    footinbib,
    floatfix,
    longbibliography,
    showpacs,showkeys,
    ]{revtex4-2}

\usepackage{amssymb,amsmath,graphicx,color}
\usepackage{bm}
\usepackage[tight]{subfigure}
\usepackage[export]{adjustbox}
\usepackage{float}

\usepackage[bookmarksnumbered,              
colorlinks = true,
linkcolor = black,
urlcolor  = blue,
citecolor = blue,
anchorcolor = blue]{hyperref}

\usepackage{array}
\usepackage{makecell} %
\usepackage{import} 
\usepackage[verbose]{placeins}
\usepackage{xstring}
\usepackage{bbm}
\usepackage[capitalise]{cleveref} 
\crefname{figure}{Fig.}{Figs.}
\usepackage{mathtools}
\usepackage[normalem]{ulem}
\usepackage{orcidlink}

\graphicspath{{figures}}
\newtheorem{definition}{Definition}

\newcommand{\abs}[1]{\left\lvert #1 \right\rvert}
\newcommand{\abss}[1]{{\lvert} #1 {\rvert}}
\newcommand{\ket}[1]{{\lvert} #1 {\rangle}}
\newcommand{\bra}[1]{{\langle} #1 {\rvert}}
\newcommand{\braket}[2]{{\langle} #1 {\, \vert \,}#2 {\rangle}} 
\newcommand{\expval}[3]{{\langle} #1 {\, \vert } #2 {\vert \,}#3 {\rangle}} 
\newcommand{\ep}[1]{\langle #1 \rangle}

\DeclarePairedDelimiterX\ketbra[2]{\lvert}{\rvert}{#1\delimsize\rangle\!\delimsize\langle#2}%
\DeclarePairedDelimiterX\projector[1]{\lvert}{\rvert}{#1\delimsize\rangle\!\delimsize\langle#1}
\newcommand{\vectstraight}[1]{\boldsymbol{\mathbf{#1}}}
\newcommand{\vect}[1]{\boldsymbol{#1}}

\newcommand{\kv}{\vectstraight{k}}
\newcommand{\rv}{\vectstraight{r}}
\newcommand{\qv}{\vectstraight{q}}
\newcommand{\Fv}{\vectstraight{F}}
\newcommand{\Bv}{\vectstraight{B}}
\newcommand{\vv}{\vectstraight{v}}
\newcommand{\hv}{\vectstraight{h}}
\newcommand{\dv}{\vectstraight{d}}
\newcommand{\av}{\vectstraight{a}}
\newcommand{\bv}{\vectstraight{b}}
\newcommand{\zv}{\vect{0}}

\newcommand{\Mc}{\mathcal{M}}
\newcommand{\Hc}{\mathcal{H}}
\newcommand{\Kc}{\mathcal{K}}

\newcommand{\Tr}[1]{\mathrm{Tr} \left\{ {#1} \right\} }

\newcommand{\colvec}[1]{\begin{pmatrix}#1 \end{pmatrix}}
\newcommand{\rowvec}[1]{\begin{pmatrix}#1 \end{pmatrix}}

\newcommand{\ld}{\vect{\lambda}} %
\newcommand{\lds}{\lambda} %

\newcommand{\del}{\partial}
\newcommand{\dg}{\dagger}

\newcommand{\qa}{\ket{\qv A}}
\newcommand{\qb}{\ket{\qv B}}
\newcommand{\qp}{\ket{\qv +}}
\newcommand{\qm}{\ket{\qv -}}
\newcommand{\qpm}{\ket{\qv \pm}}

\newcommand{\Mqc}{\mathcal{M}_{\qv}}
\newcommand{\Mp}{\mathcal{M}_{+}}
\newcommand{\Wq}{W_{\qv}}
\newcommand{\Wqd}{W^{\dagger}_{\qv}}
\newcommand{\Wpd}{W^{\dagger}_{+}}
\newcommand{\Wp}{W_{+}}
\newcommand{\diag}{\text{diag}\colvec}

\newcommand{\re}{\mathrm{Re}} %
\newcommand{\im}{\mathrm{Im}} %
\newcommand{\vac}{\ket{\text{vac}}}
\newcommand{\phq}{\varphi_{\qv}}
\newcommand{\thq}{\theta_{\qv}}

\newcommand{\ccite}[1]{%
\IfSubStr{#1}{,}{refs.}{ref.~}\cite{#1}%
}
\newcommand{\Ccite}[1]{%
\IfSubStr{#1}{,}{Refs.~}{Ref.~}\cite{#1}%
}

\makeatletter
\def\maketitle{
    \@author@finish
	\title@column\titleblock@produce
	\suppressfloats[t]}
    \makeatother
    
\makeatletter
\begin{document}

\title{Quantum geometry of bosonic Bogoliubov quasiparticles}
\newcommand{\ITPTUB}{Institut f\"{u}r Physik und Astronomie, Technische Universit\"{a}t Berlin,
Hardenbergstr.~36, D-10623 Berlin, Germany}

\author{Isaac Tesfaye\,\orcidlink{0009-0001-4194-3916}}
\email{i.tesfaye@tu-berlin.de}
\author{Andr\'e Eckardt\,\orcidlink{0000-0002-5542-3516}}
\email{eckardt@tu-berlin.de}
\affiliation{\ITPTUB}

\begin{abstract}
    Bosonic Bogoliubov de Gennes (BBdG) Hamiltonians describe the excitations of weakly interacting Bose condensates as well as photonic systems under parametric driving. Their
    topological features have been studied mainly by utilizing a generalized symplectic version of the Berry curvature and related Chern numbers. 
    However, a full characterization of geometrical features in BBdG systems is still lacking. 
    Here, we propose a symplectic quantum geometric tensor (SQGT),
    whose imaginary part leads to the previously studied symplectic Berry curvature, 
    while the real part gives rise to a symplectic quantum metric, 
    providing a natural distance measure in the space of bosonic Bogoliubov modes. The SQGT is directly related to observable properties of BBdG systems.  
    We show how to measure all components of the SQGT by extracting excitation rates 
    in response to periodic modulations of the systems'
    parameters.
    Moreover, we connect the symplectic Berry curvature to a generalized 
    symplectic anomalous velocity term for Bogoliubov-Bloch wave packets.
    We test our results for a bosonic Bogoliubov-Haldane model. 
\end{abstract}

\date{\today}

\maketitle

\hypersetup{linkcolor=blue} %
\emph{Introduction.}---Bosonic Bogoliubov-de Gennes (BBdG) Hamiltonians naturally arise 
    in systems of weakly interacting bosons~\cite{Bogolyubov1947,Colpa1978,Kawaguchi2012,Ozeri2005}, 
    or in photonic systems subjected to parametric driving, generating squeezing of light~\cite{Clerk2010, Peano2016,Peano2016a,Smirnova2020,Shi2017,Ozawa2019}.
    They also describe magnonic~\cite{Shindou2013,Shindou2013a,McClarty2022}, phononic~\cite{Zhang2010a}, 
    and mechanical systems~\cite{Wollman2015}.
    Different from their fermionic (BCS) counterparts, BBdG systems are diagonalized in a paraunitary or symplectic manner~\cite{Colpa1978,Xiao2009}. 
    Their topological features~\cite{Hasan2010,Qi2011} have first been studied in the context of magnonic crystals~\cite{Shindou2013a,Shindou2013} 
    by utilizing a generalized symplectic Berry curvature and Chern number for Bogoliubov bands. 
    A generalization of the conventional bulk-boundary correspondence~\cite{Hatsugai1993,Hatsugai1993a} to the BBdG case can even lead to novel chiral edge states with properties distinct from their particle-number-conserving counterparts~\cite{Peano2016,Barnett2013,Galilo2015,Engelhardt2015,Furukawa2015}. 
    Recent works, building on the ten-fold Altland-Zirnbauer (AZ) classification of topological insulators~\cite{Altland1997,Ryu2010,Schnyder2008,Kitaev2009,Chiu2016}, %
    corroborate the notions of the symplectic Berry curvature and Chern number~\cite{Peano2018,Lein2019} 
    and provide a topological classification of quadratic BBdG systems~\cite{Zhou2020,Gong2018,Kawabata2019,Zhou2019,Lieu2018,Bernard2002}.
    
    Although topological features of BBdG systems have recently received some attention
    ~\cite{Wang2021a,Jalali-mola2023,Furukawa2015,Engelhardt2015,Kondo2019,Lieu2018,Ohashi2020,Bardyn2016,Peano2016a,McDonald2018,Okuma2023a,Wan2021,Chaudhary2021,Xu2020,Massarelli2022,DiLiberto2016a,Flynn2020a,Huang2021a,Huang2022a,Chen2023a,Ravets2025,Julku2021,Julku2021a,Julku2023,Salerno2023}, 
    the geometry of BBdG systems is still lacking a complete characterization. 
    In particle-number-conserving systems, the Berry curvature is related to the imaginary part of the so-called quantum geometric tensor (QGT)~\cite{Resta2011,Kolodrubetz2017}
    and the real part of the QGT, known as quantum metric, provides a natural distance measure in state space~\cite{Provost1980,Kolodrubetz2017}. 
    \begin{figure}[bt!]
        \begin{center}
            \includegraphics{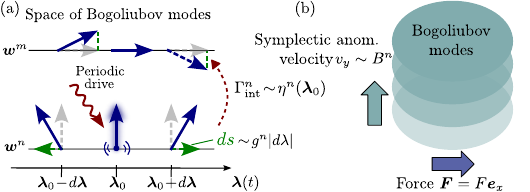}
            \caption{(a) Illustration of the symplectic quantum metric providing a distance 
                    measure $ds$ between infinitesimally close Bogoliubov modes $\vect{w}(\ld_0)$ and $\vect{w}(\ld_0+d\ld)$
                    [cf.~\cref{eq:distance-def-sympl-final-SQGT}]. 
                    The integrated excitation rate $\Gamma^n_{\text{int}}$ of a 
                    Bogoliubov system in response to (weak) periodic perturbations is 
                    governed by the symplectic quantum geometry
                    [cf.~\cref{eq:ETDPT-integrated-rate-SQG,eq:ETDPT-integrated-rate-offdiag}].
                    (b) Illustration of the transverse velocity a Bogoliubov Bloch wave packet 
                    experiences in response to an external force,
                    which is proportional to the symplectic Berry curvature [cf.~\cref{eq:ADPT-anomal-corretion-velocity-final}].
                    }
            \label{fig:Intro-Sketchp}
        \end{center}
    \end{figure}
    Here, we generalize these concepts to the non-number-conserving BBdG case and propose 
    a symplectic quantum geometric tensor (SQGT)~
    \footnote{We note that non-hermitian generalizations of the quantum geometric tensor have also recently been put forward~\cite{Zhang2019,Zhu2021,ChenYe2024}, 
    although not explicitly applied to the case of bosonic BdG systems.}. 
    We show that the imaginary part of the SQGT is related to the 
    symplectic Berry curvature investigated previously and that its real part, which we dub symplectic quantum metric, defines a natural distance measure in the space of bosonic Bogoliubov modes.
    Previous proposals for measurable signatures of the topology of BBdG systems have only focused on global properties, such as probing bosonic chiral edge modes for non-zero symplectic Chern numbers~\cite{Peano2016,Furukawa2015}.
    Here, inspired by~\Ccite{Tran2017,Ozawa2018}, we propose, for the first time, to measure the full local SQGT
    from excitation rates to other Bogoliubov modes in response to weak periodic driving [\cref{fig:Intro-Sketchp}(a)].
    As a further probe, we also show that, 
    analogous to the number-conserving case~\cite{Xiao2010,Sundaram1999,Chang1995}, 
    the symplectic Berry curvature naturally arises as a transverse velocity term, 
    when a Bogoliubov Bloch wave packet is subjected to an external force [\cref{fig:Intro-Sketchp}(b)]. 

\emph{Framework.}---
    Consider a 
    BBdG Hamiltonian
        \begin{align}
            \hat{H}&=\sum_{i,j=1}^{N}  K_{ij}\hat{a}_i^\dagger \hat{a}_j 
            + \frac{1}{2}\left(G_{ij}\hat{a}_i^\dagger\hat{a}_j^\dagger +G_{ji}^{*}\hat{a}_i\hat{a}_j \right)
            \nonumber \\
            &=\frac{1}{2}\hat{\Psi}^\dagger\mathcal{M}\hat{\Psi} 
            -\frac{1}{2}\Tr{K}, \  
            \mathcal{M}=\begin{pmatrix}
                    K & G\\
                    G^* & K^*
                \end{pmatrix},
            \label{eq:BdG-Ham}
        \end{align}
   with $K_{ij} = K_{ji}^* \in \mathbb{C}$ (hermitian),  $G_{ij} = G_{ji} \in \mathbb{C}$ (symmetric) forming 
   the hermitian $2N\times 2N$ matrix $\mathcal{M}$~\footnote{Throughout this letter, we only consider thermodynamically stable systems, i.e., 
    the coefficient matrix $\mathcal{M}$ is positive definite~\cite{Peano2018},
    such that a diagonalization of the BdG Hamiltonian~\eqref{eq:BdG-Ham} 
    satisfying the bosonic commutation relations 
    with strictly positive eigenvalues 
    is always possible~\cite{Flynn2020,Colpa1978,Simon1999,Rossignoli2005}.}. 
   and Nambu spinor $\hat{\Psi}=(\hat{a}_1 \, \cdots \, \hat{a}_N \, \hat{a}_1^\dagger \, \cdots \, \hat{a}_N^\dagger)^T$ 
   of bosonic annihilation  (creation) operators $\hat{a}^{(\dg)}_i$~\cite{Colpa1978,Xiao2009,Flynn2020}.
    The Hamiltonian can be diagonalized~\cite{Colpa1978,Xiao2009} (see also supplementary material (SM)~\cite{Note5}),
        \begin{align}
            \hat{H}=\frac{1}{2}\hat{\Phi}^\dagger W^{\dg} \mathcal{M} W\hat{\Phi}-\frac{1}{2}\Tr{K}
            =\sum_{n=1}^{N}\omega_n \hat{b}_n^\dg \hat{b}_n+ C , %
            \label{eq:BdG-Ham-diag}  
        \end{align}
     by a Bogoliubov transformation 
    $\hat{\Psi}=W\hat{\Phi}$, 
    where
    $\hat{\Phi}=(\hat{b}_1 \, \cdots \, \hat{b}_N \, \hat{b}_1^\dagger \, \cdots \, \hat{b}_N^\dagger)^T$
    and
  \begin{align}
        W=\begin{pmatrix} U & V^* \\ V & U^* \end{pmatrix},
        \text{ with }
        \left\{\begin{array}{l}
        W^\dg \tau_z W=W\tau_z W^\dg =\tau_z,\\
        W^\dg \mathcal{M} W = \Omega.\end{array}\right.
        \label{eq:paraunitary-cond}
    \end{align}
    Here, $C\equiv \sum_n \omega_n/2 - \Tr{K}/2$; $U$ and $V$ are complex $N\times N$ matrices, 
    so that $\hat{a}_i=\sum_{n}[U_{in} \hat{b}_n+(V^*)_{in}\hat{b}^\dg_n]$. 
    We only consider thermodynamically stable systems where $\Mc$ is positive definite such that 
    $\Omega =\text{diag}(\omega_1,\dots,\omega_N,\omega_1,\dots,\omega_{N})$ consists of strictly positive excitation energies, $\omega_n >0$~\cite{Peano2018,Flynn2020}. 
    The transformation $W$ is paraunitary, as indicated by the relation involving the diagonal matrix $\tau_z$, where $\tau_i=\sigma_i \otimes \mathbbm{1}_N$ with Pauli matrices $\sigma_i,\,i\in\{x,y,z\}$ and the $N\times N$ identity matrix $\mathbbm{1}_N$.
    This ensures that the new operators $\hat{b}^{(\dg)}_n$ fulfill bosonic commutation relations. 
    Representing the BdG Hamiltonian~\eqref{eq:BdG-Ham} via position-momentum quadratures $(\hat{q}_j,\hat{p}_j)=((\hat{a}_j+\hat{a}_j)/\sqrt{2},\,i(\hat{a}_j^\dg-\hat{a}_j)/\sqrt{2})$ leads to symplectic transformation matrices $\tilde{W} \in \mathrm{Sp}(2N,\mathbb{\mathbb{R}})$~\cite{Ferraro2005,Peano2018}, thence the commonly used label ``symplectic''~\cite{Engelhardt2015,Peano2016} (see also SM~\cite{Note5}).
    The Bogoliubov ground state $|\psi_0\rangle$ is the quasiparticle vacuum, $\hat{b}_n \ket{\psi_0}=0 \,\forall n$. It is generally a multi-mode squeezed state with a fluctuating non-zero number of $\hat{a}_i$ bosons~\cite{Xu2020,Blaizot1986,Peano2016}. 
    Here we will focus on the geometrical properties of the quasiparticle excitations on top of $|\psi_0\rangle$~\cite{Peano2016}.

    \emph{Symplectic quantum geometric tensor.}---
    The Bogoliubov modes are associated with the vectors $\vect{w}_n$
    given by the columns of $W=(\vect{w}^1,\cdots,\vect{w}^{2N})$. 
    For $n\le N$, they describe quasiparticles, $\hat{b}^\dg_n= \hat{\Psi}^\dg \tau_z \vect{w}^n$, and for $N>n$ to quasiholes, $\hat{b}_n= -\hat{\Psi}^\dg \tau_z \vect{w}^{N+n}$~\cite{Barnett2013,Xu2020,Note5}. 
    They play a role akin to single-particle wavefunctions~\cite{Peano2016}
    and solve the eigenvalue equation~\cite{Note5},
        \begin{align}    
             D\vect{w}^n =\tilde{\Omega}_{n} \vect{w}^n,
             \text{ with } D\equiv \tau_z\mathcal{M} 
             \text{ and } \tilde{\Omega}_n= (\tau_z\Omega)_{nn},       
            \label{eq:gen-eigenvect-eq-vector}
        \end{align} 
where the dynamical matrix $D$ is pseudo-hermitian, $D=\tau_z D^\dg \tau_z$~\cite{Schulz-Baldes2017,Peano2018,Lein2019,Xiao2009},
and obeys particle-hole symmetry (PHS) $D=-\tau_x D^* \tau_x$~\footnote{We note that the particle-hole symmetry (PHS) should correctly be called a particle-hole \emph{constraint} as it does not describe an actual symmetry of BBdG systems~\cite{Lein2019} (see also SM~\cite{Note5}). Nevertheless, we still use the terminology PHS to connect to previous literature~\cite{Peano2018,Peano2016,Engelhardt2015,Furukawa2015}}, so that the excitation energies $\tilde{\Omega}_{n}= s_n\omega_{n}$ are positive ($s_n=+1$) for quasiparticles ($n\le N$) and negative ($s_n=-1$) for quasiholes ($n>N$), with $\omega_{n+N}=-\omega_n$. 
The Bogoliubov modes $\bm{w}^n$ obey the generalized $\tau_z$ inner product structure~\cite{Schulz-Baldes2017} $(\vect{w}^m)^{\dg}\tau_z \vect{w}^n=(\tau_z)_{mn}=\delta_{mn}s_n$~\cite{Note5}. 
Thus, the projector $P^n$ onto the $n$th Bogoliubov mode and its orthogonal complement $Q^n$ are naturally defined as~\cite{Shindou2013}
        \begin{align}
            P^n=W \Gamma^n W^{-1}=W \Gamma^n \tau_z W^\dg \tau_z, 
            \;
            Q^n=\mathbbm{1}_{2N}-P^n,
            \label{eq:projector}    
        \end{align}
where $(\Gamma^n)_{lm}=\delta_{nl}\delta_{lm}$. 
Then $P^n P^m=\delta_{nm} P^n$ and $\sum_{n=1}^{2N} P^n=\mathbbm{1}_{2N}$~\cite{Shindou2013}.
Both $P^n$ and $Q^n$ are pseudo-hermitian, $\tau_z X^n \tau_z=(X^n)^\dg$ for $X^n\in \{P^n,Q^n\}$.
The projectors onto the particle and hole parts are not independent but also related via the PHS, $P^n=\tau_x(P^{n+N})^*\tau_x$~\cite{Note5}.

Let the Hamiltonian~\eqref{eq:BdG-Ham} now depend on 
a set of dimensionless parameters, $\ld =(\lds^1,\ldots,\lds^r)$, $\hat{H}=\hat{H}(\ld)$ and have, for simplicity, a non-degenerate spectrum $\tilde{\Omega}_{n}\ne\tilde{\Omega}_{m}, \forall \lds$ 
(the generalization to the degenerate case is straightforward).
Then, motivated by the definition of the 
conventional quantum geometric tensor (QGT)
~\cite{Resta2011,Provost1980},
we define the %
symplectic quantum geometric tensor (SQGT) as
        \begin{align}
            \eta^n_{\mu \nu}=\Tr {\del_{\mu} P^n Q^n \del_{\nu} P^n}, 
            \label{eq:SQGT-def}
        \end{align}
where $\del_{\mu} \equiv \del / \del \lambda^{\mu}$.
The SQGT is $U(1)$ gauge-invariant, since the projectors $P_n$ and $Q_n$ 
do not depend on the phase of the modes $\vect{w}^n(\ld)$. 
The pseudo-hermiticity of $P^n$ and $Q^n$~\eqref{eq:projector}, 
implies that the SQGT is hermitian,
$(\eta^n_{\mu \nu})^* =\eta^n_{\nu \mu}$. 
    By employing completeness, $\sum_n P^n=\mathbbm{1}_{2N}$, and orthogonality,
    $(\vect{w}^m)^{\dg}\tau_z \vect{w}^n=\delta_{mn}s_n$, it can be expressed as~\cite{Note5}
        \begin{align}
            \eta^n_{\mu \nu}=s_n (\del_{\mu}\vect{w}^n)^{\dg} \tau_z Q^n\del_{\nu} \vect{w}^n.
            \label{eq:SQGT-Ketform}
        \end{align} 
    In the SM we also derive the spectral representation~\cite{Note5}
        \begin{align}
            \eta^n_{\mu \nu}=s_n \sum_{m\ne n}s_m
            \frac{(\vect{w}^n)^{\dg} (\del_{\mu}\mathcal{M})\vect{w}^m
             (\vect{w}^m)^\dg (\del_{\nu}\mathcal{M})\vect{w}^n}
            { (\tilde{\Omega}_n - \tilde{\Omega}_m)^2}.
            \label{eq:SQGT-SpectralForm}
        \end{align} 

A first indication that the so-defined SQGT~\eqref{eq:SQGT-def} is meaningful, 
is that its imaginary part is proportional to the previously defined \emph{symplectic Berry-curvature} 
$B_{\mu \nu}^n=is_n[(\del_{\mu} \vect{w}^n)^\dg {\tau_z}(\del_{\nu} \vect{w}^n)-(\del_{\nu} \vect{w}^n)^\dg {\tau_z}(\del_{\mu} \vect{w}^n)]$~\cite{Shindou2013,Shindou2013a,Furukawa2015},
    \begin{align}
        B^n_{\mu \nu}=-2 \im [\eta^n_{\mu\nu} ]=-2\im \left[ \Tr{\del_{\mu} P^n Q^n \del_{\nu} P^n } \right],
        \label{eq:Sympl-Berry-curv}
    \end{align}
which is real-valued and antisymmetric, $B^n_{\mu \nu}=-B^n_{\nu \mu}$.
From this quantity, a symplectic (first) Chern number $C^n \equiv\frac{1}{2\pi}\int_S B^n$ 
was defined for closed two-dimensional parameter spaces $S$~\cite{Shindou2013,Shindou2013a} and found to describe a generalized bulk-boundary correspondence for BBdG systems 
~\cite{Peano2016,Barnett2013,Galilo2015,Engelhardt2015,Furukawa2015}. 
We show in the SM~\cite{Note5} that the symplectic Berry curvature admits a local conservation law, i.e., $\sum_n B^n_{\mu\nu}=0$. 
Different from the conservation law for particle-number conserving systems, 
this implies that the sum of the Berry curvature over the particle subspace can be non-zero, and may only be compensated by the hole contributions. 
This situation typically occurs in driven photonic BBdG systems, where the Hamiltonian~\eqref{eq:BdG-Ham} 
is dynamically stable \footnote{A BBdG system is called dynamic stable if all eigenvalues of the dynamical matrix $D$ are real-valued, which implies a bounded time evolution of the system.
Note that thermodynamic stability, which follows from a positive definite Hamiltonian 
\eqref{eq:BdG-Ham}, is a stronger condition which implies dynamical stability \protect\cite{Peano2018,Chaudhary2021}}.
As a consequence, the sum of the Chern numbers over the particle subspace can be non-zero, which is distinct from the conventional case~\cite{Chaudhary2021}.
We, moreover, show below that the symplectic Berry curvature 
describes the anomalous (transverse) velocity induced by an external force in Bogoliubov Bloch bands. 

We now demonstrate that also the real part of the SQGT,
        \begin{align}
            g^n_{\mu \nu}=\re[\eta^n_{\mu\nu} ]=\re \left[\Tr {\del_{\mu} P^n Q^n \del_{\nu} P^n } \right],
            \label{eq:SQG-def}
        \end{align}
carries physical meaning. We dub it \emph{symplectic quantum metric}, since it directly generalizes the conventional quantum metric to the BBdG case. It is symmetric, $g^n_{\mu \nu}=g^n_{\nu \mu}$.
Note that the PHS relates the particle and hole parts of the SQGT to each other via $\eta^n_{\mu \nu}=(\eta^{n+N}_{\mu \nu})^*$~\cite{Note5}, implying that the symplectic quantum metric~\eqref{eq:SQG-def} is the same for particles and holes, $g^n_{\mu \nu}=g^{n+N}_{\mu \nu}$, whereas the symplectic Berry curvature~\eqref{eq:Sympl-Berry-curv} switches sign, $B^n_{\mu \nu}=-B^{n+N}_{\mu \nu}$.

Unlike the symplectic Berry curvature, 
which is only non-zero for systems with broken time-reversal or inversion symmetry~\cite{Xiao2010}, 
the symplectic quantum metric can already be non-zero without these constraints. 

The conventional quantum metric (Fubini-Study metric)~\cite{fubini1904sulle,Provost1980,Aharonov1987,Anandan1990,Resta2011,Kolodrubetz2017} is given by the real part of the conventional QGT 
and defines a natural distance measure in the (projective) Hilbert space. 
It is used to characterize quantum phase transitions~\cite{Zanardi2007,CamposVenuti2007,Rezakhani2010,Carollo2020,Liu2020,Ma2013,Hauke2016}, excitations rates, and quantum fluctuations~\cite{Kolodrubetz2017,Ozawa2018,Ozawa2019a} 
(properties which have been exploited to measure the quantum metric in various experimental setups 
~\cite{Yi2023,Yu2022,Yu2024b,Tan2019,Yu2020,Gianfrate2020,Zheng2022,Cuerda2024}).

The symplectic quantum metric $g_{\mu\nu}$ naturally describes the distance $ds^2$ between two infinitesimally close Bogoliubov modes $\vect{w}(\vect{\lambda})$ and $\vect{w}(\vect{\lambda}+d\vect{\lambda})$, 
\begin{align}
            ds^2 %
            =1-\abs{\vect{w}^\dg(\ld) \tau_z \vect{w}(\ld+d\ld) }^2
 =g_{\mu \nu} d\lambda^{\mu}d\lambda^{\nu},
        \label{eq:distance-def-sympl-final-SQGT}
\end{align}
with summation over repeated indices. 
Here we first defined $ds^2$ via the fidelity $|\vect{w}^\dg(\ld) \tau_z \vect{w}(\ld\!+\!d\ld)|^2$  between both modes corresponding to the inner-product structure described above. The second identity follows from expanding $\vect{w}(\vect{\lambda}\!+\!d\vect{\lambda})$ to second order in $d\ld$, giving
$ds^2=\{(\vect{w}^\dg \tau_z \vect{w}) [(\del_{\mu}\vect{w})^\dg  \tau_z (\del_{\nu}\vect{w})]-(\del_{\mu}\vect{w})^\dg \tau_z \vect{w}\vect{w}^\dg \tau_z (\del_{\nu}\vect{w})\} d\lambda^{\mu}d\lambda^{\nu}=\eta_{\mu \nu} d\lambda^{\mu}d\lambda^{\nu}
=\frac{1}{2}(\eta_{\mu \nu}+ \eta_{\nu\mu} )d\lambda^{\nu}d\lambda^{\mu}
=g_{\mu\nu}d\lambda^{\mu}d\lambda^{\nu}$~\cite{Note5}.

\emph{Relation to linear response and probing.}---
    We will now show that the SQGT determines the transition rates of Bogoliubov excitations 
    between different modes in linear response to weak external forcing. 
    This provides another physical interpretation of this quantity and 
    opens the door for measuring its components similarly to the protocols that have been used to extract 
    the quantum geometry of free fermions~\cite{Tran2017, Ozawa2018, Ozawa2019a, Asteria2019}.

Assuming a quasiparticle in a superposition of (unperturbed) Bogoliubov modes, 
$\vect{\psi}(t)=\sum_{n=1}^N\psi_n(t)\vect{w}^n(\vect{\lambda_0})\in \mathbb{C}^{2N}$, 
its dynamics is described by the effective time-dependent Schr{\"o}dinger equation
$ -i \del_t \vect{\psi}(t)=D(t) \vect{\psi}(t) $
~\cite{Xu2020} 
(see SM
    \footnote{See supplementary material, which includes Refs.
   ~\cite{Colpa1978,Xiao2009,
    Flynn2020,Ueda2010,Pethick2008,Derezinski2017,Bognar1974,Ferraro2005,Resta2011,Schulz-Baldes2017,Lein2019,Xu2020,Shindou2013,Xiao2009,MacKay1987,Provost1980,Kolodrubetz2017,Zhang2006,Diu2019,Tran2017,Ozawa2018,Engelhardt2015,Karplus1954,Altland1997,Chiu2016,Zhou2020,Rigolin2008,DeGrandi2010,Berry1984,Peano2016,Peano2016a,Kolodrubetz2013,Ohashi2020,Bracci1975,Sundaram1999,Chang1995,Price2016,Xiao2010,Furukawa2015,Haldane1988,Grossmann2016,Dalfovo1999,Wu2017,Blaizot1986,Kawaguchi2012,Simon1999,Peano2018,Schulz-Baldes2017,Williamson1936,Nicacio2021,Hauke2014,Stamper-Kurn1999,Brunello2000,Bloch2008,Vogels2002,Gurarie2003,Serafini2023,Xiao2005}, 
    where we provide details regarding: the paraunitary Bogoliubov transformation, 
    the symmetries in BBdG systems and Krein space structure,
    the associated indefinite inner product space, 
    the derivation of alternative representations of the SQGT, 
    the comparison between the SQGT and QGT,
    the PHS of the SQGT,
    the symplectic quantum metric as an infinitesimal distance measure, 
    the local conservation law for the symplectic Berry curvature, 
    the derivation of the equation of motion for Bogoliubov modes 
    and its solution using time-dependent perturbation theory, 
    the application of the measurement protocol to BBdG lattice systems, 
    the multi-quasiparticle case in Fock space, adiabatic perturbation theory of the equation of motion, 
    the derivation of the symplectic anomalous velocity, 
    the Bogoliubov-Haldane model including its topological classification and symmetries, 
    and observable signatures for the original particles.}),
with $D(t)$ being the dynamical matrix~\eqref{eq:gen-eigenvect-eq-vector}, which becomes time-dependent in response to external forcing 
(external forces generally also induce terms in the Hamiltonian that do not conserve the number of Bogoliubov quasiparticles, 
which are not captured by $D(t)$, which will be discussed below).

We assume driving in the form of a sinusoidal modulation 
$\lambda^\nu=\lambda^\nu_0+2(A_{\nu}/\omega)\cos(\omega t -\varphi_{\nu})$ 
of the system parameters around a mean value $\vect{\lambda}_0$. 
For weak driving ($A_{\nu}/\omega\ll1$), the resulting time dependence of the dynamical matrix, $D(t)=D[\vect{\lambda}(t)]$, 
is captured by the linear approximation $D[\ld(t)]\simeq D(\ld_0)+\sum_\nu\del_{\nu}D(\ld_0) 2(A_\nu/\omega)\cos(\omega t-\varphi_\nu)$. 
For the sake of simplicity inspired by~\Ccite{Ozawa2018}, 
we focus on a two-parameter drive with relative phase $\varphi$, for which 
        \begin{align}
            D[\ld(t)]=&D(\ld_0)+\del_{1}D(\ld_0) 2(A_1/\omega)\cos(\omega t) \nonumber \\
            &\pm\del_{2}D(\ld_0) 2(A_2/\omega)\cos(\omega t-\varphi).
            \label{eq:dynamical-matrix-periodic-modulation}
        \end{align}
The driving amplitudes $A_\nu$ shall be switched on suddenly at time $t=0$, 
and we assume, for simplicity, that the system is prepared in mode $n$, 
$\vect{\psi}(0)=\vect{w}^n(\vect{\lambda}_0)$.
Applying time-dependent perturbation theory in first order (linear response theory), we obtain the probabilities 
        \begin{align}
                p^{\pm}_{mn}(\omega, t)=&\abs{(\vect{w}^m)^\dg 
                ( A_1\del_1\mathcal{M}\pm A_2 e^{i\varphi}\del_2\mathcal{M} )\vect{w}^n}^2
                \nonumber \\
                &\times ( 2\pi t/\omega^2)\delta(\tilde{\Omega}_{mn}-\omega).
            \label{eq:ETDPT-probability}
        \end{align}
for finding the system in state $\vect{w}^m$ at time $t$, where $\tilde{\Omega}_{mn} \equiv \tilde{\Omega}_{m}-\tilde{\Omega}_n$. Note that all quantities are evaluated for $\vect{\lambda}=\vect{\lambda}_0$, whenever the argument $(\vect{\lambda})$ is dropped, i.e., $\vect{w}^m\equiv\vect{w}^m(\vect{\lambda}_0)$ and $\tilde{\Omega}_{m}\equiv\tilde{\Omega}_{m}(\vect{\lambda}_0)$.
The total excitation rate  $\Gamma(\omega)$ into all other Bogoliubov modes $m\ne n$ is then given by 
$\Gamma^{n,\pm}(\omega)=t^{-1}\sum_{m\neq n}s_m p^{\pm}_{mn}(\omega,t)$. 
For an initial quasiparticle state, $n\le N$, the sum can be restricted to 
$m\le N$, since there is no quasiparticle-to-quasihole transition for a single-quasiparticle initial state~\cite{Note5}.  

To probe the SQGT, we consider the total excitation rate integrated over all (relevant) 
probing frequencies $\omega$~\cite{Tran2017, Ozawa2018}, $\Gamma^n_{\text{int}}=\int d \omega\, \Gamma^{n,\pm}(\omega)$. 
For the modulation of a single parameter ($A_1\equiv A$, $A_2=0$), we find
        \begin{align}
            \Gamma^n_{\text{int}}
            =2\pi A^2 
            \sum_{m\neq n}          \frac{\abs{(\vect{w}^m)^\dg(\del_1\mathcal{M})\vect{w}^n}^2}
            {s_m  (\tilde{\Omega}_{m}-\tilde{\Omega}_{n})^2} %
            =2\pi A^2 g^n_{1 1}. %
        \label{eq:ETDPT-integrated-rate-SQG}
        \end{align}
For the second identity, we have used the spectral representation~\eqref{eq:SQG-def} of the SQGT, 
whose diagonal elements correspond to the symmetric part given by the symplectic quantum metric, $\eta_{\nu\nu}^n=g_{\nu\nu}^n$. 
Thus, we can probe these diagonal elements by measuring the frequency integrated response 
to a modulation of $\lambda^\nu$. 

Next, we now consider the driving of both parameters $\lambda^1$ and $\lambda^2$ 
with two different phase lags $\varphi$
and equal amplitudes $A_1=A_2=A$, to obtain the frequency integrated transition rates
    \begin{align}
            \Gamma^{n,\pm}_{\text{int}}
            =2\pi A^2 
                \begin{cases}
                g^n_{1 1}\pm 2 g^n_{1 2} +g^n_{22},\!\!\! &\, \varphi =0\\
                g^n_{1 1}\mp B^n_{1 2} +g^n_{22},\!\!\! &\, \varphi =\pi/2,
            \end{cases}
    \end{align}
the differences of which,
        \begin{align}
            \Delta\Gamma^n_{\text{int}}\equiv \Gamma^{n,+}_{\text{int}}-\Gamma^{n,-}_{\text{int}}=\begin{cases}
                    \ \ 8\pi A^2 g^n_{1 2},\!\!\! &\, \varphi =0\\
                    -4\pi A^2 B^n_{1 2},\!\!\! &\, \varphi =\pi/2,
                \end{cases}
            \label{eq:ETDPT-integrated-rate-offdiag}
        \end{align}
are directly proportional to the real and imaginary part of the off diagonal element $\eta^n_{12}$ of the SQGT, 
given by $g^n_{12}$ and $-B_{12}^n/2$. 
In this way, all off-diagonal components $\eta^n_{\mu\nu}=g^n_{\mu\nu}-iB^n_{\mu\nu}/2$ can be extracted 
allowing for a complete reconstruction of the SQGT 
from quasiparticle scattering rates.

We can generalize the above results to initial states of multiple quasiparticles. 
A generic perturbation, 
$V(t)=\sum_{ij} \{A_{ij}(t) \hat{a}_i^\dagger \hat{a}_j + \frac{1}{2}[B_{ij}(t)\hat{a}_i^\dagger\hat{a}_j^\dagger+\text{h.c.}] \} + \sum_i[C_i(t)\hat{a}_i^\dagger+\text{h.c.}]$, 
takes the form $V(t)= \sum_{nm} \{A'_{nm}(t) \hat{b}^\dg_n \hat{b}_m+\frac{1}{2}[B'_{nm}(t) \hat{b}^\dg_n \hat{b}^\dg_m +\text{h.c.}]\} + \sum_n[C'_n(t)\hat{b}_n^\dagger+\text{h.c.}]+ D'(t)$ with respect to the unperturbed Bogoliubov operators $\hat{b}^{(\dg)}_n $. Here terms $\propto A'$ correspond to scattering between quasiparticle (or hole) states, whereas terms $\propto B'$ are associated with the scattering of a quasiparticle into a quasihole or vice versa. 
The terms $\propto C'$, which correspond to the creation of a single quasiparticle (or hole), 
can result from the scattering of an $\hat{a}_i$ boson out of (or into) a Bose condensate, 
which is described by a macroscopic wavefunction that is not in equilibrium with respect to the perturbation. 
All these processes are associated with corresponding bosonic enhancement factors.
To reconstruct the full SQGT, it is sufficient to consider the scattering of single quasiparticles. 
For discrete energies $\omega_n$, this can be achieved by using that the different processes 
can be addressed selectively by appropriate driving frequencies~\cite{Stamper-Kurn1999}. 
As illustrated in~\cref{fig:Ebog-Sp}(a), starting from the ground state, one can first exploit the $B'$ or $C'$ terms to populate the state $n$ 
and then modulate the system resonantly with to transitions $n\to m$ of $A'$ type 
to extract information about the SQGT.
\begin{figure}[bt!]
        \begin{center}
            \includegraphics{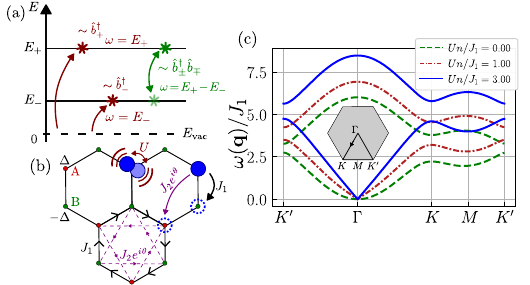}
            \caption{
            (a) Sketch of the measurement scheme: Selecting transitions by the frequency of the drive, 
            first the Bogoliubov mode $n$ is populated (red), second, a transition to another mode $m$ is induced (green). 
            (b) Illustration of the Bogoliubov-Haldane model.
            (c) Bogoliubov energy spectrum along high-symmetry path in the first Brillouin zone (inset) 
            for $J_2=0.1J_1$, $\theta=\pi/2$, $\Delta=3\sqrt{3}J_2/2$ and different $Un/J_1$.}
            \label{fig:Ebog-Sp}
        \end{center}
    \end{figure}

\emph{Application to Bogoliubov-Haldane model.}---
    As a concrete example, we study weakly interacting bosons in a two-dimensional hexagonal tight-binding lattice with two sublattice states, $s=A, B$, per unit cell. 
    The single-particle terms correspond to the 
    Haldane model~\cite{Haldane1988}, as it was recently realized with ultracold atoms in optical lattices~\cite{Jotzu2014, Asteria2019, Tarnowski2019}. 
    They are characterized by real (complex) tunneling matrix elements $-J_1$ ($-J_{2}e^{i\theta}$) between nearest (next-nearest) neighbors, 
    and on-site potentials $\Delta$ ($-\Delta$) on $A$ ($B$) sites [\cref{fig:Ebog-Sp}(b)]. 
    Due to the lattice symmetry, quasimomentum $\qv=(q_x,q_y)$ is conserved 
    by these processes, which is treated as a parameter in the following.
    By treating a weak repulsive on-site interaction $U$ at a filling of $n$ bosons per site in a 
    self-consistent Bogoliubov approximation 
    with quantum fluctuations on top of a Gross-Pitaevskii mean-field condensate, 
    we obtain a Bogoliubov-Haldane model~\cite{Note5}, 
    $\hat{H}^{(B)}=\sum_{\qv\neq\zv}\sum_ {s,s'} \hat{a}^{\dg}_{\qv s}[\mathcal{H}(\qv)+\mathcal{H}_1]_{ss'}\hat{a}^{\dg}_{\qv s'}+(\hat{a}^{\dg}_{\qv s}[\mathcal{H}_2]_{ss'}\hat{a}^{\dg}_{-\qv s'} +\text{h.c.})$, 
    Here, $\mathcal{H}(\qv)+\mathcal{H}_1$ contains the single-particle tight-binding Hamiltonian 
    including a mean-field interaction contribution, 
    and $\mathcal{H}_2$ stems exclusively from the interactions which 
    in experiments with ultracold bosons are tunable via Feshbach resonances~\cite{Chin2010}. 
    The Bogoliubov spectrum is shown in~\cref{fig:Ebog-Sp}(c).
\begin{figure}[bt!]
        \begin{center} 
            \includegraphics[scale = 1.0]{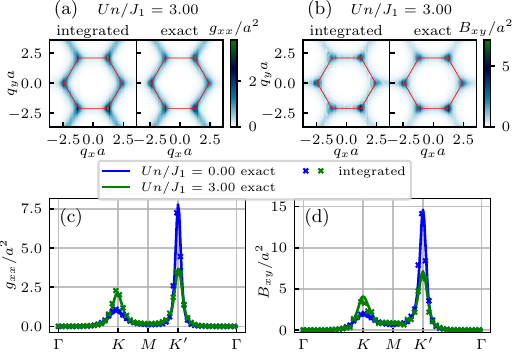}
            \caption{Symplectic quantum metric component $g_{xx}(\qv)$ 
            and Berry curvature $B_{xy}(\qv)$ for the upper particle band of the Bogoliubov-Haldane model~\protect\cite{Note5} obtained via integrated excitation rates~\eqref{eq:ETDPT-integrated-rate-SQG}-\eqref{eq:ETDPT-integrated-rate-offdiag} and via exact diagonalization~\eqref{eq:SQGT-def}.
            (a) Symplectic quantum metric component $g_{xx}(\qv)$ for $Un/J_1=3$ shown in 2D quasimomentum space from numerical integration of excitation rates (left) and exact diagonalization (right).
            (b) Same as in (a) but for the symplectic Berry curvature $B_{xy}(\qv)$.
            (c) Exact (solid line) and integrated (crosses) results for the symplectic quantum metric component $g_{xx}(\qv)$ along a high-symmetry path (cf.~inset in~\cref{fig:Ebog-Sp}(c)) for two interactions strengths 
            $Un/J_1=0$ (non-interacting Haldane model) and $Un/J_1=3$.
            (d) Same as in (c) but for the symplectic Berry curvature $B_{xy}(\qv)$.
            Simulation parameters: $A/J_1=0.01$, $J_1 t/\hbar=10$ (final integration time), 
            $\hbar\omega/J_1\in [0.1 (0.2),5.5]$ for $Un/J_1=3\,(0)$ 
            with a frequency spacing of $\delta \omega=0.05$.
            Other parameters are given in~\cref{fig:Ebog-Sp}.
            }
            \label{fig:Int-Rate-GxxBxy}
        \end{center}
    \end{figure}

    We test our proposal for extracting the symplectic quantum geometric tensor from excitation rates to other Bogoliubov modes by simulating the time evolution 
    for initially populated Bogoliubov modes corresponding to the upper particle band.
    The parameter modulation~\eqref{eq:dynamical-matrix-periodic-modulation}
    is induced by periodically shaking the Bogoliubov-Haldane model along the lattice directions $x$ and $y$. 
    In the reference frame co-moving with the lattice, the corresponding inertial forces %
    directly induce modulations of the parameters $q_x$ and $q_y$, respectively~\cite{Note5}. 
    In Figs.~\ref{fig:Int-Rate-GxxBxy}(a)-(b) we present the exact symplectic quantum metric 
    $g_{xx}^+(\qv )$ and Berry curvature $B_{xy}^+(\qv )$ of the upper quasiparticle for $Un/J_1=3$
    together with the numerically integrated excitation rates obtained from the time evolution~\cite{Note5}. 
    In Figs.~\ref{fig:Int-Rate-GxxBxy}(c)-(d) we compare the exact and integrated values for the same quantities along a high-symmetry path in quasimomentum space for $Un/J_1=0$ (corresponding to the bare number-conserving Haldane model) and $Un/J_1=3$.
    We find very good agreement between the exact and integrated symplectic quantum geometric quantities,
    validating our scheme for extracting the components of the SQGT. 
    Results for %
    $g^+_{xx}(\qv )$ and $g^+_{xy}(\qv )$ are presented in the SM~\cite{Note5}.
    Experimentally, the Bogoliubov excitation fraction can be measured by relating 
    the presence of Bogoliubov quasiparticles to the quasimomentum distribution of the original particles, 
    which can be observed using time-of-flight imaging techniques~\cite{Vogels2002,Brunello2000,Note5}.

    \emph{Symplectic anomalous velocity.}---
    We show that, like in the number-conserving case~\cite{Xiao2010,Sundaram1999,Chang1995,Zhang2006}, 
    a Bogoliubov lattice system acquires an anomalous transverse velocity proportional to the symplectic Berry curvature and transverse to an applied homogeneous force $\Fv$.
    Consider a quasiparticle prepared in the Bogoliubov-Bloch mode $\vect{w}^n(\qv_0)$. 
    In response to the force, the quasimomentum becomes $\qv (t)=\qv_0+\Fv t$, inducing a modulation of the dynamical matrix $D[\qv (t)]$. In first-order adiabatic perturbation theory with respect to $\Fv$, the time evolved state is given by~\cite{Note5} 
     \begin{align}
               \vect{w}^n(\qv ) 
                +i\sum_{m \neq n}
                \vect{w}^m(\qv )
                \frac{(\vect{w}^m(\qv ) )^\dg (\nabla_{\qv }\mathcal{M})\vect{w}^n(\qv )\cdot \Fv}
                {s_m[\tilde{\Omega}_{n}(\qv )-\tilde{\Omega}_{m}(\qv )]^2}.
            \label{eq:ADPT-anomal-corretion-state}
        \end{align} 
 The expectation value of the Bogoliubov velocity operator $\vv(\qv )=\nabla_{\qv}D(\qv)$ then reads 
        \begin{align}
                \ep{\vv(\qv )}
                =\nabla_{\qv} \tilde{\Omega}_n(\qv)+\Bv_n(\qv)\times \Fv,
            \label{eq:ADPT-anomal-corretion-velocity-final}
        \end{align}    
    with $\Bv_n(\qv)\equiv\epsilon_{\mu\nu\kappa}B^n_{\mu\nu}(\qv)\vect{e}_\kappa$.
    The first term corresponds to the slope of the occupied Bogoliubov energy band, 
    and the second term provides the symplectic anomalous transverse velocity.
    This result is complementary to that of Ref.~\cite{Zhang2006}, where the dynamics of Bogoliubov Bloch wave packets has also been studied using a time-dependent variational principle, but without the connection to the symplectic Berry curvature.
    The effect of the symplectic anomalous velocity can be verified in semiclassical dynamics of Bogoliubov-Bloch wave packets or by measuring the modified thermal depletion of a BEC due to the presence of the symplectic Berry curvature~\cite{Zhang2006}.

\emph{Conclusion.}---
    We have introduced the manifestly gauge-invariant symplectic quantum geometric tensor (SQGT) for BBdG systems, whose imaginary part is proportional to the symplectic Berry curvature.
    The real part of the SQGT defines a symplectic quantum metric, which is shown to provide a natural distance measure in the space of Bogoliubov modes.
    All components of the symplectic quantum geometric tensor can be measured via tracking Bogoliubov excitation rates 
    and the symplectic Berry curvature describes an anomalous velocity in Bogoliubov Bloch systems. 
    State-of-the-art experiments with bosonic ultracold atoms~\cite{Cooper2019, Gross2017, Schafer2020, Bloch2008}
    are well-suited to test our proposals.
    For future work, it will be interesting to connect recent topological notions for mixed quantum states~\cite{Bardyn2018,Wawer2021a,Budich2015} 
    to thermal BBdG systems and 
    to see whether Euler invariants~\cite{Bouhon2020a,Unal2020,Ahn2019} 
    may be generalized to the BBdG context. 
    
    \makeatletter %
    \let\oldaddcontentsline\addcontentsline
    \def\addcontentsline#1#2#3{}
    \makeatother
    
    \begin{acknowledgments} %
    
    We thank Christof Weitenberg, Luca Asteria, Nathan Harshman and Alexander Schnell for fruitful 
    and insightful discussions.
    Part of the work has been conducted using the QuSpin project~\cite{Weinberg2017a,Weinberg2019}.
    This research was funded by the Deutsche Forschungsgemeinschaft via 
    the Research Unit FOR 2414 under Project No. 277974659.
    I.\ T.\ acknowledges support from the Studienstiftung des deutschen Volkes.
    \end{acknowledgments} %
    
    \makeatletter
    \let\addcontentsline\oldaddcontentsline
    \makeatother

\bibliography{main}

\hypersetup{linkcolor=blue}
\clearpage
\newpage

\makeatletter
\renewcommand{\theequation}{S\arabic{equation}}
\renewcommand{\thetable}{S\arabic{table}}
\renewcommand{\thefigure}{S\arabic{figure}}
\renewcommand{\thesection}{S\arabic{section}}
\renewcommand{\theHequation}{supp.\arabic{equation}}
\renewcommand{\theHtable}{supp.\arabic{table}}
\renewcommand{\theHfigure}{supp.\arabic{figure}}
\renewcommand{\theHsection}{supp.\arabic{section}}

\makeatother
\setcounter{equation}{0}
\setcounter{figure}{0}
\setcounter{table}{0}
\setcounter{page}{1}
\title{Supplementary Material for \\ ``Quantum geometry of bosonic Bogoliubov quasiparticles''}
\maketitle

\onecolumngrid
\tableofcontents
\newpage

\hypersetup{linkcolor=blue} %

\section{Paraunitary Bogoliubov transformation from bosonic commutation relations}
{\label{app:BCR-constraints}} 
    First, for ease of reference, we restate the bosonic BdG Hamiltonian [\cref{eq:BdG-Ham} in the main text] here 
        \begin{align}
            \hat{H}=\sum_{i,j=1}^{N}  K_{ij}\hat{a}_i^\dagger \hat{a}_j 
            + \frac{1}{2}\left(G_{ij}\hat{a}_i^\dagger\hat{a}_j^\dagger +G_{ji}^{*}\hat{a}_i\hat{a}_j \right)
            =\frac{1}{2}\hat{\Psi}^\dagger\mathcal{M}\hat{\Psi} 
            -\frac{1}{2}\Tr{K}, \  \mathcal{M}=\begin{pmatrix}
                K & G\\
                G^* & K^*
            \end{pmatrix}.
            \label{eq:supp:BdG-Ham}
        \end{align}
    The hermiticity of $\hat{H}=\hat{H}^\dg$ imposes $K$ to be hermitian, i.e., $K_{ij} = K_{ji}^* \in \mathbb{C}$, 
    and the bosonic commutation relations for $\hat{a}_i^{(\dg)}$ lead to $G$ being symmetric, i.e., 
    $G_{ij} = G_{ji} \in \mathbb{C}$~\cite{Colpa1978,Xiao2009,Flynn2020}.
    In the second equality, the BdG Hamiltonian is reformulated in terms of the bosonic Nambu spinors $\hat{\Psi}^{(\dg)}$,
    \begin{align}
        \hat{\Psi}^\dg=\rowvec{\hat{a}_1^\dg  & \cdots & \hat{a}_N^\dg & \hat{a}_1 & \cdots & \hat{a}_N}, \quad
        \hat{\Psi}=\rowvec{
            \hat{a}_1 & \cdots & \hat{a}_N & \hat{a}_1^\dagger & \cdots & \hat{a}_N^\dagger}^T, 
            \label{eq:supp:Nambu-spinor}
        \end{align}
    and the $2N\times2N$ coefficient matrix $\mathcal{M}$ consisting of $N\times N$ matrix blocks $K$ and $G$.
    Here and in the following we denote hermitian conjugation by $^\dg$, the transposition by $^T$, and the complex conjugation by $^*$.

    The bosonic operators $\hat{a}^{(\dg)}_i$ in~\eqref{eq:supp:BdG-Ham} preserve the bosonic commutation relations (BCR), 
        \begin{align}
            [\hat{a}_i,\hat{a}_j]=0, \quad [\hat{a}_i,\hat{a}^{\dagger}_j]=\delta_{ij}, 
            \Leftrightarrow \,
            [\hat{\Psi}_i,\hat{\Psi}^{\dagger}_j]=(\tau_z)_{ij}, \quad
            \tau_z = \begin{pmatrix}
                \mathbbm{1}_N & 0_N\\
                0_N & -\mathbbm{1}_N
             \end{pmatrix},  
             \ 
            \label{eq:supp:BCR}
        \end{align}
    where in the last equation, we have reformulated the bosonic commutation relations 
    via the Nambu spinors~\eqref{eq:supp:Nambu-spinor} 
    with $\tau_z = \sigma_z \otimes \mathbbm{1}_N$, where $\sigma_z=\diag{1,-1}$ is the Pauli-z matrix and 
    $\mathbbm{1}_N$ ($0_N$) is the $N\times N$ identity (zero) matrix.
    Now, our goal is to find a (Bogoliubov) transformation from the original set of bosonic operators $\hat{a}^{(\dg)}_i$ 
    to a new set of bosonic operators $\hat{b}^{(\dg)}_n$ that diagonalizes the Hamiltonian~\eqref{eq:supp:BdG-Ham}.

    Let us first discuss the constraints on this transformation. 
    Importantly, this transformation must preserve the BCR~\eqref{eq:supp:BCR}, i.e., the BCR must also hold 
    for the new bosonic operators $\hat{b}^{(\dg)}_n$, 
        \begin{align}
            [\hat{b}_n,\hat{b}_m]=0, \quad [\hat{b}_n,\hat{b}^{\dg}_m]=\delta_{nm}, 
            \Leftrightarrow \
            [\hat{\Phi}_n,\hat{\Phi}^{\dagger}_m]=(\tau_z)_{nm},
            \label{eq:supp:BCR-b}
        \end{align}
    where we defined $\hat{\Phi}=(\hat{b}_1\, \cdots \,\hat{b}_N \, \hat{b}^{\dagger}_1 \,\cdots \, \hat{b}^{\dagger}_N)^T$ and 
    $\hat{\Phi}^\dg=(\hat{b}^\dg_1\, \cdots\,\hat{b}^\dg_N \, \hat{b}_1 \, \cdots \,\hat{b}_N)$
    as the bosonic Nambu spinors~\eqref{eq:supp:Nambu-spinor} for the new bosonic operators $\hat{b}^{(\dg)}_n$.

    We parametrize such a general Bogoliubov transformation from $\hat{a}^{(\dg)}_i$ to $\hat{b}^{(\dg)}_n$ by~\cite{Blaizot1986,Colpa1978,Kawaguchi2012}
        \begin{align}
            &\hat{a}_i=\sum_{n}U_{in} \hat{b}_n+(V^*)_{in}\hat{b}^\dg_n,\quad
            \hat{a}_i^\dg=\sum_{n}\hat{b}^\dg_n(U^\dg)_{ni} +\hat{b}_n(V^T)_{ni},
            \label{eq:supp:paraunitary-matrix-bog-transf-gen-0}\\
            &\hat{\Psi}= W \hat{\Phi}, \ \hat{\Psi}^\dg= \hat{\Phi}^\dg W^\dg ,
            \quad  W=\begin{pmatrix} U & V^* \\ V & U^* \end{pmatrix}, \ 
            W^\dg=\begin{pmatrix} U^\dg & V^\dg \\ V^T & U^T \end{pmatrix},
            \label{eq:supp:paraunitary-matrix-bog-transf-gen}
        \end{align}
    where in the second line we have reformulated the Bogoliubov transformation via a transformation 
    between the original and new Nambu spinors, $\hat{\Psi}$ and $\hat{\Phi}$ respectively.

    The requirement that also the $\hat{b}^{(\dg)}_n$ are bosonic operators~\eqref{eq:supp:BCR-b} will 
    impose constraints on the complex $2N\times 2N$ transformation matrix $W$,
    or equivalently on its constituents, the $N\times N$ complex matrices $U$ and $V$.
    Using~\eqref{eq:supp:paraunitary-matrix-bog-transf-gen-0} 
    these constraints from the BCR~\eqref{eq:supp:BCR}-\eqref{eq:supp:BCR-b} are given by
        \begin{align}
                \delta_{ij}=&[\hat{a}_i,\hat{a}^{\dg}_j]=
                \sum_{n,n'}[(U_{in} \hat{b}_n+(V^*)_{in}\hat{b}^\dg_n),
                (\hat{b}^\dg_{n'}(U^\dg)_{n'j} +\hat{b}_{n'}(V^T)_{n'j})]
                =\sum_{n}(U_{in}(U^\dg)_{nj}-(V^*)_{in}(V^T)_{nj}),\\
                0=&[\hat{a}_i,\hat{a}_j]=
                \sum_{n,n'}[(U_{in} \hat{b}_n+(V^*)_{in}\hat{b}^\dg_n),
                (U_{jn'} \hat{b}_{n'}+(V^*)_{jn'}\hat{b}^\dg_{n'})]
                =\sum_{n,}(U_{in}(V^*)_{jn}-(V^*)_{in}(U)_{jn}).
                \label{eq:paraunitary-cond-UV-comp}
            \end{align}
In matrix form, these conditions read 
        \begin{align}
            U U^\dg - (V^*)V^T = \mathbbm{1}_N, \qquad V^*U^T  - U V^\dg =0_N,
            \label{eq:supp:paraunitary-cond-UV}
        \end{align}
    where $\mathbbm{1}_N$ ($0_N$) is again the $N\times N$ identity (zero) matrix.
    We can also formulate the conditions for 
    the entire transformation matrix $W$~\eqref{eq:supp:paraunitary-matrix-bog-transf-gen} 
    by using BCR in terms of the Nambu spinors~\eqref{eq:supp:BCR}-\eqref{eq:supp:BCR-b}
        \begin{align}
            (\tau_z)_{ij}= [\hat{\Psi}_i,\hat{\Psi}^{\dagger}_j]=
            \sum_{m,m'}[(W_{im}\hat{\Phi}_m),(\hat{\Phi}^{\dagger}_{m'}(W^\dg)_{m'j})]=
            \sum_{m,m'}W_{im}(W^\dg)_{m'j} [\hat{\Phi}_m,\hat{\Phi}^{\dg}_{m'}]=
            \sum_{m,m'}W_{im}(\tau_z)_{mm'}(W^\dg)_{m'j}
            \label{eq:supp:paraunitary-cond-from-BCR-comp}.
        \end{align}
    In matrix form these conditions for the full transformation matrix $W$ read
        \begin{align}
            \tau_z = W\tau_z W^\dg =W^\dg \tau_z W,
            \label{eq:supp:paraunitary-cond-from-BCR}
        \end{align}
    which lead to~\cref{eq:paraunitary-cond} in the main text.
    Note that we have added a second equality in~\eqref{eq:supp:paraunitary-cond-from-BCR} 
    in addition to the first from~\eqref{eq:supp:paraunitary-cond-from-BCR-comp}.
    To see that the second equality holds, 
    note that the first equality, $\tau_z = W\tau_z W^\dg$, 
    is equivalent to $W^\dg = \tau_z W^{-1} \tau_z$.
    Then we immediately see that $W^\dg \tau_z W=(\tau_z W^{-1} \tau_z) \tau_z W =\tau_z$ 
    proving the second equality in~\eqref{eq:supp:paraunitary-cond-from-BCR}.
    The Bogoliubov transformation matrices $W$ which fulfill the constraints of~\cref{eq:supp:paraunitary-cond-from-BCR} are called
    \emph{paraunitary} matrices~\cite{Colpa1978}.

    Conditions~\eqref{eq:supp:paraunitary-cond-from-BCR} 
    further allow us to obtain the useful relations for the inverse transformation 
    of~\eqref{eq:supp:paraunitary-matrix-bog-transf-gen}, from the new 
    bosonic operators $\hat{b}^{(\dg)}_n$ to the original bosonic operators $\hat{a}^{(\dg)}_i$ 
    via 
        \begin{align}
            \hat{b}_n&=\sum_{i}(U^\dg)_{ni}\hat{a}_{i}-(V^\dg)_{ni}\hat{a}^\dg_{i}, \quad  
            \hat{b}^{\dg}_n=\sum_{i}\hat{a}^{\dg}_{i}U_{in}-\hat{a}_{i}V_{in},
            \label{eq:supp:paraunitary-matrix-bog-transf-INV-gen-0}\\
            \hat{\Phi}&= W^{-1} \hat{\Psi}, \ 
            \hat{\Phi}^\dg= \hat{\Psi}^\dg (W^{\dg})^{-1}=\hat{\Psi}^\dg \tau_z W\tau_z , 
            \label{eq:supp:paraunitary-matrix-bog-transf-INV-gen} \\
            W^{-1}&=\tau_z W^{\dg}\tau_z=\begin{pmatrix}
                U^{\dg} & -V^{\dg}\\
                -V^T & U^T
                \end{pmatrix},\ 
                (W^{\dg})^{-1}=\tau_z W \tau_z=\begin{pmatrix}
                    U & -V^{*}\\
                    -V & U^*
                    \end{pmatrix}.
            \label{eq:supp:paraunitary-matrix-INV-rel-2}
        \end{align}
    If we denote the $n$-th column of $W=(\vect{w}^1,\cdots,\vect{w}^{2N})$ by $\vect{w}^n$, 
    we can directly see from~\eqref{eq:supp:paraunitary-matrix-bog-transf-INV-gen-0}
    that the new bosonic quasiparticle operators $\hat{b}^{(\dg)}_n$ can be compactly expressed 
    as 
        \begin{alignat}{3}
            \hat{b}_n&=(\vect{w}^n)^\dg \tau_z \hat{\Psi}, \quad
            &&\hat{b}^\dg_n= \hat{\Psi}^\dg \tau_z \vect{w}^n,\quad &&\text{for}\ n\leq N, \ \text{or equivalently} \\
            \hat{b}_n&= -\hat{\Psi}^\dg \tau_z \vect{w}^{N+n}, \quad
            &&\hat{b}^\dg_n= -(\vect{w}^{N+n})^\dg \tau_z \hat{\Psi}, \quad &&\text{for}\ n > N.
            \label{eq:supp:bog-operators-compact}
        \end{alignat}

    Having discussed the constraints on the Bogoliubov transformation matrix $W$, we turn to the diagonalization
    of the BdG Hamiltonian~\eqref{eq:supp:BdG-Ham}.
    With the Bogoliubov transformation~\eqref{eq:supp:paraunitary-matrix-bog-transf-gen} 
    the Hamiltonian~\eqref{eq:supp:BdG-Ham} can be written in terms of the
    new bosonic quasiparticle operators $\hat{b}^{(\dg)}_n$,
        \begin{align}
            \hat{H}
            =\frac{1}{2}\hat{\Phi}^\dg W^\dg \mathcal{M} W \hat{\Phi}  -\frac{1}{2}\Tr{K}.
            \label{eq:supp:BdG-Ham-diag-pre}
        \end{align}
    For the Hamiltonian~\eqref{eq:supp:BdG-Ham} to become diagonal 
    in $\hat{b}^{(\dg)}_n$, the coefficient matrix $\mathcal{M}$
    must become diagonal via the Bogoliubov transformation,
            \begin{align}
                W^\dg \mathcal{M} W=\Omega =\begin{pmatrix}
                    \vect{\omega} & 0\\
                    0 & \vect{\omega}
                    \end{pmatrix}, \quad
                    \vect{\omega} = \diag{\omega_1,\ldots,\omega_N},
            \label{eq:supp:BdG-Ham-diag-cond}
            \end{align}
    where $\Omega$ is the $2N\times 2N$ diagonal matrix consisting of the $N\times N$ 
    diagonal matrix $\vect{\omega}$ with Bogoliubov excitation energies $\omega_n$.
    As dictated by an adapted version of Williamson's theorem~\cite{Simon1999,Williamson1936,Nicacio2021} to the paraunitary case, such a diagonalization~\eqref{eq:supp:BdG-Ham-diag-cond} 
    via Bogoliubov matrices $W$ fulfilling~\eqref{eq:supp:paraunitary-cond} 
    is always possible for a positive definite coefficient matrix $\mathcal{M}$~\cite{Colpa1978,Blaizot1986}.
    In this work, we only consider \emph{thermodynamically stable} systems, 
    where the coefficient matrix $\mathcal{M}$ is strictly positive definite~\cite{Peano2018}.

   \cref{eq:supp:BdG-Ham-diag-cond} leads to a generalized eigenvalue equation~\cite{Colpa1978,Xiao2009}
        \begin{align}    
        \begin{split}
            \mathcal{M} W&=(W^{\dg})^{-1}\Omega=\tau_z W\tau_z \Omega=\tau_z W\tilde{\Omega}, \
            \text{with}\ \tilde{\Omega} \equiv \tau_z\Omega =
            \begin{pmatrix}
                \vect{\omega} & 0 \\
                0 & -\vect{\omega}
            \end{pmatrix},
        \end{split}
        \\
        \Leftrightarrow \ 
        D W &= W\tilde{\Omega} , \
        \text{with}\ D \equiv \tau_z \mathcal{M},
        \label{eq:supp:gen-eigenvect-eq-matrix}
        \end{align}
    which, rewritten in terms of the Bogoliubov modes $\vect{w}^n$ (given by the $n$th column of $W$), reads
        \begin{align}
            D\vect{w}^n =\tilde{\Omega}_n \vect{w}^n, \ \text{with} \ 
            D=\tau_z \mathcal{M} = \begin{pmatrix}
                K & G\\
                -G^* & -K^*
            \end{pmatrix},
            \label{eq:supp:gen-eigenvect-eq-vector}
        \end{align} 
    Here $D\equiv \tau_z \mathcal{M}$ is the so-called dynamical matrix of the system and $\vect{w}^n$
    is the $n$-th Bogoliubov mode corresponding 
    to the excitation energy $\tilde{\Omega}_n=s_n\omega_{n \bmod \! N}$~\cite{Shindou2013,Xiao2009}.
    The dynamical matrix $D$ also determines the time evolution in BBdG system (cf.~\cref{app:Effective-TDSE}).
    From~\cref{eq:supp:gen-eigenvect-eq-vector} we see that the spectrum $\Omega$ is entirely determined by the eigenvalues of the \emph{pseudo-hermitian} matrix $D=\tau_z D^\dg \tau_z$.

    Inserting~\eqref{eq:supp:BdG-Ham-diag-cond} into~\eqref{eq:supp:BdG-Ham-diag-pre} then eventually leads to the diagonal form of the full Hamiltonian~\eqref{eq:supp:BdG-Ham}
        \begin{align}
            \hat{H}
            =\frac{1}{2}\hat{\Phi}^\dg \Omega \hat{\Phi}-\frac{1}{2}\Tr{K} 
            =\sum_{n=1}^{N} \omega_{n}
            \left(\hat{b}_{n}^\dg \hat{b}_{n} + \frac{1}{2}\right) -\frac{1}{2}\Tr{K}
            = \sum_{n=1}^{N} \omega_{n} \hat{b}_{n}^\dg \hat{b}_{n} + C,
            \label{eq:supp:BdG-Ham-diag}
        \end{align}
    with $C\equiv \sum_n \omega_n/2 - \Tr{K}/2$.
    The ground state $\ket{\psi_0}$ of this Hamiltonian~\eqref{eq:supp:BdG-Ham-diag}
    is generally given by a multi-mode squeezed state with a non-zero number of original 
    $\hat{a}^{(\dg)}_i$ bosons, which is also called the Bogoliubov vacuum.
    This Bogoliubov vacuum is defined by the condition that it contains 
    no Bogoliubov quasiparticle excitations, 
    i.e., $\hat{b}_n\ket{\psi_0}=0$ for all $n$~\cite{Ueda2010,Pethick2008}.

\section{Symmetries and Krein space perspective}
\label{app:Krein-space}
    Here, we briefly discuss the symmetries and the inherent (real) Krein space structure of 
    the BBdG system~\eqref{eq:supp:BdG-Ham}, mainly following \Ccite{Peano2018,Schulz-Baldes2017}.
    First, we note that the BdG Hamiltonian $\hat{H}$~\eqref{eq:supp:BdG-Ham} acts on 
    the entire bosonic Fock space $\mathcal{F}(\Hc)$, where $\Hc$ is the finite-dimensional complex single-particle Hilbert space~\cite{Peano2018}.
    The Hamiltonian $\hat{H}$ can be cast into a block diagonal (Nambu spinor) representation, giving rise to the linear coefficient matrix $\mathcal{M}$~\eqref{eq:supp:BdG-Ham}, which acts 
    on a so-called particle-hole Hilbert space $\mathcal{H}_{\text{ph}}=\mathcal{H}\otimes \mathbb{C}^2$.
    The matrix $\mathcal{M}$, consists matrices $K$ and $G$~\eqref{eq:supp:BdG-Ham}, which themselves act on $\Hc$~\cite{Peano2018}.
    
    To analyze the symmetries of BBdG systems, we first introduce three additional operators which act on $\mathcal{H}_{\text{ph}}$,
            \begin{align}
                \tau_x=\sigma_x \otimes \mathbbm{1}_N = \begin{pmatrix}
                    0 & \mathbbm{1}_N\\
                    \mathbbm{1}_N & 0
                \end{pmatrix}, \quad
                \tilde{\tau}_y=-i\tau_y=-i \sigma_y \otimes \mathbbm{1}_N = \begin{pmatrix}
                    0 & -\mathbbm{1}_N\\
                    \mathbbm{1}_N & 0
                \end{pmatrix}, \quad
                c = \frac{1}{\sqrt{2}}\begin{pmatrix}
                    \mathbbm{1}_N & -i\mathbbm{1}_N\\
                    \mathbbm{1}_N & i\mathbbm{1}_N
                \end{pmatrix}.
            \label{eq:supp:aux-symm-op}
            \end{align}
    All operators of~\eqref{eq:supp:aux-symm-op} together with $\tau_z$ are unitary.
    The operators $\tau_x$ and $\tau_z$ square to the identity, 
    whereas $\tilde{\tau}_y$ squares to minus the identity, i.e., 
    $\tau_x^2=\tau_z^2=\mathbbm{1}_{2N}=-\tilde{\tau}_y^2$.
    The operators $\tau_i$ are just like generalized Pauli matrices, with the same properties, i.e.,
    $\tau_i \tau_j = \delta_{ij}\mathbbm{1}_{2N}+i\epsilon_{ijk}\tau_k$.
    The last operator $c$ is called a Cayley transform~\cite{Peano2018} and 
    is used to translate between different representations of the BdG Hamiltonian~\eqref{eq:supp:BdG-Ham}, which will become clearer 
    once the position-momentum quadrature representation of the BdG Hamiltonian is introduced 
    (cf.~\cref{subsec:symp-origin}).

    We know that the bosonic BdG spectrum of~\eqref{eq:supp:BdG-Ham} does not follow 
    from the spectrum of the hermitian coefficient matrix $\Mc=\Mc^\dg$ 
    but rather from the non-hermitian dynamical matrix $D\equiv \tau_z \mathcal{M}$, 
    which also determines the dynamics of the system~\cite{Peano2018,Lein2019}.
    Hence, we will analyze the symmetries of the dynamical matrix $D$.

    The coefficient matrix $\Mc$~\eqref{eq:supp:BdG-Ham} possesses the \emph{partice-hole symmetry} (PHS) 
    $\Mc=\tau_x \Mc^* \tau_x$, which translates to the PHS of the dynamical matrix
        \begin{align}
            \tau_x D^*\tau_x =-D.
            \label{eq:supp:PHS-DynMat}
        \end{align}
    We note that the PHS should, more correctly, actually be called a particle-hole constraint, as it does not describe an actual symmetry of the system~\cite{Lein2019}. 
    This is further discussed in \ref{app:BH-model-symm-classf}.
    However, in the following we will still use the term PHS to connect to the previous literature~\cite{Peano2016,Peano2018}.
    Moreover, the dynamical matrix is also $\tau_z$-\emph{pseudo-hermitian},
            \begin{align}
                D^\dg=\tau_z D \tau_z,
                \label{eq:supp:PseudoHerm-DynMat}
            \end{align}
    which acts on the so-called \emph{Krein space} $(\mathcal{H}_{\text{ph}},\tau_z)$.
    Before we continue, we briefly collect some definitions and properties of Krein spaces 
   ~\cite{Schulz-Baldes2017,Peano2018}.
        \begin{definition}[Krein space]
            A Krein space $(\Kc,J)$ is a complex Hilbert space $\Kc$ equipped with a 
            so-called fundamental symmetry, a self-adjoint unitary $J$, which 
            fulfills $J^2=\mathbbm{1}$ and $J=J^\dg$.
            \label{def:Krein-space}
        \end{definition}
        Note that the fundamental symmetry $J$ further induces an indefinite scalar product 
        of the form $v,w \in \Kc \mapsto v^\dg J w$.
        Given an additional real structure $C$ (complex conjugation) on $\Kc$, 
        one can also define a \emph{real} Krein space as follows~\cite{Schulz-Baldes2017} 
        (a linear operator $A$ is called real if $CAC\equiv A^* = A$ holds).
        \begin{definition}[Real Krein space]
            A Real Krein space $(\Kc,J,S)$ of kind $(\eta,\zeta)=(\pm1,\pm1)$ 
            is a complex Krein space $(\Kc,J)$ with a real fundamental symmetry $J=J^*$ 
            and an additional real unitary $S$.
            The real unitary $S$ squares to $\eta \mathbbm{1}$
            and it either commutes ($\zeta=+1$) or anti-commutes ($\zeta=-1$) with $J$, i.e., $JS=\zeta SJ$.
            \label{def:Real-Krein-space}
        \end{definition}
    Here, we have we deal with a real Krein space $(\mathcal{H}_{\text{ph}},\tau_z,\tau_x)$ 
    of kind $(1,-1)$~\cite{Schulz-Baldes2017}, as $\tau_z$ is a real hermitian unitary and 
    $\tau_x$ is a real unitary which anti-commutes with $\tau_z$, i.e., $\tau_z\tau_x=-\tau_x\tau_z$.
    Hence, in our finite-dimensional case, $\text{dim}(\mathcal{H}_{\text{ph}})=2N$, the dynamical matrix $D$ is an element of the 
    Lie algebra
        \begin{align}
            \mathbb{D}(\Hc_{\text{ph}},\tau_z,\tau_x)&=
            \{D\in \mathbb{C}^{2N\times 2N} \,|\,
            D^\dagger = \tau_z D \tau_z, \, 
            \tau_x D^* \tau_x = -D\},\nonumber \\
            &=\left\{\begin{pmatrix}
                K & G\\
                -G^* & -K^*
            \end{pmatrix}\, | \,
            K=K^\dg\in \mathbb{C}^{N\times N}, \, G=G^T \in \mathbb{C}^{N\times N}\right\},
            \label{eq:supp:Lie-alg-DynMat}
        \end{align}
    where in the second equality, we have explicitly expressed the conditions 
    using the form of the coefficient matrix $\Mc=\tau_z D$~\eqref{eq:supp:BdG-Ham}.

    The $\tau_z$-\emph{paraunitary} (Bogoliubov) transformation matrices $W$ 
   ~\eqref{eq:supp:paraunitary-cond-from-BCR} are elements of the associated Lie group which are obtained by exponentiation $W=e^{iD}$,
        \begin{align}
            \mathbb{W}(\Hc_{\text{ph}},\tau_z,\tau_x)&=
            \{ W \in \text{GL}(2N,\mathbb{C}) \,|\, W^\dg \tau_z W = \tau_z, 
            \, \tau_x W^* \tau_x = W\},\nonumber \\
            &=\left\{\begin{pmatrix}
                U & V^*\\
                V & U^*
            \end{pmatrix}\, | \, U,V\in \text{GL}(N,\mathbb{C}),\
            UU^\dg - V^*V^T = \mathbbm{1}_N, \, V^*U^T - U V^\dg = 0_N\right\},
            \label{eq:supp:Lie-group-ParaUnit-DynMat}                
        \end{align}
    where in the second equality above we have explicitly expressed the conditions (cf.~\eqref{eq:supp:paraunitary-cond-UV}) in terms of elements of the general linear group $\text{GL}(2N,\mathbb{C})=\{A\in\mathbb{C}^{2N\times 2N} \,|\, \det(A)\neq 0\}$.

    Note that both properties of the Lie group $\mathbb{W}(\Hc_{\text{ph}},\tau_z,\tau_x)$ directly follow from the properties of the Lie algebra, i.e., 
    $W^\dg \tau_z W = e^{-iD^\dg} \tau_z e^{iD} = \tau_z e^{-iD}\tau_z \tau_z e^{iD} =\tau_z$ 
    and $\tau_x W^* \tau_x = \tau_x e^{-iD^*} \tau_x = e^{iD} = W$.

    Both the particle-hole-symmetry and the pseudo-hermiticity of the dynamical matrix $D$ also 
    determine the spectrum of $D$.
    From the particle-hole-symmetry~\eqref{eq:supp:PHS-DynMat} we can deduce that for every eigenvalue $\lambda$ there exists a corresponding negative complex conjugated eigenvalue $-\lambda^*$. 
    The pseudo-hermiticity~\eqref{eq:supp:PseudoHerm-DynMat} 
    implies that the eigenvalues of $D$ always come in complex conjugated pairs.
    Thus, when $D$ has complex eigenvalues they always come in quadrupoles ($-\lambda^*,\lambda^*,\lambda,-\lambda$). 
    However, for a spectrum that is purely real, as for a positive definite matrix $\Mc=\tau_z D$, they always come in pairs ($\lambda,-\lambda$)~\cite{Schulz-Baldes2017}.
    
    \subsection{On the origin of the term \emph{symplectic}}
    \label{subsec:symp-origin}
     
    We briefly explain the origin of the terminology \emph{symplectic} which has been established by previous literature e.g., for the symplectic polarization, symplectic Berry curvature and Chern number, and symplectic orthonormalization (cf.~\Ccite{Engelhardt2015,Peano2016,Peano2016a,Peano2018,Shindou2013}).

    To this end, we first note that the BdG Hamiltonian $\hat{H}$~\eqref{eq:supp:BdG-Ham} given in the $\hat{a}_j$ bosonic mode representation via the operators can also be equivalently expressed in the position-momentum quadrature representation $(\hat{q}_j,\hat{p}_j)$ via 
        \begin{align}
            \hat{a}_{j}^{\dagger}&=\frac{1}{\sqrt{2}}(\hat{q}_j+ i \hat{p}_j), \quad
            \hat{a}_{j}=\frac{1}{\sqrt{2}}(\hat{q}_j- i \hat{p}_j), \quad
            \hat{q}_{j}=\frac{1}{\sqrt{2}}(\hat{a}_j+  \hat{a}_j^{\dg}), \quad
            \hat{p}_{j}=\frac{i}{\sqrt{2}}(\hat{a}^\dg_j-  \hat{a}_j).
        \end{align}
    The position and momentum quadratures fulfill the canonical commutation relation $[\hat{q}_k,\hat{p}_l]=i\delta_{kl}$ (whereas the mode operators $\hat{a}_{k}^{(\dagger)}$ fulfill bosonic commutation relations $[\hat{a}_{i},\hat{a}_{j}^{\dagger}]=\delta_{ij}$~\eqref{eq:supp:BCR}).
    By writing the vector of mode operators (bosonic Nambu spinor) via $\hat{\Psi}^\dg\equiv (\hat{\bm{a}}^\dg,\hat{\bm{a}})$~\eqref{eq:supp:Nambu-spinor} and the $\hat{S}^T=(\hat{q}_1,\ldots,\hat{q}_N,\hat{p}_1,\ldots,\hat{p}_N)\equiv (\hat{\bm{q}},\hat{\bm{p}})$, we can see that the Cayley transform $c$~\eqref{eq:supp:aux-symm-op} allows us to switch between these two representations, 
    i.e., 
    \begin{align}
        \hat{\Psi}^\dg=({\bm{a}}^\dg,\hat{\bm{a}})=(\hat{\bm{q}},\hat{\bm{p}})\frac{1}{\sqrt{2}}\begin{pmatrix}
            \mathbbm{1}_N & \mathbbm{1}_N\\
            i\mathbbm{1}_N & -i\mathbbm{1}_N
        \end{pmatrix}=\hat{S}^Tc^\dg
        \label{eq:supp:Cayley-transform-Nambu}
    \end{align}
    or equivalently $\hat{\Psi}=c\hat{S}$.
    Similar to the commutation relations for the vector of mode operators reading $[\hat{\Psi}_i,\hat{\Psi}^{\dagger}_j]=(\tau_z)_{ij}$~\eqref{eq:supp:BCR}, we also have commutations relations for the vector of 
    position-momentum quadratures 
    \begin{align}
        [\hat{S}^T_k,\hat{S}_l]=-i(\tilde{\tau}_y)_{kl}=-i \begin{pmatrix}
            0 & -\mathbbm{1}_N\\
            \mathbbm{1}_N & 0
        \end{pmatrix}_{kl},
        \label{eq:supp:comm-rel-Sympl-pos-mom}
    \end{align} 
    with being $\tilde{\tau}_y$~\eqref{eq:supp:aux-symm-op} the standard symplectic form~\cite{Ferraro2005}.
    Using~\eqref{eq:supp:Cayley-transform-Nambu}, the BdG Hamiltonian~\eqref{eq:supp:BdG-Ham} in the position-momentum quadrature representation reads~\cite{Gurarie2003}
    \begin{align}
        \hat{H}&=\frac{1}{2}\hat{\Psi}^\dg\Mc\hat{\Psi}
        =\frac{1}{2}\hat{S}^T c^\dg \Mc c \hat{S}=\frac{1}{2}\hat{S}^T A \hat{S}, \quad A=\begin{pmatrix}
            B & E\\
            E^T & F
        \end{pmatrix},\nonumber \\
        &\text{with}\
        B=B^T=\re[K+G], \ E=\im[K-G], \ F=F^T=\re[K-G].
        \label{eq:supp:BdG-Ham-pos-mom-rep}
    \end{align}
    Here, $A=A^T \in \mathbb{R}^{2N\times2N}$ denotes the symmetric coefficient matrix 
    in the $(\hat{\bm{q}},\hat{\bm{p}})$-quadrature representation where $B$ and $F$ are real symmetric matrices and $E$ is a general real matrix.

    The paraunitary matrices $W$ diagonalizing the BdG Hamiltonian in the mode representation~\eqref{eq:supp:BdG-Ham} preserve the commutations relations of the mode operators~\eqref{eq:supp:BCR}.
    Here, to diagonalize $\hat{H}$ in this representation~\eqref{eq:supp:BdG-Ham-pos-mom-rep} we transform into a new set of quadrature operators $\hat{R}^T=(\hat{\bm{Q}},\hat{\bm{P}})$ via 
    $(\hat{\bm{q}},\hat{\bm{p}})=\hat{S}^T=\hat{R}^T\tilde{W}^T$.
    Importantly, the new quadrature operators must also fulfill $[\hat{R}^T_k,\hat{R}_l]=-i(\tilde{\tau}_y)_{kl}$~\eqref{eq:supp:comm-rel-Sympl-pos-mom}.
    Similarly, here, the transformation matrices $\tilde{W}$ diagonalizing the BdG Hamiltonian in the position-momentum quadrature representation~\eqref{eq:supp:BdG-Ham-pos-mom-rep} 
    must preserve the canonical (position-momentum) commutation relations~\eqref{eq:supp:comm-rel-Sympl-pos-mom}.
    
    This directly leads to
    \begin{align}
        -i(\tilde{\tau}_y)_{kl}=[\hat{S}^T_k,\hat{S}_l]=\sum_{m,n} \tilde{W}^T_{km}[\hat{R}^T_m,\hat{R}_n]\tilde{W}_{nl}=-i\sum_{m,n} \tilde{W}^T_{km}(\tilde{\tau}_y)_{mn}\tilde{W}_{nl},
    \end{align}
    which is equivalent to $\tilde{\tau}_y=\tilde{W}^T\tilde{\tau}_y\tilde{W}$, namely, 
    the defining properties of the real symplectic group $\tilde{W}\in \text{Sp}(2N,\mathbb{R})$~\eqref{eq:supp:Lie-group-DynMat-Cayley-Sp2N}.
    
    Now, the Cayley transform $c$ allows us to directly switch between paraunitary Lie group 
    $\mathbb{W}(\Hc_{\text{ph}},\tau_z,\tau_x)$~\eqref{eq:supp:Lie-group-ParaUnit-DynMat} 
    and the Lie group $\text{Sp}(2N,\mathbb{R})$ of the symplectic transformations.
    To see this, we apply the Cayley transform $c$~\eqref{eq:supp:aux-symm-op} 
    directly onto $\mathbb{W}(\Hc_{\text{ph}},\tau_z,\tau_x)$
        \begin{align}
            \tilde{\mathbb{W}}=c^\dg \mathbb{W} c=
            \left\{ \tilde{W} \in \text{GL}(2N,\mathbb{R}) \,|\, 
            \tilde{W}^\dg \tilde{\tau}_y \tilde{W} = \tilde{\tau}_y,\, \tilde{W}^*=\tilde{W} \right\}=\text{Sp}(2N,\mathbb{R}),
            \label{eq:supp:Lie-group-DynMat-Cayley-Sp2N}
        \end{align}
    which coincides with the real symplectic group $\text{Sp}(2N,\mathbb{R})$~\cite{Peano2018}.
    The elements of the real symplectic group, $\tilde{W}\in \text{Sp}(2N,\mathbb{R})$, 
    directly follow from applying the Cayley transform onto elements of 
    $\mathbb{W}(\Hc_{\text{ph}},\tau_z,\tau_x)$, 
    i.e., $\tilde{W}=c^\dg W c$ with $W\in \mathbb{W}(\Hc_{\text{ph}},\tau_z,\tau_x)$.
    Moreover, the defining properties of the symplectic group $\text{Sp}(2N,\mathbb{R})$ are
    directly inherited from the properties paraunitary group 
    $\mathbb{W}(\Hc_{\text{ph}},\tau_z\tau_x)$, i.e., 
    $\tilde{W}^\dg \tilde{\tau}_y \tilde{W} =(c^\dg W^\dg c)\tilde{\tau}_y(c^\dg W c)=-i c^\dg W^\dg \tau_z W c=-ic^\dg \tau_z c = \tilde{\tau}_y$ and 
    $\tilde{W}^*=c^T W^* c^* = c^T \tau_x W \tau_x c^*=c^\dg W c = \tilde{W}$.
    Here, we have used that the Cayley transform $c$~\eqref{eq:supp:aux-symm-op} 
    is unitary and
    allows us to cycle between different $\tau_i$ operators 
   ~\cite{Grossmann2016}, 
        \begin{align}
            c\tau_xc^\dg =\tau_y = c^\dg \tau_z c, \quad
            c \tau_y c^\dg =\tau_z = c^\dg \tau_x c, \quad
            c \tau_z c^\dg =\tau_x = c^\dg \tau_y c.
            \label{eq:supp:Pauli-Cayley-properties}
        \end{align}
    This explains the common use of the term \emph{symplectic} in the literature of BBdG systems, even though we and many in the literature~\cite{Engelhardt2015,Peano2016,Chaudhary2021} work in the bosonic mode 
    representation~\eqref{eq:supp:BdG-Ham}, where only paraunitary conditions 
   ~\eqref{eq:supp:paraunitary-cond} are present.
   
   We briefly discuss how, in the absence of pairing terms $G=0$ in~\eqref{eq:supp:BdG-Ham}, the paraunitary~\eqref{eq:supp:Lie-group-ParaUnit-DynMat} and symplectic structures~\eqref{eq:supp:Lie-group-DynMat-Cayley-Sp2N} reduce to simple unitary structures.
   For $G=0$, the coefficient matrix $\Mc$~\eqref{eq:supp:BdG-Ham} becomes block-diagonal, $\Mc=\begin{psmallmatrix} K & 0 \\ 0 & K^* \end{psmallmatrix}$, with $K=K^\dg$, such that doubling of the degrees of freedom to the particle-hole space $\mathcal{H}_{\text{ph}}$ is purely artificial.
   In this case, the problem simply reduces to a simple unitary diagonalization of the hermitian matrix $K=K^\dg$ via the unitary $U\in U(N,\mathbb{C})$.
   This can be seen by noting that the paraunitary transformation matrix $W$~\eqref{eq:supp:Lie-group-ParaUnit-DynMat} reduces to a block-diagonal unitary transformation 
   $W=\begin{psmallmatrix} U & 0 \\ 0 & U^* \end{psmallmatrix}$ with $V=0$ blocks vanishing~\eqref{eq:supp:paraunitary-matrix-bog-transf-gen-0} such that $W$ becomes a simple enlarged unitary matrix $W^\dg W=\mathbbm{1}_{2N}$~\eqref{eq:supp:paraunitary-cond-UV}.
    
   Now, for the $(\hat{\bm{q}},\hat{\bm{p}})$-quadrature representation of the bosonic BBdG system~\eqref{eq:supp:BdG-Ham-pos-mom-rep} with $G=0$, the coefficient matrix $A$ becomes 
        \begin{align}
            A=\begin{pmatrix} \re[K] & \im[K] \\ -\im[K] & \re[K] \end{pmatrix}=\begin{pmatrix} X & Y \\ -Y & X \end{pmatrix},
        \label{eq:supp:BdG-Ham-pos-mom-rep-G0}
        \end{align}
    which is just the embedding of the hermitian matrix $K=X+iY=K^\dagger$ into the real-valued symmetric matrix $A=A^T\in \mathbb{R}^{2N\times 2N}$, since $X=X^T$ and $Y=-Y^T$.
    Moreover, the symplectic transformation $\tilde{W}$ in the quadrature representation 
    can be computed via the Cayley transform $\tilde{W}=c^\dg W c$ and reads
        \begin{align}
            \tilde{W}=c^{\dagger}Wc=\frac{1}{2}\begin{pmatrix}
                \mathbbm{1}_N & \mathbbm{1}_N\\
                i\mathbbm{1}_N & -i\mathbbm{1}_N
            \end{pmatrix}
            \begin{pmatrix}
                U & 0\\
                0 & U^*
            \end{pmatrix}
            \begin{pmatrix}
                \mathbbm{1}_N & -i\mathbbm{1}_N\\
                \mathbbm{1}_N & i\mathbbm{1}_N
            \end{pmatrix}
            =
            \begin{pmatrix}
                \re[U] & \im[U]\\
                -\im[U] & \re[U]
            \end{pmatrix}.
            \label{eq:supp:Sympl-matrix-bog-transf-gen-0}
        \end{align}
    This is again a real embedding, here for the complex unitary $U=\alpha+i\beta \in U(N,\mathbb{C})$. 
    We note that since $U$ is unitary, $U^\dg U=\mathbbm{1}_N$, the real and imaginary parts $\alpha=\re[U]$ and $\beta=\im[U]$ fulfill $\alpha^T\alpha+\beta^T\beta=\mathbbm{1}_N$ and $\alpha^T\beta=\beta^T\alpha$.
    From this we can see that the real-valued symplectic transformation 
    $\tilde{W}$~\eqref{eq:supp:Sympl-matrix-bog-transf-gen-0} fulfills $\tilde{W}^T\tilde{W}=\mathbbm{1}_{2N}$, i.e., $\tilde{W}\in \mathrm{Sp}(2N,\mathbb{R})\cap \mathrm{O}(2N)$, which is the orthogonal symplectic group, a compact subgroup of the real symplectic group $\mathrm{Sp}(2N,\mathbb{R})$.
    Moreover, it follows directly from the map $f:U(N,\mathbb{C})\rightarrow \mathrm{Sp}(2N,\mathbb{R})\cap \mathrm{O}(2N)$ defined via $f(U)=\tilde{W}=c^{\dagger}W(U)c$~\eqref{eq:supp:Sympl-matrix-bog-transf-gen-0} that this real orthogonal symplectic group is isomorphic to the complex unitary group $\mathrm{Sp}(2N,\mathbb{R})\cap \mathrm{O}(2N)\simeq U(N,\mathbb{C})$~\cite{Serafini2023}.
    Hence, in the absence of pairing terms $G=0$, both the paraunitary transformation $W$ in the bosonic mode representation and the symplectic transformation $\tilde{W}$ in the position-momentum quadrature representation reduce to simple unitary transformations, as expected.
    
    \section{Indefinite Bogoliubov inner product space}
{\label{app:indefinite-inner-product}} 
    We collect some useful relations for the indefinite inner product space spanned by
    the Bogoliubov modes $\vect{w}^n$. 
    For a more rigorous introduction into these concepts pertaining indefinite inner product spaces, we refer to~\Ccite{Bognar1974}.
    Given a Bogoliubov transformation matrix $W\in \mathbb{C}^{2N\times 2N}$ which satisfies 
   ~\eqref{eq:supp:paraunitary-cond-from-BCR}
        \begin{align}
            W\tau_z W^\dg=W^\dg \tau_z W=\tau_z
            \label{eq:supp:paraunitary-cond}
        \end{align}
    and whose $n$-th column corresponds to the Bogoliubov mode $\vect{w}^n$, 
    we can define a projector $P^n$ onto the $n$-th non-degenerate Bogoliubov eigenspace 
    (and its complement $Q^n$) [\cref{eq:projector} in the main text] 
    \begin{align}
        P^n=W \Gamma^n W^{-1}=W \Gamma^n\tau_z W^\dg \tau_z=s_n\vect{w}^n(\vect{w}^n)^\dg\tau_z, \quad
        Q^n= \mathbbm{1}_{2N}-P^n,
        \label{eq:supp:projectors}
    \end{align}
    with $(\Gamma^n)_{ml}=\delta_{nm}\delta_{ml}$ being the $2N\times 2N$ matrix whose only non-zero element
    is the $n$-th diagonal element being equal to $+1$ 
    and where $s_n=1\,(-1)\, \text{for}\ n\leq N\, (n>N)$ for the so-called particle (hole) states.
    We note that these projectors $X^n$ are necessarily invariant 
    under (local) $U(1)$ gauge transformations for the Bogoliubov modes of the form
    $\vect{w}^n \rightarrow e^{i\phi}\vect{w}^n$, i.e., $X^n \rightarrow X^n$ 
    for $X^n\in\{P^n,Q^n\}$~\eqref{eq:supp:projectors}.

    Both $P^n$ and $Q^n$ are Krein-hermitian (pseudo-hermitian) 
    $(X^n)^\dg=\tau_z X^n \tau_z$ for $X^n\in \{ P^n,Q^n\}$~\cite{Schulz-Baldes2017,Lein2019}, 
    which directly follows from~\eqref{eq:supp:paraunitary-cond}.
    Moreover, the projectors are idempotent $P^n P^k=\delta_{nk} P^n$ and
    span the entire inner product space $\sum_n P^n=\mathbbm{1}_{2N}$.
    This allows for a resolution of the identity in the indefinite Bogoliubov inner product space,
        \begin{align}
            \mathbbm{1}_{2N}=\sum_{n=1}^{2N} P^n=\sum_{n=1}^{2N} s_n\vect{w}^n(\vect{w}^n)^\dg\tau_z 
            =\sum_{l=1}^N (P^l_{+}+P^l_{-})=\sum_{l=1}^N\left(\vect{w}^l_{+}(\vect{w}^l_+)^\dg-\vect{w}^l_{-}(\vect{w}^l_{-})^\dg\right)\tau_z,
            \label{eq:supp:res-identity-proj}
        \end{align}
    where in the third equality we have separated the projectors onto particle parts $P^l_{+}=\vect{w}^l_{+}(\vect{w}^l_+)^\dg\tau_z$ and hole parts $P^l_{-}=-\vect{w}^l_{-}(\vect{w}^l_{-})^\dg\tau_z$,
    which correspond to the first $N$ and last $N$ elements of the projector $P^n$ respectively.
    Furthermore, we note that the particle and hole parts are not independent but 
    are directly related via the particle-hole symmetry (PHS)~\eqref{eq:supp:Lie-group-ParaUnit-DynMat} 
    $\vect{w}^l_{-}=\tau_x (\vect{w}^l_{+})^*$, with $\tau_x$ defined in~\eqref{eq:supp:aux-symm-op}.
    Thus, the PHS also relates the particle and hole parts of the projectors via 
        \begin{align}
            P_{+}^l=\tau_x (P_{-}^l)^*\tau_x.
            \label{eq:supp:proj-PHS}
        \end{align}
   \cref{eq:supp:proj-PHS} also follows directly from~\eqref{eq:supp:projectors}, when using $\tau_x W^*\tau_x =W$~\eqref{eq:supp:Lie-group-ParaUnit-DynMat}, $\tau_x\tau_z=-\tau_z\tau_x$, and $\tau_x\Gamma^n\tau_x=\Gamma^{n+N}$.
    
    The Bogoliubov modes $\vect{w}^n$ adopt a generalized (pseudo) orthonormalization
    condition~\cite{Xu2020,Shindou2013,Xiao2009,Colpa1978} following from~\eqref{eq:supp:paraunitary-cond},
    \begin{align}
        (\vect{w}^n)^\dg \tau_z \vect{w}^m =s_n \delta_{nm}, \,
        \Leftrightarrow (\vect{w}^l_{\pm})^\dg{\tau_z}{\vect{w}^k_{\pm}}=\pm \delta_{lk}, \
        (\vect{w}^l_{\pm})^\dg{\tau_z}{\vect{w}^k_{\mp}}=0.
        \label{eq:supp:gen-eigenvect-normalization}
        \end{align}
    
    Next, note that the trace of a matrix $A\in \mathbb{C}^{2N\times 2N}$ can be evaluated 
    in any basis which spans the entire $\mathbb{C}^{2N}$.
    The Bogoliubov modes are such a basis which can be seen from 
    the resolution of the identity~\eqref{eq:supp:res-identity-proj}.
    Thus, the trace of a matrix $A$ can be expressed in terms of the Bogoliubov modes as
        \begin{align}
            \Tr{A}&=\sum_{n}\Tr{P^nA}=\sum_{n}\Tr{P^nAP^n}
            =\sum_{n,i}
            [(\vect{e}_i)^{\dg}\vect{w}^n (\vect{w}^n)^\dg \tau_z A \vect{w}^n (\vect{w}^n)^\dg \tau_z \vect{e}_i]
            \nonumber \\
            &=
            \sum_{n}
            [(\vect{w}^n)^\dg \tau_z A  \vect{w}^n (\vect{w}^n)^\dg \tau_z
            \big(\sum_i \vect{e}_i(\vect{e}_i)^{\dg} \big) \vect{w}^n ]
            =\sum_{n}s_n[(\vect{w}^n)^\dg \tau_z A \vect{w}^n],
            \label{eq:supp:trace-def}
        \end{align}
    where $\vect{e}_i$ are the standard basis vectors in $\mathbb{C}^{2N}$.
    In the second equality, we have used the cyclic property of the trace and the idempotence of the projector $P^nP^n=P^n$. 
    In the last equality of~\eqref{eq:supp:trace-def} we have further used that 
    $\mathbbm{1}_{2N}= \sum_i \vect{e}_i(\vect{e}_i)^\dg$ and 
    the orthonormalization condition of the Bogoliubov modes~\eqref{eq:supp:gen-eigenvect-normalization}.

    Any vector $\vect{\psi}\in \mathbb{C}^{2N}$ can be expanded 
    in the basis of the Bogoliubov modes by
    making use of the resolution of identity~\eqref{eq:supp:res-identity-proj},
    such that
        \begin{align}
            \vect{\psi}=\sum_{n=1}^{2N}P^n \vect{\psi}
            =\sum_{n=1}^{2N} s_n\vect{w}^n(\vect{w}^n)^\dg\tau_z \vect{\psi} \equiv 
            \sum_{n} c_n^{\psi}\vect{w}^n,
            \label{eq:supp:indefinite-inner-product-expansion}
        \end{align}
    where $c_n^{\psi}=s_n (\vect{w}^n)^\dg\tau_z \vect{\psi}$.
    Moreover, since $\vect{\psi}$ lives in the indefinite inner product space, its norm is only correctly defined via
    the generalized inner product norm $(\vect{\psi})^\dg\tau_z \vect{\psi}$.
    Using the generalized orthonormality condition~\eqref{eq:supp:gen-eigenvect-normalization}
    this norm can consequently also be expressed as
        \begin{align}
            (\vect{\psi})^\dg\tau_z \vect{\psi}
            =\sum_{nm}(c_n^{\psi})^{*}c_m^{\psi}(\vect{w}^n)^{\dg}\tau_z\vect{w}^m
            =\sum_{nm}(c_n^{\psi})^{*}c_m^{\psi}s_m\delta_{nm}
            =\sum_{n=1}^{2N} s_n\abs{c_n^{\psi}}^2.
            \label{eq:supp:indefinite-inner-product-norm}
        \end{align}
    Lastly, we note that the sign of the $\tau_z$ norm $(\vect{\psi})^\dg\tau_z \vect{\psi}$ is called the Krein signature of $\vect{\psi}$~\cite{Schulz-Baldes2017,Xu2020}.
    Here, for the (thermodynamically stable) pseudo-hermitian matrix $D$~\eqref{eq:supp:gen-eigenvect-eq-vector} all eigenvectors $\vect{w}^n$ associated to an eigenvalue $\tilde{\Omega}_n$ of $D$ either have a definite Krein signature, namely $+1(-1)$ Krein signature for the particle (hole) modes.

\section{Symplectic quantum geometric tensor}
{\label{app:SQGT}} 
    In addition to the manifestly gauge-invariant form of the symplectic quantum geometric tensor
    $\eta^n_{\mu \nu}=\Tr {\del_{\mu} P^n Q^n \del_{\nu} P^n}$ 
    [\cref{eq:SQGT-def} in the main text],
    we derive two equivalent representations of the SQGT, which are more suitable for practical calculations.
    We again assume implicitly that the BdG Hamiltonian~\eqref{eq:supp:BdG-Ham} depends on a set of 
    $\ld$ parameters and all derivatives 
    should be understood as derivates 
    with respect to these parameters, i.e., $\del_{\mu}\equiv \frac{\del}{\del \lambda^{\mu}}$.

    Using the definition 
    of the trace via the Bogoliubov modes~\eqref{eq:supp:trace-def}, 
    we see that the SQGT 
        \begin{align}
            \eta^n_{\mu \nu}=\Tr {\del_{\mu} P^n Q^n \del_{\nu} P^n}
            \label{eq:supp:SQGT-def}
        \end{align}
    can equivalently be expressed as 
        \begin{align}
            \eta^n_{\mu \nu}&=\sum_m s_m[(\vect{w}^m)^\dg \tau_z  
            (\del_{\mu} P^n Q^n \del_{\nu} P^n)  \vect{w}^m]
            =\sum_m s_m
            [(\vect{w}^m)^\dg \tau_z  \vect{w}^n
            \ (\del_{\mu} \vect{w}^n)^\dg \tau_z Q^n (\del_{\nu} \vect{w}^n) (\vect{w}^n)^\dg
             \tau_z   \vect{w}^m]\nonumber \\
             &=s_n (\del_{\mu} \vect{w}^n)^\dg {\tau_z Q^n}(\del_{\nu} \vect{w}^n).
            \label{eq:supp:SQGT-Ketform}
            \end{align} 
    Here, in the second equality we have used that $Q^n \vect{w}^n=(\vect{w}^n)^{\dg}\tau_z Q^n=0$ 
    which directly follows from~\eqref{eq:supp:projectors}.
    In the final equality, the orthonormality conditions~\eqref{eq:supp:gen-eigenvect-normalization} 
    has been used. 
   \cref{eq:supp:SQGT-Ketform} is the first equivalent representation of the SQGT.
    
    To obtain the spectral representation of the SQGT, we first derive a 
    symplectic version of the Hellmann-Feynman theorem. 
    To this end, we first observe from eigenvalue equation~\eqref{eq:supp:gen-eigenvect-eq-vector}
    that 
         \begin{align}
            (\del_{\mu}D)\vect{w}^n+D\del_{\mu}\vect{w}^n 
            =(\del_{\mu}\tilde{\Omega}_n) \vect{w}^n+\tilde{\Omega}_n \del_{\mu}\vect{w}^n,
            \label{eq:supp:Hellmann-Feynman-deriv1}
         \end{align}   
    Then we project a distinct ($m\neq n$)
    Bogoliubov mode $(\vect{w}^m)^\dg \tau_z$ from the left onto~\eqref{eq:supp:Hellmann-Feynman-deriv1} and 
    obtain with use of the orthonormality conditions 
   ~\eqref{eq:supp:gen-eigenvect-normalization} the symplectic 
    version of the Hellmann-Feynman theorem,
        \begin{align}
            ({\vect{w}^m})^\dg (\del_{\mu}\mathcal{M}){\vect{w}^n}=(\tilde{\Omega}_n - \tilde{\Omega}_m)
            ({\vect{w}^m})^\dg \tau_z(\del_{\mu}\vect{w}^n).
            \label{eq:supp:sympl-relation-derivative-eigenvectors}
        \end{align}
    Next, by inserting two resolutions of the identity~\eqref{eq:supp:res-identity-proj} into~\cref{eq:supp:SQGT-Ketform} and using the idempotence property of the projectors $P^n$, i.e., 
    $P^n P^m = \delta_{nm}P^n$~\eqref{eq:supp:projectors},
    one obtains the spectral representation of the SQGT,
        \begin{align}
            \eta^n_{\mu \nu}&=s_n (\del_{\mu} \vect{w}^n)^\dg {\tau_z Q^n}(\del_{\nu} \vect{w}^n)
            =\sum_{m,m'}s_n (\del_{\mu} \vect{w}^n)^\dg {\tau_z P^m Q^nP^{m'}}(\del_{\nu} \vect{w}^n)
            \nonumber \\
            &=\sum_{m,m'}s_n (\del_{\mu} \vect{w}^n)^\dg {\tau_z P^m (\mathbbm{1}_{2N}-P^n)P^{m'}}(\del_{\nu} \vect{w}^n)
            =\sum_{m,m'}s_n (\del_{\mu} \vect{w}^n)^\dg
             {\tau_z (\delta_{mm'}P^m-\delta_{nm'}\delta_{nm}P^m)}(\del_{\nu} \vect{w}^n)
            \nonumber \\
            &=s_n\sum_{m}\left[ (\del_{\mu} \vect{w}^n)^\dg
             \tau_z (P^m)(\del_{\nu} \vect{w}^n)\right]-s_n (\del_{\mu} \vect{w}^n)^\dg
             \tau_z (P^n)(\del_{\nu} \vect{w}^n)
            \nonumber \\
            &=s_n\sum_{\substack{m \neq n\\ m=1}}^{2N}\left[ (\del_{\mu} \vect{w}^n)^\dg
            \tau_z (P^m)(\del_{\nu} \vect{w}^n)\right]=s_n \sum_{\substack{m \neq n\\ m=1}}^{2N} s_m 
            \frac{(\vect{w}^n)^{\dg} (\del_{\mu}\mathcal{M})\vect{w}^m
             \times (\vect{w}^m)^\dg (\del_{\nu}\mathcal{M})\vect{w}^n}
            {(\tilde{\Omega}_n - \tilde{\Omega}_m)^2},
            \label{eq:supp:SQGT-SpectralForm}
        \end{align}
    where in the last step, 
    the symplectic Hellmann-Feynman theorem~\eqref{eq:supp:sympl-relation-derivative-eigenvectors}
    has been used.
   \cref{eq:supp:SQGT-SpectralForm} is the spectral representation of the SQGT.
    
    \subsection{Comparison SQGT and QGT}
    \label{app:Comp-SQGT-QGT}
    To highlight the differences and similarities, 
    we compare the properties related to BBdG systems~\eqref{eq:supp:BdG-Ham} 
    and the symplectic quantum geometric tensor (SQGT) $\eta^n_{\mu \nu}$ and compare them to particle-number-conserving (PNC) systems and the conventional quantum geometric tensor (QGT).
    The results are summarized in~\cref{tab:SQGT-QGT-comparison}.
    \begin{table}[hbt!]
        \centering
        \begin{tabular}{|c|c|c|}
            \hline
            \textbf{Relations} & \textbf{BBdG Systems} & 
            \textbf{PNC Systems} \\
            \hline
            Eigenvalue equation &
            \makecell{$D\vect{w}^n =\tilde{\Omega}_n \vect{w}^n$, \\ $D=\tau_z \Mc=\tau_z D^\dg \tau_z$, pseudo-hermitian~\eqref{eq:supp:gen-eigenvect-eq-vector}}
            & \makecell{$K\vect{u}^n =\omega_n \vect{u}^n$, \\ $K=K^\dg$, hermitian} 
            \\ \hline
            Transformation group & \makecell{$W^\dg \tau_z W=\tau_z$, paraunitary~\eqref{eq:supp:paraunitary-cond} \\ $(\vect{w}^n)^\dg\tau_z\vect{w}^m=s_n\delta_{nm}$~\eqref{eq:supp:gen-eigenvect-normalization}}&
            \makecell{$U^\dg U=\mathbbm{1}_N$, unitary \\ 
            $(\vect{u}^n)^\dg\vect{u}^m=\delta_{nm}$} \\ \hline
            Projector & 
            \makecell{$P^n=W\Gamma^n W^{-1}=s_n \vect{w}^n(\vect{w}^n)^\dg\tau_z$, \\
            $(P^n)^\dg=\tau_z P^n \tau_z$, pseudo-hermitian~\eqref{eq:supp:projectors}}
            &\makecell{$P^n=U\Gamma^n U^{-1}=\vect{u}^n(\vect{u}^n)^\dg$,\\ $(P^n)^\dg=P^n$, hermitian} 
             \\
            \hline
            \textbf{Property} & \textbf{SQGT} & 
            \textbf{QGT} \\
            \hline
            Mode representation & $\eta^{n}_{\mu \nu}= 
            s_n (\del_{\mu} \vect{w}^n)^\dg {\tau_z Q^n}(\del_{\nu} \vect{w}^n), 
            \, Q^n=\mathbbm{1}_{2N}-P^n$~\eqref{eq:supp:SQGT-Ketform}
            & $\eta^{n}_{\mu \nu}=(\del_{\mu} \vect{u}^n)^\dg Q^n (\del_{\nu} \vect{u}^n), 
            \, Q^n=\mathbbm{1}_N-P^n$ 
            \\ \hline
            Projector representation &
            $\eta^n_{\mu \nu}=\Tr {\del_{\mu} P^n Q^n \del_{\nu} P^n}$~\eqref{eq:supp:SQGT-def} 
            & $\eta^n_{\mu \nu}=\Tr {\del_{\mu} P^n Q^n \del_{\nu} P^n}$ 
            \\ \hline
            Spectral representation & 
            $\eta^n_{\mu \nu}=\sum_{\substack{m \neq n\\ m=1}}^{2N} s_m \frac{(\vect{w}^n)^{\dg} (\del_{\mu}\mathcal{M})\vect{w}^m  (\vect{w}^m)^\dg (\del_{\nu}\mathcal{M})\vect{w}^n}{(\tilde{\Omega}_n - \tilde{\Omega}_m)^2}$~\eqref{eq:supp:SQGT-SpectralForm}
            & $\eta^n_{\mu \nu}=\sum_{\substack{m \neq n\\ m=1}}^{N} \frac{(\vect{u}^n)^{\dg} (\del_{\mu}K)\vect{u}^m  (\vect{u}^m)^\dg (\del_{\nu}K)\vect{u}^n}{(\omega_n - \omega_m)^2}$ 
            \\ \hline
            $U(1)$ gauge invariance & $\vect{w}^n(\ld)\to e^{i\phi(\ld)}\vect{w}^n(\ld)$
            & $\vect{u}^n(\ld)\to e^{i\phi(\ld)}\vect{u}^n(\ld)$ 
            \\ \hline
            Hermiticity & $\eta^n_{\mu \nu}=(\eta^n_{\nu \mu})^*$ 
            & $\eta^n_{\mu \nu}=(\eta^n_{\nu \mu})^*$
            \\ \hline
            Metric 
            & \makecell{$g^n_{\mu \nu}=\re[\eta^n_{\mu\nu} ]=g^n_{\nu \mu}$\\ 
            $ds^2=1-\abs{(\vect{w}(\ld))^\dg {\tau_z}{\vect{w}(\ld \!+\!d\ld)}}^2\approx g_{\mu \nu}d\lambda^\mu d\lambda^\nu $~\eqref{eq:supp:distance-def-sympl-2-SQGT} }
            & \makecell{$g^n_{\mu \nu}=\re[\eta^n_{\mu\nu} ]\!=\!g^n_{\nu \mu}$\\
            $ds^2=1-\abs{(\vect{u}(\ld))^\dg{\vect{u}(\ld\!+\!d\ld)}}^2\approx g_{\mu \nu}d\lambda^\mu d\lambda^\nu$}
            \\ \hline
            Berry curvature & $B^n_{\mu \nu}=-2\im[\eta^n_{\mu \nu}]=-B^n_{\nu \mu}$
            & $B^n_{\mu \nu}=-2\im[\eta^n_{\mu \nu}]=-B^n_{\nu \mu}$\\
            \hline
        \end{tabular}
        \caption{Comparison of the symplectic quantum geometric tensor (SQGT) 
        arising in bosonic Bogoliubov-de Gennes systems (BBdG) and 
        the conventional quantum geometric tensor (QGT) arising in particle-number-conserving (PNC) systems.}
        \label{tab:SQGT-QGT-comparison}
        \end{table}
        \FloatBarrier
        Here, for particle-number-conserving systems (bosonic or fermionic), 
        we consider a Hamiltonian $\hat{H}=\sum_{ij}K_{ij} \hat{a}^\dg_i \hat{a}_j$ 
        ($G=0$ in~\eqref{eq:supp:BdG-Ham}) with $K^\dg=K$ and $\hat{a}_i^{(\dg)}$ the annihilation (creation) operator of the $i$-th mode.
        Such systems are diagonalized via by a unitary transformation matrix $U$, i.e., 
        $\sum_{ij}(U^{\dg})_{ni}K_{ij}U_{jn}=\omega_n$ with $UU^{\dg}=\mathbbm{1}_N$.
        In this case, the projector onto the $n$-th eigenspace is simply given by 
        $P^n=U\Gamma^n U^{-1}=\vect{u}^n(\vect{u}^n)^{\dg}$,
        where $\vect{u}^n$ is the $n$-th column of $U$, namely the eigenvector of $K$ corresponding to $\omega_n$.
        Here, $(\Gamma^n)_{lm}=\delta_{nl}\delta_{lm}$ is the $N\times N$ matrix with only the $n$-th diagonal element equal to $1$.
        For the properties and different representations of the QGT presented in~\cref{tab:SQGT-QGT-comparison}, we refer to \Ccite{Resta2011}.

    \subsection{Particle-hole symmetry of the SQGT}
    \label{app:PHS-SQGT}
    We use the particle-hole symmetry of the BBdG system~\eqref{eq:supp:PHS-DynMat} to relate 
    the particle parts of the SQGT, $\eta^{l,+}_{\mu \nu}$, to the hole parts, $\eta^{l,-}_{\mu \nu}$.
    Here, $l$ runs from $1$ to $N$, where $\eta^{l,+}$ corresponds to SQGT $\eta^n$ of the particle bands $n<N$ and $\eta^{l,-}$ to the SQGT $\eta^n$ of the hole bands $n>N$.
    Using the PHS of the projectors $P^l_{+}=\tau_x (P^l_{-})^*\tau_x$ 
    (which follows from~\eqref{eq:supp:Lie-group-ParaUnit-DynMat}), we can see that
        \begin{align}
            \eta^{l,+}_{\mu \nu}&=\Tr{\del_{\mu} P^l_{+} Q_{+}^l \del_{\nu} P^l_{+}}
                =\Tr{\del_{\mu} (\tau_x (P^l_{-})^*\tau_x) \tau_x (Q_{-}^l)^*\tau_x \del_{\nu} (\tau_x (P^l_{-})^*\tau_x)}\nonumber \\
                &=\Tr{ \del_{\mu} (P^l_{-})^* (Q_{-}^l)^*\del_{\nu} (P^l_{-})^*}
                =(\eta^{l,-}_{\mu \nu})^*
                \label{eq:supp:PHS-SQGT-proof}
        \end{align}
    Thus, the particle part of the SQGT is equal to the complex conjugate of the hole part of the SQGT.
    This implies that the particle and hole parts of symplectic quantum metric 
    $g^n_{\mu \nu}=\re[\eta^n_{\mu\nu} ]$ are equal, whereas the particle and hole parts of the symplectic Berry curvature $B^n_{\mu, \nu}=-2\im[\eta^n_{\mu \nu}]$ are opposite in sign
        \begin{align}
            g^{l,+}_{\mu \nu}=g^{l,-}_{\mu \nu}, \quad B^{l,+}_{\mu \nu}=-B^{l,-}_{\mu \nu}.
            \label{eq:supp:PHS-SQM-SBC}
        \end{align}
            
    \subsection{Symplectic quantum metric}
    {\label{app:Sympl-Metric}} 
    We show that the symplectic quantum metric naturally arises when considering the distance $ds^2$
    between two infinitesimally close Bogoliubov modes.
    In analogy to~\Ccite{Provost1980,Kolodrubetz2017}, we define the distance $ds^2$ between nearby states 
    $\vect{w}^n(\ld)$ and $\vect{w}^n(\ld+d\ld)$ (omitting the index $n$ in the following) via
        \begin{align}
            ds^2 =1-f^2=1-\abs{(\vect{w}(\ld))^\dg {\tau_z}{\vect{w}(\ld+d\ld)}}^2,
            \label{eq:supp:distance-def-sympl}
        \end{align}
    with the fidelity $f=\abs{(\vect{w}(\ld))^\dg {\tau_z}{\vect{w}(\ld+d\ld)}}$.
    We first expand $\vect{w}(\ld+d\ld)$ to second order in $d\ld$,
        \begin{align}
            \vect{w}(\ld+d\ld)=\vect{w}(\ld)+\del_{\mu}\vect{w}(\ld) d\lambda^{\mu}
            +\frac{1}{2}\del_{\mu}\del_{\nu}\vect{w}(\ld)d\lambda^{\mu} d\lambda^{\nu},
            \label{eq:supp:Bogoliubov-mode-expansion}
        \end{align}
    where we use the Einstein summation convention over Greek indices in the following. 
    We then insert this expression~\eqref{eq:supp:Bogoliubov-mode-expansion} 
    into~\eqref{eq:supp:distance-def-sympl}, 
    using orthonormalization conditions~\eqref{eq:supp:gen-eigenvect-normalization}.
    Collecting only terms up to second order in $d\ld$ then yields 
        \begin{align}
            ds^2=&1-\Big\{(\vect{w}^\dg \tau_z \vect{w})^2
            +(\vect{w}^\dg \tau_z \vect{w})[((\del_{\mu}\vect{w})^\dg  \tau_z \vect{w})
            +(\vect{w}^\dg  \tau_z (\del_{\mu}\vect{w}))]d\lambda^{\mu}
             +[((\del_{\mu}\vect{w})^\dg  \tau_z \vect{w})(\vect{w}^\dg  \tau_z (\del_{\nu}\vect{w}))]
            d\lambda^{\mu}d\lambda^{\nu}
            \nonumber \\
            & \ \ +\frac{1}{2}(\vect{w}^\dg \tau_z \vect{w})
            [((\del_{\nu}\del_{\mu}\vect{w})^\dg  \tau_z \vect{w})
            +(\vect{w}^\dg  \tau_z (\del_{\nu}\del_{\mu}\vect{w}))]d\lambda^{\mu}d\lambda^{\nu}\Big\}
            +\mathcal{O}(\abs{d\ld}^3)
            \nonumber \\
            =&\{(\vect{w}^\dg \tau_z \vect{w}) [(\del_{\mu}\vect{w})^\dg  \tau_z (\del_{\nu}\vect{w})]
            -(\del_{\mu}\vect{w})^\dg \tau_z \vect{w} \  
            \vect{w}^\dg \tau_z (\del_{\nu}\vect{w}) \}
            d\lambda^{\mu}d\lambda^{\nu}+\mathcal{O}(\abs{d\ld}^3)
            \nonumber \\
            =&\left[(\vect{w}^\dg \tau_z \vect{w})(\del_{\mu} \vect{w})^\dg {\tau_z Q}(\del_{\nu} \vect{w}) \right]d\lambda^{\mu}d\lambda^{\nu}+\mathcal{O}(\abs{d\ld}^3)
            \label{eq:supp:distance-def-sympl-2-SQGT}
        \end{align}
    where the explicit parameter dependence of the states has been omitted for clarity.
    In the second equality above, we have on the one hand used $(\vect{w}^\dg \tau_z \vect{w})^2=1$~\eqref{eq:supp:gen-eigenvect-normalization} 
    and on the other hand that the linear terms ($\sim d\lambda^{\mu}$) vanish, which follows from 
    the relation $(\vect{w})^\dg {\tau_z}(\del_{\mu} \vect{w})=-(\del_{\mu}\vect{w})^\dg {\tau_z}(\vect{w})$ 
   ~\eqref{eq:supp:gen-eigenvect-normalization}.
    From this relation, it moreover follows that 
    $((\del_{\nu}\del_{\mu}\vect{w})^\dg  \tau_z \vect{w})+(\vect{w}^\dg  \tau_z (\del_{\nu}\del_{\mu}\vect{w}))=-[(\del_{\mu}\vect{w})^\dg  \tau_z (\del_{\nu}\vect{w})+(\del_{\nu}\vect{w})^\dg  \tau_z (\del_{\mu}\vect{w})]$ 
    which has also been used in the second equality above~\eqref{eq:supp:distance-def-sympl-2-SQGT}.
    Finally, comparing this expression to the definition of the symplectic quantum geometric tensor in~\cref{eq:supp:SQGT-Ketform},
    we see that the symplectic quantum geometric tensor (SQGT) naturally emerges as a distance measure 
    in the space of the Bogoliubov modes
    \begin{align}
        ds^2=\eta_{\mu \nu} d\lambda^{\mu}d\lambda^{\nu}+\mathcal{O}(\abs{d\ld}^3)
        =\frac{1}{2}\left(\eta_{\mu \nu}+\eta_{\nu \mu}\right) d\lambda^{\mu}d\lambda^{\nu}+\mathcal{O}(\abs{d\ld}^3)
        =g_{\mu \nu} d\lambda^{\mu}d\lambda^{\nu}+\mathcal{O}(\abs{d\ld}^3).
        \label{eq:supp:distance-def-sympl-final}
    \end{align}
    More precisely, since the distance measure $ds^2$~\eqref{eq:supp:distance-def-sympl-2-SQGT} is a symmetric bilinear form, 
    the antisymmetric part of the SQGT does not contribute to $ds^2$, i.e.,
    only the symmetric part, i.e., the symplectic quantum metric 
    $g_{\mu \nu}=\frac{1}{2}\left(\eta_{\mu \nu}+\eta_{\nu \mu}\right)=\re[\eta_{\mu\nu} ]$, 
    determines the distance measure between two infinitesimally close Bogoliubov states,
    hence, the second equality in~\cref{eq:supp:distance-def-sympl-final}.
        
\subsection{Local conservation law for symplectic Berry curvature}
{\label{app:conservation-law-berry-curv}}
    We show that the symplectic Berry curvature, admits 
    a local conservation law over all subspaces (Bogoliubov subspaces), 
    including both the particle and hole subspaces~\cite{Shindou2013}.
    The first simple proof directly makes use of the particle-hole symmetry 
    of the symplectic Berry curvature~\eqref{eq:supp:PHS-SQM-SBC}, i.e.
        \begin{align}
            \sum_{n=1}^{2N}B^n_{\mu \nu}=\sum_{l=1}^N (B^{l,+}_{\mu \nu}+B^{l,-}_{\mu \nu})
            =\sum_{l=1}^{N}(B^{l,+}_{\mu \nu}-B^{l,+}_{\mu \nu})=0.
        \end{align}
    For completeness, we also present a second proof without explicitly utilizing the PHS.
    For this, we note that using~\cref{eq:supp:SQGT-Ketform}, 
    the symplectic Berry curvature of the subspace $n$, 
    $B_{\mu \nu}^n=-2\im\,[\eta^n_{\mu\nu}]$ can be written as 
        \begin{align}
            B_{\mu \nu}^n=is_n((\del_{\mu} \vect{w}^n)^\dg {\tau_z}(\del_{\nu} \vect{w}^n)
            -(\del_{\nu} \vect{w}^n)^\dg {\tau_z}(\del_{\mu} \vect{w}^n)),
            \label{eq:supp:Sympl-Berry-curv-Ketform}
        \end{align}
    where we omit the explicit parameter dependence for clarity.
    Next, inserting the resolution of the identity 
   ~\eqref{eq:supp:res-identity-proj} into~\eqref{eq:supp:Sympl-Berry-curv-Ketform},
    we see that we may also write $B_{\mu \nu}^n$ as 
        \begin{align}
            B_{\mu \nu}^n&=is_n \left\{
                \sum_m s_m (\del_{\mu} \vect{w}^n)^\dg {\tau_z}{\vect{w}^m}
                (\vect{w}^m)^\dg {\tau_z}(\del_{\nu} \vect{w}^n)
                -(\del_{\nu} \vect{w}^n)^\dg {\tau_z}(\del_{\mu} \vect{w}^n)
                \right\} \nonumber \\
                &=is_n \left\{
                    \sum_m \left[ s_m
                    (\del_{\nu} \vect{w}^m)^\dg {\tau_z}{\vect{w}^n}
                    (\vect{w}^n)^\dg {\tau_z}(\del_{\mu} \vect{w}^m)
                    \right]
                    -(\del_{\nu} \vect{w}^n)^\dg {\tau_z}(\del_{\mu} \vect{w}^n)
                    \right\},
        \label{eq:supp:Sympl-Berry-curv-conserv-law-step1} 
        \end{align}
    where in the second equality we used
    $(\vect{w}^m)^\dg {\tau_z}(\del_{\nu} \vect{w}^n)=-(\del_{\nu}\vect{w}^m)^\dg {\tau_z}(\vect{w}^n)$ 
    twice, which itself simply follows from~\cref{eq:supp:gen-eigenvect-normalization}.
    Now, by summing~\cref{eq:supp:Sympl-Berry-curv-conserv-law-step1} over all subspaces $n$,
        \begin{align}
            \sum_n B_{\mu \nu}^n &= i \left\{
                \sum_m \left[ s_m
            (\del_{\nu} \vect{w}^m)^\dg {\tau_z}\left(\sum_n s_n{\vect{w}^n}
            (\vect{w}^n)^\dg {\tau_z}\right)(\del_{\mu} \vect{w}^m)
            \right]
            -\sum_n s_n(\del_{\nu} \vect{w}^n)^\dg {\tau_z}(\del_{\mu} \vect{w}^n)
                \right\},
            \nonumber \\
            &=i \left\{
                \sum_m \left[ s_m
                (\del_{\nu} \vect{w}^m)^\dg {\tau_z}
                (\del_{\mu} \vect{w}^m)
                \right]
                -\sum_n\left[s_n(\del_{\nu} \vect{w}^n)^\dg {\tau_z}(\del_{\mu} \vect{w}^n)
                \right]
                    \right\}=0,
        \label{eq:supp:Sympl-Berry-curv-conserv-law-step2} 
        \end{align}
    we immediately see that the symplectic Berry curvature
    admits a local conservation law for all parameter values $\ld$,
        \begin{align}
            \sum_n B^n_{\mu \nu}(\ld) =0, \ \forall \ld.
            \label{eq:supp:Sympl-Berry-curv-conserv-law}
        \end{align}
    One immediate consequence of this conservation law is that the total (first) Chern number over 
    both the particle \emph{and} hole subspaces is zero.
    We stress that this is distinct from the conventional, particle-number-conserving case,
    where both the sum of the Berry curvature and hence the sum of the Chern numbers over all particle bands is zero. 
    In particular, there are situations where the sum of the Berry curvature over all particle bands becomes non-zero. 
    A consequence of this is the weaker statement that the total Chern number over the particle subspace can become non-zero, which has already been noticed in Ref.~\cite{Chaudhary2021}.
    The situation occurs in BBdG systems which are dynamically stable, but non-positive definite, such as 
    photonic driven systems in a rotating frame governed by an external pump frequency.
    Dynamical stability requires that the eigenvalues of the dynamical matrix $D$ are real, 
    although the Hamiltonian $\hat{H}$~\eqref{eq:supp:BdG-Ham} itself must not be positive definite 
   ~\cite{Chaudhary2021}.
    We note that thermodynamic stability is a stronger condition than dynamical stability, 
    as the former implies the latter, but not vice versa~\cite{Peano2018}.

\section{Time-dependent Schr{\"o}dinger equation for Bogoliubov modes}
{\label{app:Effective-TDSE}} 
    Starting from the Heisenberg equations of motions (EOMs) for the original bosonic operators, 
    we derive an effective time-dependent 
    single-particle Schr{\"o}dinger equation for the Bogoliubov modes $\vect{w}^n$.
    Using~\cref{eq:supp:BdG-Ham}, we see that the Heisenberg EOM 
    ($\hbar \equiv 1$ throughout) for the original bosonic operators read 
        \begin{align}
            i\del_t \hat{a}_m &= [\hat{a}_m,\hat{H}]=
            \sum_{j} K_{mj} \hat{a}_j + \frac{1}{2}[ \hat{a}_j^\dg G_{jm} + \hat{a}^\dg_jG_{jm} ],\\
            i \del_t \hat{a}_m^\dg &= [\hat{a}_m^\dg,\hat{H}]=
            -\sum_{j}  \hat{a}_j^\dg K_{jm} - \frac{1}{2}[ G_{mj}^*\hat{a}_j + G_{mj}^* \hat{a}_j ].
            \label{eq:supp:Heisenberg-eq}
        \end{align}
    These EOMs can also be compactly written in terms of
    the bosonic Nambu spinor $\hat{\Psi}=(\hat{a}_1,\dots,\hat{a}_N,\hat{a}_1^\dagger,\dots,\hat{a}_N^\dagger)^T$~\cite{Xiao2009,Flynn2020}
        \begin{align}
            i\del_t \hat{\Psi}= [\hat{\Psi},\hat{H}]=D \hat{\Psi},\ \text{with}\ D=\tau_z \mathcal{M},
            \label{eq:supp:Heisenberg-eq-Nambu}
        \end{align}
    where $D$ is again the dynamical matrix~\eqref{eq:supp:gen-eigenvect-eq-vector} and 
    $\mathcal{M}$ is the coefficient matrix of the BdG Hamiltonian consisting of
    matrix blocks $K$ and $G$~\eqref{eq:supp:BdG-Ham}.
   ~\cref{eq:supp:Heisenberg-eq-Nambu} is easily solved by 
        \begin{align}
            \hat{\Psi}(t)= e^{-iDt}\hat{\Psi}(0)
            \label{eq:supp:original-Nambu-spinor-sol}
        \end{align}
    which itself leads to
    the following time-dependence for the Nambu spinor of the new bosonic operators
    $\hat{\Phi}(t)=W^{-1}\hat{\Psi}(t)$~\eqref{eq:supp:paraunitary-matrix-bog-transf-gen}
        \begin{align}
            \hat{\Phi}(t)= e^{-i\tilde{\Omega} t}\hat{\Phi}(0)=e^{-i\tilde{\Omega} t} W^{-1}\hat{\Psi}(0),
            \label{eq:supp:Heisenberg-eq-Nambu-Eigmod-sol}
        \end{align}
    where we have utilized that 
    $W^{-1}DW=\tau_zW^{\dg}\mathcal{M}W=\tau_z\Omega\equiv \tilde{\Omega}$ which follows from 
   ~\cref{eq:supp:gen-eigenvect-eq-matrix}. 
    Thus, each new bosonic operator acquires a phase factor associated to eigenenergy
    $\hat{b}_n(t) =e^{-i\omega_n t} \hat{b}_n(0)$~\cite{Xiao2009,Flynn2020}.

    Next, we note that we may also absorb the time-dependence of the original operators $\hat{\Psi}(t)$ 
    into the (inverse) Bogoliubov matrix, 
    $\hat{\Phi}(t)=W^{-1}\hat{\Psi}(t)\equiv W^{-1}(t)\hat{\Psi}(0)$, i.e., 
    in~\eqref{eq:supp:Heisenberg-eq-Nambu-Eigmod-sol} 
    $W^{-1}(t)=e^{-i\tilde{\Omega}t}W^{-1}(0)=W^{-1}(0)e^{-iDt}$. 
    This leads to the immediate question, what kind of equation of motion $W^{-1}(t)$ 
    fulfills.
    In compact matrix form, the answer is given by the following set of 
    equations for the Bogoliubov modes, 
        \begin{align}
            \del_tW^{-1}(t)=-iW^{-1}(t)D \  \Leftrightarrow \del_t W(t)=iDW(t).
            \label{eq:supp:Effective-TDSE-InvMatrix}
        \end{align}
    Since we know that $W^{-1}D=\tilde{\Omega}W^{-1}$, 
    we see that~\eqref{eq:supp:Effective-TDSE-InvMatrix} is solved by
    $W^{-1}(t)=e^{-i\tilde{\Omega}t}W^{-1}(0)$, 
    which is in agreement with~\eqref{eq:supp:Heisenberg-eq-Nambu-Eigmod-sol}.
    This may also be expressed in as an effective Schr{\"o}dinger equation for the 
    individual Bogoliubov modes $\vect{w}^n(t)$~\cite{Flynn2020,Zhang2006}
        \begin{align}
            -i \del_t \vect{w}^n(t)=D \vect{w}^n(t),
            \label{eq:supp:Effective-TDSE-Vector}
        \end{align}
    which is the sought after effective time-dependent Schr{\"o}dinger equation.
    From~\cref{eq:supp:Effective-TDSE-Vector}, for a time-independent dynamical matrix 
    $D$, we can define the paraunitary 
    time evolution operator $P(t)$,
        \begin{align}
            P(t)=\exp \left(iDt\right),
            \label{eq:supp:Effective-TDSE-Vector-Operator}
        \end{align}
    such that $\vect{w}^n(t)=P(t)\vect{w}^n(0)$.
    For a time-dependent dynamical matrix $D(t)$ in~\eqref{eq:supp:Effective-TDSE-Vector}, 
    the paraunitary time evolution operator 
    reads 
        \begin{align}
            P(t)=\mathcal{T}\exp \left(i\int_0^t dt' D(t')\right),
            \label{eq:supp:Effective-TDSE-Vector-Operator-TD}
        \end{align}
    where $\mathcal{T}$ denotes the time-ordering operator.
    Note that the paraunitarity of $P(t)$, i.e., 
    $\tau_z P^{\dg}(t)\tau_z=P^{-1}(t)$, directly follows from the pseudo-hermiticity of the dynamical matrix,
    $D^\dg(t) =\tau_z D(t) \tau_z$.
    Thus, the time evolution of the Bogoliubov modes $\vect{w}^n$~\eqref{eq:supp:Effective-TDSE-Vector} 
    directly specifies the time-dependence of the Heisenberg-picture quasiparticle operators 
    $\hat{b}^{\dg}_n(t)=\hat{\Psi}^{\dg}(t)\tau_z \vect{w}^n$~\eqref{eq:supp:bog-operators-compact},
        \begin{align}
            \hat{b}^{\dg}_n(t)=\hat{\Psi}^{\dg}(t)\tau_z \vect{w}^n
            =\hat{\Psi}^{\dg}(0)e^{i(D^\dg) t}\tau_z \vect{w}^n
            =\hat{\Psi}^{\dg}(0)\tau_z e^{iDt} \vect{w}^n
            =\hat{\Psi}^{\dg}(0)\tau_z \vect{w}^n(t),
        \end{align}
    where we have used~\eqref{eq:supp:original-Nambu-spinor-sol} 
    and the pseudo-hermiticity of the dynamical matrix $D^\dg =\tau_z D \tau_z$.

\section{Time-dependent perturbation theory for the effective Schr{\"o}dinger equation}{\label{app:TDPT}}
    Consider the effective time-dependent single-particle Schr{\"o}dinger equation~\eqref{eq:supp:Effective-TDSE-Vector}
    for some state $\vect{\psi}(t)\in\mathbb{C}^{2N}$ in the indefinite inner product space spanned 
    by the Bogoliubov modes $\vect{w}^n(t)$
    with the addition of a small time-dependent perturbation $V(t)$ added to the dynamical matrix, i.e., 
        \begin{align}
            -i \del_t \vect{\psi}(t)=D(t) \vect{\psi}(t)=\left[D+V(t)\right] \vect{\psi}(t)
            \label{eq:supp:Effective-TDSE-Vector-Pert}
        \end{align}
    where $V(t)$ shall only be present for $t>0$. 
    Adapting standard time-dependent perturbation theory~\cite{Diu2019} to the space of Bogoliubov modes, 
    we now derive the effect of the perturbation $V(t)$ on the population of Bogoliubov modes.

    At $t=0$, where $V(0)=0$, we consider the initial state $\vect{\psi}(0)$ to be 
    given by the Bogoliubov eigenstate $\vect{w}^n$, i.e., 
    $D\vect{\psi}(0)=\tilde{\Omega}_n\vect{w}^n$~\eqref{eq:supp:gen-eigenvect-eq-vector}.
    At this point, $\vect{w}^n$ adheres to the effective time-dependent Schr{\"o}dinger equation without the perturbation~\eqref{eq:supp:Effective-TDSE-Vector}, 
    which is easily solved by 
    $\vect{w}^n(t)=e^{iDt}\vect{w}^n(0)=e^{i\tilde{\Omega}_n t}\vect{w}^n(0)$~\eqref{eq:supp:Effective-TDSE-Vector}.
    To first order in the perturbation $V(t)$, we now compute the resulting excitation probability 
    to other distinct Bogoliubov modes $\vect{w}^m(t)$
    at time $t$.

    First, we introduce the interaction picture
        \begin{align}
            \vect{\psi}_{I}(t)=\tilde{P}(t)\vect{\psi}(t)= e^{-iD t}\vect{\psi}(t).
            \label{eq:supp:InteractionPicture-ket}
        \end{align}
    The transformation is explicitly given by 
    $\tilde{P}(t)=e^{-iD t}$ where $D$ is the static part in  
   ~\cref{eq:supp:Effective-TDSE-Vector-Pert}.
    Due to the pseudo-hermiticity of the dynamical matrix, $D^\dg =\tau_z D \tau_z$, 
    in the equation of motion~\eqref{eq:supp:Effective-TDSE-Vector}, 
    the transformation $\tilde{P}(t)$ must be paraunitary, 
    i.e., $\tilde{P}^{\dg}(t)=\tau_z \tilde{P}^{-1}(t)\tau_z$.
    This directly follows from 
    $\tau_z\tilde{P}^\dg(t)\tau_z=\tau_z e^{iD^\dg t}\tau_z=e^{iD t}=\tilde{P}^{-1}(t)$.
    The effective time-dependent Schr{\"o}dinger equation in the interaction picture then reads
        \begin{align}
            -i \del_t \vect{\psi}_I(t)=D_I(t)\vect{\psi}_I(t),
            \label{eq:supp:ETDSE-InteractionPicture}
        \end{align}
    where we used~\cref{eq:supp:Effective-TDSE-Vector} and introduced $D_I(t)$, 
    the dynamical matrix in the interaction picture given by
        \begin{align}
            D_{I}(t)= \tilde{P}(t)D(t)\tilde{P}^{-1}(t)-i(\del_t\tilde{P}(t))\tilde{P}^{-1}(t).
            \label{eq:supp:ETDSE-InteractionPicture-DynMat}
        \end{align}
    With $\tilde{P}(t)=e^{-iD t}$,~\cref{eq:supp:ETDSE-InteractionPicture} then becomes
        \begin{align}
            -i \del_t \vect{\psi}_I(t)=V_I(t)\vect{\psi}_I(t),
            \label{eq:supp:ETDSE-InteractionPicture-appl}
        \end{align}
    where $V_I(t)\equiv \tilde{P}(t)V(t)\tilde{P}^{-1}(t) = e^{-iD t}V(t)e^{iD t}$ 
    is the perturbation in the interaction picture.
    The solution to~\eqref{eq:supp:ETDSE-InteractionPicture-appl} is then given by a Dyson-type series,
        \begin{align}
            \vect{\psi}_{I}(t)=\vect{\psi}_{I}(0)
            +i\int_0^t dt' V_I(t')\vect{\psi}_{I}(0)
            +i^2\int_0^t dt' \int_0^{t'} dt'' V_I(t')V_I(t'')\vect{\psi}_{I}(0)+\dots,
            \label{eq:supp:ETDSE-InteractionPicture-Dyson}
        \end{align}
    where we are only interested in the first order contribution.
    Now, given an initial eigenstate $\vect{\psi}_{I}(0)=\vect{w}^n(0)$, we compute 
    the transition amplitude $c^{\psi}_{n}(t)$ to some other arbitrary mode $\vect{w}^m(t)$ to first order at time $t$ by 
    first transforming back to the original picture~\eqref{eq:supp:InteractionPicture-ket}
        \begin{align}
            \vect{\psi}(t)=e^{iDt}\vect{\psi}_{I}(t)=e^{i\tilde{\Omega}_nt}\vect{w}^n(0)
            +e^{iDt}i\int_0^t dt' V_I(t')\vect{w}^n
        \end{align}
    and then projecting onto
    $s_m(\vect{w}^m )^\dg\tau_z$ [cf.~\cref{eq:supp:indefinite-inner-product-expansion}], 
        \begin{align}
            c^{\psi}_{m}(t)&=s_m(\vect{w}^m)^\dg\tau_z \vect{\psi}(t)
            =s_m s_n\delta_{nm} e^{i(\tilde{\Omega}_n-\tilde{\Omega}_m)t}
            +is_m\int_0^t dt' 
            (\vect{w}^m)^\dg\tau_z V(t')\vect{w}^n e^{i(\tilde{\Omega}_n-\tilde{\Omega}_m)t'},
            \label{eq:supp:ETDSE-InteractionPicture-Dyson-TransitionAmp}
        \end{align}
    where we have utilized that the dynamical matrix is pseudo-hermitian, $D^{\dg}=\tau_zD \tau_z$,
    which implies that $\exp(iDt)^{\dg}=\tau_z\exp(-iDt)\tau_z$.
    Eventually, the probability for the system to transition into some distinct orthogonal state $m\neq n$ is given by
        \begin{align}
            p_{mn}\equiv \abs{c^{\psi}_{m}(t)}^2
            =\abs{\int_0^t dt' ((\vect{w}^m)^\dg\tau_z V(t')\vect{w}^n)e^{i(\tilde{\Omega}_n-\tilde{\Omega}_m)t'}}^2 .
            \label{eq:supp:ETDSE-InteractionPicture-Dyson-TransitionAmp-Prob}
        \end{align}
    Next, we consider a periodic perturbation of the form $V(t)=F\cos(\omega t-\varphi)=\frac{1}{2}F(e^{i(\omega t-\varphi)}+e^{-i(\omega t-\varphi)})$,
    where $F$ is some constant real-valued amplitude. 
    Thus, using standard assumptions of $t \gg 1/\omega$ and $\omega \simeq (\tilde{\Omega}_{m}-\tilde{\Omega}_n)$~\cite{Diu2019},
   ~\cref{eq:supp:ETDSE-InteractionPicture-Dyson-TransitionAmp-Prob} becomes 
    \begin{align}
        p_{mn}(\omega,t)=\frac{2\pi}{4}t\abs{((\vect{w}^m)^\dg\tau_z F e^{i\varphi} \vect{w}^n)}^2
        \delta_{t}(\tilde{\Omega}_{m}-\tilde{\Omega}_n-\omega),
        \label{eq:supp:ETDSE-InteractionPicture-Dyson-TransitionAmp-Prob-2}
    \end{align}
    where $\delta_{t}(\alpha)=\frac{\sin^2(\alpha t/2)}{\pi\alpha^2 t}$ approaches a Dirac-Delta distribution in the limit of large observation times $t$,
    which then yields 
   ~\cref{eq:ETDPT-probability} in the main text. 

\section{Measuring the symplectic quantum geometry in lattice systems}
{\label{app:SQGT-lattice-systems}}
    In this appendix, we specifically apply the general formalism above to the case of two-dimensional 
    lattice BBdG systems, described by a translationally invariant dynamical matrix $D$, 
    which is inspired by~\Ccite{Tran2017,Ozawa2018}.
    For such a system, the Bogoliubov eigenmodes of the dynamical matrix $D$
    have Bloch function form, $\vect{\psi}^n(\qv)=e^{i\qv \cdot \rv}\vect{w}^n(\qv)$~\cite{Engelhardt2015,Zhang2006},
    such that the Bogoliubov eigenvalue equation in the quasimomentum space representation 
   ~\eqref{eq:supp:gen-eigenvect-eq-vector} 
    is given by $D(\qv)\vect{w}^n(\qv)=\tilde{\Omega}_n(\qv)\vect{w}^n(\qv)$.
    Here, $\vect{w}^n(\qv)$ is the cell-periodic part of the Bogoliubov Bloch function for the $n$-th band.
    Next, in direct analogy to~\Ccite{Tran2017,Ozawa2018}, 
    we consider periodically shaking the underlying lattice along $x$-direction of the lattice,
    resulting in the following time-dependent dynamical matrix, 
        \begin{align}
            D_{x}(t)=D+2A\cos(\omega t) \mathcal{X}=D+V(t), \ V(t)=2A\cos(\omega t) \mathcal{X},
            \label{eq:supp:dynamical-matrix-periodic-shaking-x}
        \end{align}
    where $D$ is the initial dynamical matrix in position space and $\mathcal{X}=\sigma_z \otimes x$ 
    the (pseudo-hermitian) position operator for the direction $x$
    (extended to Nambu space), and with $x$ being the usual position operator.

    We note that such periodically shaken systems described via $D_x(t)$
    are in direct analogy to the general discussion following 
   ~\cref{eq:dynamical-matrix-periodic-modulation} in the main text.
    To see this, we first note that $D_x(t)$ becomes translationally invariant in the 
    co-moving frame of the lattice~\cite{Tran2017,Ozawa2018}, 
    when introducing a paraunitary rotation 
    \begin{align}
        \tilde{P}(t)=\exp\left(-i\int dt' V(t')\right)=\exp(-i(2A/\omega) \sin(\omega t)\mathcal{X}),
        \label{eq:supp:paraunitary-rotation-X-comoving}
    \end{align}
    such that the dynamical matrix
    in the co-moving frame~\eqref{eq:supp:ETDSE-InteractionPicture-DynMat} becomes 
    \begin{align}
        D_x'(t)&=\tilde{P}D_{x}(t)\tilde{P}^{-1}-i(\del_t \tilde{P})\tilde{P}^{-1}
        =\tilde{P}[D+V(t)]\tilde{P}^{-1} -i(-i V(t)\tilde{P})\tilde{P}^{-1} 
        \nonumber \\
        &=\tilde{P}D\tilde{P}^{-1}= D(\qv) + \frac{2A}{\omega}\sin(\omega t)\del_{q_x}D(\qv)+\mathcal{O}((A/\omega)^2),
        \label{eq:supp:DynMat-IntPic-periodic-shaking-x}     
    \end{align}
    which agrees with the formalism in the main text.

    As in the general discussion in the main text, given an initially populated Bogoliubov mode
    $\vect{\psi}^n=e^{i\qv_0 \cdot \rv}\vect{w}^n(\qv_0)$,
    we compute the total excitation rate into all other available Bogoliubov modes $\vect{\psi}^m$, 
        \begin{align}
            \Gamma_x^n(\omega)=\frac{1}{t}\sum_{m\neq n}s_m p_{mn}(\omega,t),
            \label{eq:supp:total-exc-rate}
        \end{align}
    with $p_{mn}(\omega,t)$ being given by
   ~\eqref{eq:supp:ETDSE-InteractionPicture-Dyson-TransitionAmp-Prob-2}.
    We note that the factor $s_m$ comes from the indefinite inner product structure
    of the Bogoliubov modes [cf.~\cref{eq:supp:indefinite-inner-product-norm}].
    The total excitation rate can then be expressed as 
        \begin{align}
            \begin{split}
                \Gamma^n_x(\omega)
                =&2\pi A^2 \sum_{m\neq n} s_m
                \abs{(\vect{w}^m(\qv))^\dg e^{-i\qv \cdot \rv} \tau_z \mathcal{X} e^{i\qv_0 \cdot \rv}\vect{w}^n(\qv_0)}^2 
                \times \delta\left(\tilde{\Omega}_{m}(\qv)-\tilde{\Omega}_n(\qv_0)-\omega\right)\\
                =&2\pi A^2 \sum_{m\neq n} s_m
                \abs{(\vect{w}^m(\qv_0))^\dg \tau_z i\del_{q_x} \vect{w}^n(\qv_0)}^2 
                \times \delta\left(\tilde{\Omega}_{m}(\qv_0)-\tilde{\Omega}_n(\qv_0)-\omega\right).
            \end{split}
            \label{eq:supp:ETDPT-excitation-rate-lattice}
        \end{align}
    Here, we have used that the position operator acts as a derivate in the quasimomentum space representation,
    i.e., 
    $(\vect{w}^m(\qv))^\dg e^{-i\qv \cdot \rv} \tau_z \mathcal{X} e^{i\qv_0 \cdot \rv}\vect{w}^n(\qv_0)=i\delta_{\qv,\qv_0}(\vect{w}^m(\qv))^{\dg}{\tau_z \del_{q_x}}\vect{w}^n(\qv)$~\cite{Karplus1954,Ozawa2018}.
    By integrating the excitation rate~\eqref{eq:supp:ETDPT-excitation-rate-lattice} over $\omega$, we arrive at 
        \begin{align}
            \Gamma^n_{x,\text{int}}=&\int d\omega \, \Gamma^n_{x}(\omega)=
            2\pi A^2
            \sum_{m\neq n}s_m 
                \abs{(\vect{w}^m(\qv_0))^\dg \tau_z \del_{q_x} \vect{w}^n(\qv_0)}^2 
                \nonumber \\
            =&\sum_{\substack{m \neq n}}\left[ (\del_{q_x} \vect{w}^n(\qv_0))^\dg
            \tau_z (P^m)(\del_{q_x} \vect{w}^n(\qv_0))\right]=2\pi A^2 g^n_{xx}(\qv_0),
            \label{eq:supp:ETDPT-integrated-rate-lattice}
        \end{align}
    where in the last step we have used~\eqref{eq:supp:SQGT-SpectralForm} to identity
    the symplectic quantum metric.
    Thus, the integrated excitation rate after shaking the lattice along the $x$ direction 
    is directly proportional 
    to the diagonal component $g^n_{xx}(\qv_0)$ of the symplectic quantum metric tensor.
    By similarly shaking the lattice along the $y$ direction, the diagonal component $g^n_{yy}(\qv_0)$
    is obtained.

    To obtain the remaining off-diagonal components of the symplectic quantum geometric tensor, 
    i.e., the symplectic quantum metric $g^n_{xy}(\qv_0)t$ and the symplectic Berry curvature $B^n_{xy}(\qv_0)$,
    we now imagine shaking the lattice along both the $x$ and $y$ directions~\cite{Tran2017,Ozawa2018}, 
    which is described by
        \begin{align}
            D_{x\pm y}(t)=D+2A(\cos(\omega t)\mathcal{X} \pm \cos(\omega t-\varphi)\mathcal{Y})
            =D+V(t), \ V(t)=2A(\cos(\omega t)\mathcal{X} \pm \cos(\omega t-\varphi)\mathcal{Y}),
            \label{eq:supp:dynamical-matrix-periodic-shaking-xy}
        \end{align}
    where here $\mathcal{Y}=\sigma_z \otimes y$ is the (pseudo-hermitian) 
    $y$-position operator (extended to Nambu space) with $y$ being the usual position operator.
    
    For completeness, we also provide the paraunitarity rotation to the co-moving frame for this case 
   ~\eqref{eq:supp:dynamical-matrix-periodic-shaking-xy}
        \begin{align}
            P(t)=\exp\left(-i\int dt' V(t')\right)=
            \exp\{-i(A/\omega) (\sin(\omega t)\mathcal{X}\pm \sin(\omega t-\varphi)\mathcal{Y})\},
        \end{align}
    which then leads to the following dynamical matrix in the co-moving frame 
   ~\eqref{eq:supp:ETDSE-InteractionPicture-DynMat}
        \begin{align}
            D_{x\pm y}'(t)=D(\qv) + \frac{2A}{\omega}
            \left(\sin(\omega t)\del_{q_x}D(\qv)\pm \sin(\omega t-\varphi)\del_{q_y}D(\qv)\right)
            +\mathcal{O}((A/\omega)^2).
            \label{eq:supp:DynMat-IntPic-periodic-shaking-xy}
        \end{align}
    This allows us again to see the connection to the general discussion in the main text.
    We emphasize that~\cref{eq:supp:DynMat-IntPic-periodic-shaking-xy} is in direct analogy 
    to the results for the conventional case in~\Ccite{Tran2017}.

    Then, in the same fashion as before 
   ~\eqref{eq:supp:total-exc-rate}-\eqref{eq:supp:ETDPT-integrated-rate-lattice}, 
    computing the integrated excitation rate leads to the emergence of the off-diagonal components of the SQGT
    depending on the phase offset $\varphi$,
        \begin{align}
            \begin{split}
                \Gamma^{n,\pm}_{\text{int}}=2\pi A^2
                \times 
                \begin{cases}
                    (g^n_{xx}(\qv_0)\pm 2 g^n_{xy}(\qv_0) +g^n_{yy}(\qv_0)),\!\!\! &\text{for}\, \varphi =0,\\
                    (g^n_{xx}(\qv_0)\mp B^n_{xy}(\qv_0) +g^n_{yy}(\qv_0)),\!\!\! &\text{for}\, \varphi =\pi/2.
                \end{cases}
            \end{split}
        \end{align}
    By taking the differential integrated rate~\cite{Tran2017}
    $\Delta \Gamma ^n_{\text{int}}= \Gamma^{n,+}_{\text{int}}-\Gamma^{n,-}_{\text{int}}$ 
    of the two protocols with the opposite sign in $D_{x\pm y}(t)$
   ~\eqref{eq:supp:dynamical-matrix-periodic-shaking-xy}, we obtain
        \begin{align}
            \Delta\Gamma^n_{\text{int}}=\begin{cases}
                \ \ 8\pi A^2 g^n_{xy}(\qv_0),\!\!\! &\text{for}\, \varphi =0,\\
                -4\pi A^2 B^n_{xy}(\qv_0),\!\!\! &\text{for}\, \varphi =\pi/2.\end{cases}
                \label{eq:supp:ETDPT-integrated-rate-offdiag-lattice}
        \end{align}
    Depending on the phase offset $\varphi$ of the periodic modulation in~\eqref{eq:supp:dynamical-matrix-periodic-shaking-xy},
    we see that the differential integrated rate is directly proportional to the off-diagonal
    symplectic quantum metric $g^n_{xy}(\qv_0)$ or the 
    symplectic Berry curvature $B^{n}_{xy}(\qv_0)$.

    All the prior formalism, which assumed a single populated Bogoliubov mode at some quasimomentum $\qv_0$,
    may also be generalized to an initial Bogoliubov Bloch wave packet, e.g., a weighted average
    of multiple Bogoliubov (particle) modes within the same band $n$, i.e., 
    $\vect{\phi}^n=\sum_{\qv} c(\qv) e^{i\qv \cdot \rv}\vect{w}^n(\qv)$,
    where the wave packet $\vect{\phi}^n$ is normalized with respect to the Bogoliubov norm 
    $(\vect{\phi}^n)^\dg {\tau_z}\vect{\phi}^n=\sum_{\qv} \abs{c(\qv)}^2=1$~\cite{Zhang2006}. 
    One may further assume that the wave packet is localized around some central quasimomentum mode 
    $\qv_0=\sum_{\qv} \abs{c(\qv)}^2\qv$, as well as around some central position $\rv_c$,
    which is given by $\rv_c=(\vect{\phi}^n)^\dg {\tau_z} \rv \vect{\phi}^n$~\cite{Zhang2006}.
    This then leads simply leads to the integrated rates being proportional 
    to the weighted average of the symplectic quantum geometric quantities, e.g., for the linear shaking 
    protocol of~\cref{eq:supp:dynamical-matrix-periodic-shaking-x}, the integrated rate becomes 
        \begin{align}
            \Gamma^n_{x,\text{int}}=2\pi A^2\sum_{\qv}\abs{c(\qv)}^2 g^n_{xx}(\qv),
            \label{eq:supp:ETDPT-integrated-rate-lattice-wavepacket}
        \end{align}
    which is in direct analogy to the proposals for probing the quantum geometry 
    in particle-number-conserving systems~\cite{Ozawa2018}.
    
\section{Time-dependent perturbation theory for many-body Bogoliubov Fock states}{\label{app:TDPT-MB-Bog-Fock}}
    In the main text we have shown that the symplectic quantum geometric tensor can be extracted from 
    integrated excitation rates to other Bogoliubov modes in response to weak periodic modulations of 
    bosonic BdG systems~\cref{eq:dynamical-matrix-periodic-modulation}.
    For this, we have used the results of first-order perturbation theory for the effective Schr{\"o}dinger equation 
   ~\eqref{eq:supp:ETDSE-InteractionPicture-Dyson-TransitionAmp-Prob-2}.
    In this appendix, we now show that this derivation is equivalent to applying 
    standard first-order time-dependent perturbation theory (Fermi's Golden rule) to 
    treat transitions between many-body Bogoliubov Fock states with quasiparticle excitations present.

    Let us first consider the general case, where, in addition to the already solved (unperturbed) problem of~\eqref{eq:supp:BdG-Ham}, 
    we apply a generic time-dependent particle-number-conserving perturbation of the form
    $V(t)=\sum_{i,j}  \hat{a}_i^\dagger A_{ij}(t) \hat{a}_j$ to the system~\eqref{eq:supp:BdG-Ham}.
    To understand how this perturbation affects the population of Bogoliubov quasiparticle 
    excitations, we express $V(t)$ via the diagonal (quasiparticle) basis of the unperturbed problem using 
   ~\eqref{eq:supp:paraunitary-matrix-bog-transf-gen-0}
        \begin{align}
            V(t)=\sum_{i,j}&  \hat{a}_i^\dagger A_{ij}(t) \hat{a}_j=\sum_{i,j}  A_{ij}(t)\sum_{m,m'}
            \left(\hat{b}^\dg_m(U^\dg)_{mi} +\hat{b}_m(V^T)_{mi}\right)A_{ij}(t)
            \left( U_{jm'} \hat{b}_{m'}+(V^*)_{jm'}\hat{b}^\dg_{m'}\right)
            \nonumber \\
            =\sum_{m,m'}& [ \hat{b}^\dg_m \hat{b}_{m'} \sum_{ij} (U^\dg)_{mi} A_{ij}(t) U_{jm'}
            +\hat{b}_m \hat{b}^\dg_{m'} \sum_{ij} (V^T)_{mi} A_{ij}(t) (V^*)_{jm'} 
            \nonumber \\
            +&\hat{b}^\dg_m \hat{b}^\dg_{m'} \sum_{ij} (U^\dg)_{mi} A_{ij}(t) (V^*)_{jm'}
            +\hat{b}_m \hat{b}_{m'} \sum_{ij} (V^T)_{mi} A_{ij}(t) U_{jm'}].
            \label{eq:supp:perturbation-gen-bogoliubov-no-linear}
        \end{align}
    Note that the ground state of the original BdG Hamiltonian~\eqref{eq:supp:BdG-Ham-diag} 
    contains zero Bogoliubov quasiparticle excitations.
    In the main text, we have argued that such a generic perturbation~\eqref{eq:supp:perturbation-gen-bogoliubov-no-linear}, 
    can be used to controllably create 
    pairs of quasiparticle excitations in desired Bogoliubov energy bands via terms $\propto \hat{b}^\dg_m \hat{b}^\dg_{m'}$ or 
    $\propto \hat{b}^\dg_m$ [cf.~\cref{fig:Ebog-Sp}(a) in the main text].
    The presence of such Bogoliubov quasiparticle excitations is the starting point 
    for the protocol to extract the symplectic quantum geometric tensor from excitation rates between Bogoliubov modes, 
    as will show in the following
    [cf.~\cref{eq:ETDPT-probability} in the main text].

    \subsection{Transitions between Bogoliubov quasiparticle excitations lead to SQGT}
    \label{app:TDPT-MB-Bog-to-SQGT}
        Now, as a first simple case, we consider the transition of single Bogoliubov quasiparticle excitation from some occupied
        mode $\beta$ to some unoccupied mode $\alpha$ via the perturbation $V(t)$~\eqref{eq:supp:perturbation-gen-bogoliubov-no-linear}.
        As an initial state $\ket{\phi_{\beta}}$, we then choose the Bogoliubov Fock state
        with a single quasiparticle excitation in some mode $\beta$ and, i.e., $\ket{\phi_{\beta}}=\ket{\ldots,(0)_{\alpha},(1)_{\beta},\ldots}$
        and as a final state the state with an excitation in some mode $\alpha$, i.e., $\ket{\psi_{\alpha}}=\ket{\ldots,(1)_{\alpha},(0)_{\beta},\ldots}$.
        Standard first-order time-dependent perturbation theory~\cite{Diu2019}, tells us that we have to look at matrix elements 
        $\expval{\psi_{\alpha}}{V(t)}{\phi_{\beta}}$ to analyze the transitions between the initial and final states.
        So the only matrix elements of the perturbation $\expval{\psi_{\alpha}}{V(t)}{\phi_{\beta}}$ that contribute to the transition
        are only the ones which are number-conserving in the Bogoliubov basis, such that using~\eqref{eq:supp:perturbation-gen-bogoliubov-no-linear} 
        we obtain 
            \begin{align}
                &\expval{\psi_{\alpha}}{V(t)}{\phi_{\beta}}=
                \bra{\psi_{\alpha}}\sum_{m,m'} \left[ \delta_{\alpha,m}\delta_{\beta,m'}\hat{b}^\dg_m \hat{b}_{m'}
                \sum_{ij} (U^\dg)_{m i} A_{ij}(t) U_{jm'}
                +\delta_{\alpha,m'}\delta_{\beta,m}\hat{b}_m \hat{b}^\dg_{m'}
                \sum_{ij} (V^T)_{m i} A_{ij}(t) (V^*)_{jm'}
                \right]
                \ket{\phi_{\beta}}
                \nonumber \\
                &=\sum_{ij} (U^\dg)_{\alpha i} A_{ij}(t) U_{j\beta} \expval{\psi_{\alpha}}{\hat{b}^\dg_{\alpha} \hat{b}_{\beta}}{\phi_{\beta}}
                + (V^T)_{\beta i} A_{ij}(t) (V^*)_{j\alpha} \expval{\psi_{\alpha}}{\hat{b}_{\beta} \hat{b}^\dg_{\alpha}}{\phi_{\beta}}
                \nonumber \\
                &=\sum_{ij} (U^\dg)_{\alpha i} A_{ij}(t) U_{j\beta} 
                + (V^\dg)_{\alpha j} (A^*)_{ji}(t) V_{i\beta} 
                =\sum_{ij}
                \begin{pmatrix}
                    U^\dg & V^{\dg}\\
                    V^T & U^{T}
                \end{pmatrix}_{\alpha i} \ 
                    \begin{pmatrix}
                        A(t) & 0\\
                        0 & A^*(t)
                    \end{pmatrix}_{ij}
                \
                \begin{pmatrix}
                    U & V^{*}\\
                    V & U^{*}
                \end{pmatrix}_{j \beta}
                \nonumber \\
                &=\sum_{ij} (W^\dg)_{\alpha i} \tilde{\mathcal{M}}_{ij}(t) W_{j \beta}=
                \vect{w}_{\alpha}^\dg \tilde{\mathcal{M}}(t) \vect{w}_{\beta},
                \ \text{with} \ \tilde{\mathcal{M}}(t)=
                \begin{pmatrix}
                    A(t) & 0\\
                    0 & A^*(t)
                \end{pmatrix},
                \label{eq:supp:transitions-single-excitation}
            \end{align}
        where in the second to last step we have introduced the $2N\times 2N$ matrix $\tilde{\mathcal{M}}(t)$ and 
        reformulated the Bogoliubov transformation in terms of the paraunitary matrix 
        $W$~\eqref{eq:supp:paraunitary-matrix-bog-transf-gen}, where both 
        the $\alpha$ and $\beta$ indices are restricted to the particle sector, i.e.,
        $1\leq \alpha,\beta\leq N$ for the equality to hold.
        In the last step of~\eqref{eq:supp:transitions-single-excitation}, we have introduced the column vector $\vect{w}_{\beta}$ given by the 
        $\beta$-th column of $W$
        and the row vector $\vect{w}_{\alpha}^\dg$ given by the $\alpha$-th row of $W^\dg$, where 
        again $1\leq \alpha,\beta\leq N$ (restriction to the particle modes).

        We can generalize this result to the case of an initial eigenstate of~\eqref{eq:supp:BdG-Ham-diag} 
        with $n_{\beta}$ quasiparticle excitations in mode $\beta$ and $n_{\alpha}$ excitations in mode $\alpha$, 
        i.e., $\ket{\phi_{\beta}}=\ket{\ldots,(n_{\alpha})_{\alpha},(n_{\beta})_{\beta},\ldots}$, and 
        a final state with $n_{\beta}-1$ quasiparticle excitations in mode $\beta$, and $n_{\alpha}+1$ 
        excitation in mode $\alpha$, i.e., $\ket{\psi_{\alpha}}=\ket{\ldots,(n_{\alpha}+1)_{\alpha},(n_{\beta}-1)_{\beta},\ldots}$.
        This will yield a similar result as in~\eqref{eq:supp:transitions-single-excitation} but with an 
        additional bosonic enhancement factor of due to 
        $\hat{b}^{\dg}_{\alpha}\hat{b}_{\beta}\ket{\phi_{\beta}}=\sqrt{n_{\beta}(n_{\alpha}+1)}\ket{\phi_{\alpha}}$,
            \begin{align}
                \expval{\psi_{\alpha}}{\mathcal{A}(t)}{\phi_{\beta}}
                =\sqrt{n_{\beta}(n_{\alpha}+1)}\vect{w}_{\alpha}^\dg \tilde{\mathcal{M}}(t) \vect{w}_{\beta}.
                \label{eq:supp:transitions-multi-excitation}
            \end{align}
        
        Now we explicitly specify the explicit form of the perturbation.
        Inspired by~\Ccite{Ozawa2018}, we assume an original parameter-dependent coefficient matrix $\mathcal{M}(\ld(t))$, where 
        $\ld(t) =(\lds^1(t),\ldots,\lds^r(t))$ is a set of time-dependent parameters.
        We first restrict ourselves only to have parameter dependence in the diagonal (particle-number-conserving) blocks of 
        the coefficient matrix
        $\mathcal{M}(\ld(t))$, i.e., $K(\ld(t))$ and $G=\text{const.}$ (cf.~\cref{eq:supp:BdG-Ham}).
        The system shall initially be prepared at some parameter values $\ld=\ld_0$.
        Let us then, for simplicity, consider a single parameter modulation $\lds^1(t)$ of the form 
        $\lambda^1(t)=\lambda^1_0+2(A/\omega)\cos(\omega t)$ with $A/\omega \ll 1$.
        The rest of the parameters are assumed to be constant, i.e., $\lds^{\nu}(t)=\lds^{\nu}_0$ for $\nu \neq 1$.
        We then Taylor-expand the relevant particle-number-conserving perturbation $K(\ld(t))$ to first order
            \begin{align}
                &K(\ld(t))=K(\ld_0)+\del_{1}K(\ld_0)\times (2A/\omega)\cos(\omega t)=K(\ld_0)+\tilde{K}(t),  
                \label{eq:supp:perturbation-K-tilde-exp} \\
                &\text{with} \ \tilde{K}(t)=f(t)\tilde{K}, \ f(t) \equiv (2A/\omega)\cos(\omega t), \ \tilde{K} \equiv \del_{1}K(\ld_0).
                \nonumber
            \end{align}
        Here, the matrix $(\tilde{K})_{ij}(t)$ plays the role of the matrix $A_{ij}(t)$ in the general perturbation considered in 
       ~\cref{eq:supp:perturbation-gen-bogoliubov-no-linear}.
        We note that this explicit form of the parameter modulations leading to~\eqref{eq:supp:perturbation-K-tilde-exp}
        is not arbitrary 
        but readily appears in the co-moving frame of periodically shaken systems 
       ~\eqref{eq:supp:dynamical-matrix-periodic-shaking-x}~\cite{Tran2017,Ozawa2018}.

        On the level of the Hamiltonian~\eqref{eq:supp:BdG-Ham}
        the perturbation provides an additional term of the form 
            \begin{align}
                V(t)= \sum_{ij} \hat{a}_i^\dagger \tilde{K}_{ij}(t) \hat{a}_j 
                =\left( \sum_{ij} \hat{a}_i^\dagger \tilde{K}_{ij} \hat{a}_j \right)f(t) \equiv
                \tilde{V} f(t),
                \label{eq:supp:perturbation-K-tilde-exp-2}
            \end{align}
        Now, initially, we assume the system to be in some eigenstate $\ket{\phi_{\beta}}$ of~\eqref{eq:supp:BdG-Ham-diag}.
        As before, we let us first discuss the case of 
        a single excitation and generalize to multiple excitations later.
        Thus, we start with a state with a single Bogoliubov quasiparticle excitation 
        in some mode $\beta$, i.e., $\ket{\phi_{\beta}}=\ket{\ldots,(0)_{\alpha},(1)_{\beta},\ldots}$ with the corresponding 
        energy eigenvalue equation $\hat{H}\ket{\phi_{\beta}}=\omega_{\beta}\ket{\phi_{\beta}}$~\eqref{eq:supp:BdG-Ham-diag}. 
        Our goal is to compute the transition probability to some final eigenstate $\ket{\psi_{\alpha}}$, which for now shall consist 
        of a single Bogoliubov quasiparticle excitation in some other mode $\alpha$, i.e., $\ket{\psi_{\alpha}}=\ket{\ldots,(1)_{\alpha},(0)_{\beta},\ldots}$
        with corresponding energy eigenvalue equation $\hat{H}\ket{\psi_{\alpha}}=\omega_{\alpha}\ket{\psi_{\alpha}}$.

        Then, given the periodic perturbation of type~\eqref{eq:supp:perturbation-K-tilde-exp}-\eqref{eq:supp:perturbation-K-tilde-exp-2}, 
        we can use Fermi's Golden rule 
        (first-order time-dependent perturbation theory)~\cite{Diu2019} to compute this transition probability
        at time $t$ with via
            \begin{align}
                n_{\alpha \beta}(\omega,t)&=2\pi t \abs{\expval{\psi_{\alpha}}{\tilde{V}}{\phi_{\beta}}}^2 \delta(\omega_{\alpha}-\omega_{\beta}-\omega)
                    =2\pi t \abs{\expval{\psi_{\alpha}}{\sum_{ij} \hat{a}_i^\dagger \tilde{K}_{ij} \hat{a}_j}{\phi_{\beta}}}^2
                    \delta(\omega_{\alpha}-\omega_{\beta}-\omega),
                    \label{eq:supp:Fermi-rule-1st-order}
            \end{align}
        where $\hbar \equiv 1$ in the following.
        Into~\cref{eq:supp:Fermi-rule-1st-order} we can insert the explicit expansion of the perturbation $\tilde{V}$ 
       ~\eqref{eq:supp:perturbation-K-tilde-exp-2} in terms of the 
        Bogoliubov quasiparticle operators~\eqref{eq:supp:perturbation-gen-bogoliubov-no-linear}-\eqref{eq:supp:transitions-single-excitation} 
        leading to 
            \begin{align}
                n_{\alpha \beta}(\omega,t)&=2\pi t  
                \abs{\sum_{ij} [ (U^\dg)_{\alpha i} \tilde{K}_{ij} U_{j\beta} \expval{\psi_{\alpha}}{\hat{b}^\dg_{\alpha} \hat{b}_{\beta}}{\phi_{\beta}}
                + (V^T)_{\beta i} \tilde{K}_{ij} (V^*)_{j\alpha} 
                \expval{\psi_{\alpha}}{\hat{b}_{\beta} \hat{b}^\dg_{\alpha}}{\phi_{\beta}}]}^2 \delta(\omega_{\alpha}-\omega_{\beta}-\omega)
                \nonumber \\
                    &=2\pi t \frac{A^2}{\omega^2} \abs{\vect{w}_{\alpha}^\dg (\del_{1}{\mathcal{M}})\vect{w}_{\beta}}^2
                    \delta(\omega_{\alpha}-\omega_{\beta}-\omega).
                \label{eq:supp:Fermi-rule-1st-order-bogoliubov}
            \end{align}
        Next, we compute the total excitation rate by summing~\cref{eq:supp:Fermi-rule-1st-order-bogoliubov}
        over all possible final states 
        $\ket{\psi_{\alpha}}$ with a single excitation in some mode $\alpha$,
            \begin{align}
                \Gamma^{\alpha}(\omega)=\frac{1}{t} \sum_{ \alpha \neq \beta } n_{\alpha \beta}(\omega,t)=
                \sum_{ \alpha \neq \beta} 2\pi  \frac{A^2}{\omega^2} \abs{\vect{w}_{\alpha}^\dg (\del_{1}{\mathcal{M}})\vect{w}_{\beta}}^2
                \delta(\omega_{\alpha}-\omega_{\beta}-\omega)
            \end{align}
        Then, inspired by~\Ccite{Tran2017}, we introduce the integrated excitation rate $\Gamma^{\alpha}_{\text{int}}$
            \begin{align}
                \Gamma^{\alpha}_{\text{int}}=\int d\omega \, \Gamma^{\alpha}(\omega)
                =2\pi A^2 
                \sum_{\alpha\neq \beta, \  
                \alpha=1}^{N}
                \frac{\abs{\vect{w}_{\alpha}^\dg (\del_{1}\mathcal{M})\vect{w}_{\beta}}^2}
                {(\omega_{\alpha}-\omega_{\beta})^2} .
                \label{eq:supp:integrated-excitation-rate-part-sector}
            \end{align}
        The last expression already resembles the spectral representation of the symplectic quantum metric 
       ~\eqref{eq:supp:SQGT-SpectralForm},
        although not exactly, since the sum over $\alpha$ in~\eqref{eq:supp:integrated-excitation-rate-part-sector} is only
        restricted to the particle Bogoliubov modes $(\vect{w}_{\alpha}^\dg)_i=(W^\dg)_{\alpha i}$ with 
        $1\leq \alpha \leq N$.
        However, since the spectral representation of the symplectic quantum metric is defined via the sum
        over all Bogoliubov modes, including the hole modes with $N+1\leq \alpha \leq 2N$, 
        we need to extend the sum in~\eqref{eq:supp:integrated-excitation-rate-part-sector}.
        This is allowed by the fact that the operator $\del_{1}\mathcal{M}$ 
        (with only particle-number-conserving parts)
        does not couple the particle and hole Bogoliubov modes.
        By this, we arrive at 
            \begin{align}
                \Gamma_{\text{int}}=\int d\omega \sum_{\alpha} \Gamma^{\alpha}(\omega)
                =2\pi A^2 
                \sum_{\alpha\neq \beta, \  
                \alpha=1}^{2N}s_{\alpha}
                \frac{\abs{\vect{w}_{\alpha}^\dg (\del_{1}\mathcal{M})\vect{w}_{\beta}}^2}
                {(\tilde{\omega}_{\alpha}-\omega_{\beta})^2}=
                2\pi A^2 g_{11}^{\beta}.
                \label{eq:supp:integrated-excitation-rate}
            \end{align}
        We observe that the integrated excitation rates between states $\ket{\phi_{\beta}}$ and $\ket{\psi_{\alpha}}$, 
        with single Bogoliubov quasiparticle excitations, give rise to the symplectic quantum metric~\eqref{eq:supp:SQGT-SpectralForm}.
        This agrees with the result derived from first-order perturbation theory 
        for the effective Schr{\"o}dinger equation 
       ~\eqref{eq:supp:ETDSE-InteractionPicture-Dyson-TransitionAmp-Prob-2} 
        [\cref{eq:ETDPT-integrated-rate-SQG} in the main text].

        As before this result can be generalized, to the case of an initial Bogoliubov Fock state
        with $n_{\beta}$ quasiparticle excitations in mode $\beta$ and $n_{\alpha}$ excitations in mode $\alpha$,
        i.e., $\ket{\phi_{\beta}}=\ket{\ldots,(n_{\alpha})_{\alpha},(n_{\beta})_{\beta},\ldots}$, and
        a final state with $n_{\beta}-1$ quasiparticle excitations in mode $\beta$, and $n_{\beta}$
        excitation in mode $\alpha$, i.e., $\ket{\psi_{\alpha}}=\ket{\ldots,(n_{\alpha}+1)_{\alpha},(n_{\beta}-1)_{\beta},\ldots}$.
        This will lead to an additional bosonic enhancement factor of $\sqrt{n_{\beta}}$ 
        in the transition matrix elements as in~\eqref{eq:supp:transitions-multi-excitation}.
        The energy difference between the initial and final state is still given by 
        $[(n_{\alpha}+1)\omega_{\alpha}+\omega_{\beta}(n_{\beta}-1)]-[\omega_{\beta}n_{\beta}-\omega_{\alpha}n_{\alpha}]=\omega_{\alpha}-\omega_{\beta}$ 
        such that we can directly generalize the Fermi's Golden rule result of 
       ~\cref{eq:supp:integrated-excitation-rate} to
            \begin{align}
                \Gamma_{\text{int}}=\int d\omega \sum_{\alpha} \Gamma^{\alpha}(\omega)
                =n_{\beta}(n_{\alpha}+1)\times  2\pi A^2 g_{11}^{\beta}.
                \label{eq:supp:integrated-excitation-rate-multi-excitation}
            \end{align}

\section{Adiabatic perturbation theory for the effective Schr{\"o}dinger equation}{\label{app:ADPT}}
    In this appendix, we derive the first order correction to the adiabatic theorem for the
    time evolution of Bogoliubov modes~\eqref{eq:supp:Effective-TDSE-Vector}.
    This is a prerequisite for deriving the symplectic anomalous velocity term in the following 
   ~\cref{app:Sympl-Anom-Vel}.
    To this end, we adapt the standard theory of adiabatic perturbation theory for conventional 
    systems~\cite{Rigolin2008,DeGrandi2010} to the symplectic Bogoliubov structure here.

    We consider a time-dependent dynamical matrix $D(t)\equiv D(\ld(t))$, where the time dependence
    is encoded via some dimensionless parameters $\ld(t)=(\lambda^1(t),\dots,\lambda^r(t))$.
    At each point in time $t$, we assume that the Bogoliubov system is thermodynamically stable 
    (the coefficient matrix $\mathcal{M}(\ld(t))$ is positive definite $\forall t$~\cite{Flynn2020}),
    such that an instantaneous eigenvalue equation of the following form holds 
   ~\eqref{eq:supp:gen-eigenvect-eq-vector}
        \begin{align}
            D(t)\vect{w}^n(t)=\tilde{\Omega}_n(t)\vect{w}^n(t),
            \label{eq:supp:gen-eigenvect-eq-vector-instantaneous}
        \end{align}
    where the time-dependence of the Bogoliubov modes 
    $\vect{w}^n(t)$ is again specified through the parameters $\ld(t)$.
    In what follows, we assume a non-degenerate spectrum 
    $\tilde{\Omega}_n\neq \tilde{\Omega}_m$ for $n\neq m$ for all times $t$.
    This assumption can be relaxed 
    both regarding states that are not coupled to $\vect{w}^n$ as a result of symmetry 
    and when considering the adiabatic theorem 
    with respect to a degenerate eigenspace, which is however beyond the scope of this work.
    We are interested in the time-evolution an arbitrary state $\vect{\psi}(t)$ in the space spanned by the Bogoliubov modes
   ~\eqref{eq:supp:indefinite-inner-product-expansion},
        \begin{align}
            \vect{\psi}(t)=\sum_{n} c_n(t)\vect{w}^n(t),
            \label{eq:supp:indefinite-inner-product-expansion-instantaneous}
        \end{align}
    which is governed by the effective time-dependent Schr{\"o}dinger equation~\eqref{eq:supp:Effective-TDSE-Vector},
        \begin{align}
            -i \del_t \vect{\psi}(t)=D(t) \vect{\psi}(t).
            \label{eq:supp:ETDSE-Vector-ADPT}
        \end{align}
    Now, under the assumption of slow parameter variations of $\ld(t)$ (adiabatic limit), 
    we would like to solve~\cref{eq:supp:ETDSE-Vector-ADPT} perturbatively.
    To this end, we first insert the instantaneous expansion of $\vect{\psi}(t)$
   ~\eqref{eq:supp:indefinite-inner-product-expansion-instantaneous} into 
    the effective Schr{\"o}dinger equation~\eqref{eq:supp:ETDSE-Vector-ADPT}, 
    then project onto a specific mode via multiplying with
    $s_m(\vect{w}^m)^\dg\tau_z$ from the left,
    by which we arrive
        \begin{align}
            \dot{c}_m +\sum_n c_n s_m(\vect{w}^m)^\dg\tau_z \del_t \vect{w}^n
            =i\tilde{\Omega}_m c_m.
        \end{align}
    The explicit time-dependence will be omitted for brevity in the following.
    Next, we perform a gauge transformation 
    $c_m(t)\rightarrow c_m(t)e^{i\xi_m(t)}$,
    where $\xi_m(t) =\int_0^t dt' \tilde{\Omega}_m(t')$.
    This removes the dynamical phase factor from the equations, which will, however, 
    obviously be reintroduced at the end.
    This leads to the following differential equation for the time-dependent coefficients $c_m(t)$
        \begin{align}
            \dot{c}_m &=-\sum_n c_n s_m(\vect{w}^m)^\dg\tau_z \del_t \vect{w}^n e^{i\xi_{nm}(t)}
            \label{eq:supp:ADPT-coefficient-1}
        \end{align}
    with $\xi_{nm}(t)\equiv \xi_{n}(t)-\xi_{m}(t)$.
    We formally integrate~\eqref{eq:supp:ADPT-coefficient-1} and obtain 
        \begin{align}
            c_m(t)=c_m(0)-\int_0^t  dt' \sum_n c_n(t') s_m
            (\vect{w}^m)^\dg\tau_z \del_{\mu} \vect{w}^n\dot{\lambda}^{\mu}e^{i\xi_{nm}(t')}
            \label{eq:supp:ADPT-coefficient-2}
        \end{align}
    where we have used that 
    $\del_t \vect{w}^n(\ld(t))=\sum_{\mu}\del_{\mu} \vect{w}^n \dot{\lambda}^{\mu} \equiv \del_{\mu} \vect{w}^n \dot{\lambda}^{\mu}$
    with $\del_{\mu} \equiv \frac{\del}{\del \lambda^{\mu}}$.
    
    Let us now consider the situation, where the system is initially 
    occupying a single Bogoliubov mode, i.e., $\vect{\psi}(0)=\vect{w}^m$ 
    in~\eqref{eq:supp:indefinite-inner-product-expansion-instantaneous},
    such that $c_m(0)=1$ and $c_n(0)=0$ for $n\neq m$.
    In the limit of $\dot{\lambda}^{\mu} \to 0$ all transition amplitudes to other modes
    are highly suppressed, and the only leading order contribution in~\eqref{eq:supp:ADPT-coefficient-2} 
    stems from the diagonal $n=m$ term in the sum~\cite{DeGrandi2010}. 
    Thus, to zeroth order, the solution for $c_m(t)$ is given by
        \begin{align}
            c_m(t)=c_m(0)e^{i\gamma^{m}(t)}. 
            \label{eq:supp:ADPT-coefficient-sol-diag}
        \end{align}
    Here, $\gamma^m$ denotes the symplectic version of the Berry phase~\cite{Berry1984}
        \begin{align}
            \gamma^m(t)=i \int_0^t dt' s_m(\vect{w}^m)^\dg\tau_z \del_{\mu} \vect{w}^m\dot{\lambda}^{\mu}
            = \int_{\ld(0)}^{\ld(t)} A_{\mu}^m(\ld) d\lambda^{\mu}=\int A^m ,
            \label{eq:supp:ADPT-sympl-Berry-phase}
        \end{align}
    which may also written as an integral over the symplectic Berry connection~\cite{Zhang2006,Shindou2013,Xu2020,Peano2016}
    \begin{align}
        A^m=A^m_{\mu} d\lambda^{\mu}=[is_m(\vect{w}^m)^\dg\tau_z \del_{\mu} \vect{w}^m] d\lambda^{\mu}.
        \label{eq:supp:ADPT-sympl-Berry-connect}
    \end{align}
    The result~\eqref{eq:supp:ADPT-coefficient-sol-diag} represents 
    the conventional adiabatic theorem (adiabatic limit) adapted to the 
    time evolution Bogoliubov modes, i.e., under slow parameter variations the system will remain
    in the initially occupied mode while acquiring an additional symplectic Berry phase $\gamma^m(t)$.
    We stress that this represents the $0$-th order contribution, since Berry phase is independent
    of the rate of change in the parameters 
    $\dot{\lambda}^{\mu}$ [cf.~\cref{eq:supp:ADPT-sympl-Berry-phase}].

    Next, we aim to solve~\eqref{eq:supp:ADPT-coefficient-2} perturbatively, to self-consistently
    obtain corrections to first order in $\dot{\lambda}^{\mu}$. 
    To this end, we insert the $0$-th order solution $c_m(t)$~\eqref{eq:supp:ADPT-coefficient-sol-diag}
    into the right-hand side of the~\cref{eq:supp:ADPT-coefficient-2}, by which we obtain
        \begin{align}
            c_m(t)& \approx c_m(0)- \int_0^t dt'\sum_n c_n(0)e^{i\gamma^n(t')} s_m
            (\vect{w}^m)^\dg\tau_z \del_{\mu} \vect{w}^n\dot{\lambda}^{\mu}e^{i\xi_{nm}(t')}
            \label{eq:supp:ADPT-coefficient-3}
        \end{align}
    Now for all other coefficients $c_k(t)$ with $k\neq m$,
    we only need to consider the $n=m$ term within the sum in~\eqref{eq:supp:ADPT-coefficient-3},
    where $m$ is again the index of the initially occupied mode.
    To first order, this leads to the following equation for $c_{k\neq m}(t)$
        \begin{align}
            c_k(t)&=-\int_0^t dt'c_m(0)e^{i\gamma^m(t')}  s_k(\vect{w}^k)^\dg\tau_z \del_{\mu}
             \vect{w}^m\dot{\lambda}^{\mu}e^{i\xi_{km}(t')}
             \nonumber \\
                &=i\int_0^t dt'e^{i\gamma^m(t')}  s_k \frac{(\vect{w}^k)^\dg\tau_z \del_{\mu}
                \vect{w}^m}{\tilde{\Omega}_{km}}\dot{\lambda}^{\mu}\frac{d}{dt'}e^{i\xi_{km}(t')},
        \label{eq:supp:ADPT-coefficient-offdiag-1}
        \end{align}
    with $\tilde{\Omega}_{km} \equiv \tilde{\Omega}_k-\tilde{\Omega}_m$ and where 
    we have used that $\frac{d}{dt'}e^{i\xi_{km}(t')}=\tilde{\Omega}_{km}e^{i\xi_{km}(t')}$.
    Due to the fast oscillating exponential factor in ~\cref{eq:supp:ADPT-coefficient-offdiag-1}, 
    we can integrate this expression by parts~\cite{DeGrandi2010} and obtain the following 
    solution for the off-diagonal coefficients
        \begin{align}
            c_k(t)=i s_k e^{i\gamma^m(t)} \frac{(\vect{w}^k)^\dg\tau_z \del_{\mu}
            \vect{w}^m}{\tilde{\Omega}_{km}}\dot{\lambda}^{\mu}e^{i\xi_{km}(t)}.
            \label{eq:supp:ADPT-coefficient-offdiag-2}
        \end{align}
    Thus, to first order in $\dot{\lambda}^{\mu}$, 
    the state $\vect{\psi}(t)$~\eqref{eq:supp:indefinite-inner-product-expansion-instantaneous}
    is finally given by
        \begin{align}
            \vect{\psi}(t)=&e^{-i(\xi_m(t)-\gamma^m(t))} \left[\vect{w}^m
            +i\sum_{k\neq m}\vect{w}^k s_k
            \frac{(\vect{w}^k)^\dg \tau_z \del_{\mu}\vect{w}^m}{(\tilde{\Omega}_{km})}
            \dot{\lambda}^{\mu} \right]
            \nonumber \\
            =&e^{-i(\xi_m(t)-\gamma^m(t))} \left[\vect{w}^m
            +i\sum_{k\neq m}\vect{w}^k s_k\frac{(\vect{w}^k)^\dg \del_{\mu}\mathcal{M}\vect{w}^m}
            {(\tilde{\Omega}_{km})^2}\dot{\lambda}^{\mu} \right],
        \label{eq:supp:ADPT-coefficient-offdiag-final}
        \end{align}
    where we have reintroduced the dynamical phase $c_n(t)\to c_n(t)e^{-i\xi_n(t)}$
    and utilized the symplectic version of the Hellmann-Feynman theorem
   ~\eqref{eq:supp:sympl-relation-derivative-eigenvectors}
    in the second equality.
   \cref{eq:supp:ADPT-coefficient-offdiag-final} constitutes the first
    order correction (in $\dot{\lambda}^{\mu}$) 
    to the adiabatic theorem for the time evolution of Bogoliubov modes.

    If required, one may also normalize the state $\vect{\psi}(t)$~\eqref{eq:supp:ADPT-coefficient-offdiag-final} 
    up to second order~\cite{Kolodrubetz2013},
    i.e., up to $(\dot{\lambda}^{\mu})^2\equiv (v^{\mu})^2$, which leads to
        \begin{align}
            \vect{\psi}(t)=e^{-i(\xi_m(t)-\gamma^m(t))} \left[(1+\beta_{\mu}^m (v^{\mu})^2 )\vect{w}^m
            +iv^{\mu}\sum_{k\neq m}\alpha^{km}_{\mu}\vect{w}^k\right], 
        \end{align}
    where have defined
    $\alpha^{km}_{\mu}\equiv s_k\frac{(\vect{w}^k)^\dg \del_{\mu}\mathcal{M} \vect{w}^m}{(\tilde{\Omega}_{km})^2}$
    with $\beta_{\mu}^m \equiv -\frac{1}{2}s_m\sum_{k\neq m} \abs{\alpha^{km}_{\mu}}^2$.

\section{Symplectic anomalous velocity}{\label{app:Sympl-Anom-Vel}}
    In this appendix, we provide a more detailed derivation of the symplectic anomalous velocity term,
    which is proportional to the symplectic Berry curvature. 
    Consider a translationally invariant BdG system such that the Bogoliubov eigenmodes of the
    dynamical matrix $D$ admit Bloch function type solution, i.e., $D\vect{\psi}^n(\qv)
    =De^{i\qv \cdot \rv}\vect{w}^n(\qv)=\tilde{\Omega}_n(\qv)e^{i\qv \cdot \rv}\vect{w}^n(\qv)$.
    Here, $\vect{w}^n(\qv)$ is the cell-periodic part of the Bogoliubov Bloch function for the $n$-th band, 
    which fulfill the Bogoliubov eigenvalue equation for the quasimomentum transformed dynamical matrix 
    $D(\qv)\vect{w}^n(\qv)=\tilde{\Omega}_n(\qv)\vect{w}^n(\qv)$.
    To make connections to the conventional anomalous velocity, we consider a two-dimensional
    time-dependent quasimomentum parameter space $\qv(t)=(q_x(t),q_y(t))$.
    We imagine a uniform force $\Fv$ to be applied to the Bogoliubov 
    system of interest, such that quasimomenta satisfy
    $\dot{\qv}=\Fv$, leading to $\qv(t)=\qv(0)+\Fv t$.
    
    Next, we consider an initially populated Bogoliubov particle mode $\vect{\psi}(0)=\vect{w}^n(\qv)$ and 
    let the applied field point in the $x$-direction, i.e., $\Fv=F_x \vect{e}_x$.
    Using the results of ~\cref{eq:supp:ADPT-coefficient-offdiag-final},
    we see that the solution to the time-dependent effective Schr{\"o}dinger equation for
    the Bogoliubov modes is, to first order in $F_x$, given by
        \begin{align}
            \vect{\psi}(\qv)=\vect{w}^n(\qv)
            +iF_x\sum_{m\neq n}\vect{w}^m(\qv)
            s_n\frac{(\vect{w}^m(\qv))^\dg \del_{q_x}\mathcal{M} \vect{w}^n(\qv)}
            {(\tilde{\Omega}_{n}(\qv)-\tilde{\Omega}_{m}(\qv))^2}+\mathcal{O}(F_x^2),
            \label{eq:supp:ADPT-anomal-corretion-state}
        \end{align}
    where the irrelevant phase factors have been omitted for clarity.
    To obtain the velocity in the $y$-direction to first order in $F_x$, we take the expectation value of 
    the quasiparticle velocity operator $v_y(\qv)=\del_{q_y}D(\qv)$
    with respect to the state of~\cref{eq:supp:ADPT-anomal-corretion-state},
    where $D(\qv)$ is again the dynamical matrix in the quasimomentum space representation.
    Note that $v_y(\qv)=\del_{q_y}D(\qv)$ 
    just follows from the Heisenberg equations of motion for the position operator $\rv$ 
    transformed into the quasimomentum space~\cite{Xiao2010} and then extended to Nambu space.
    The velocity then equates to 
    \begin{align}
        \begin{split}
            \ep{v_y(\qv)}&=(\vect{\psi}(\qv))^\dg \tau_z\del_{q_y}D(\qv)\vect{\psi}(\qv)\\
            &=(\vect{w}^n)^\dg \del_{q_y}\mathcal{M}(\qv)\vect{w}^n+\left(iF_x\sum_{m\neq n}
            s_n\frac{\vect{w}^n\del_{q_y}\mathcal{M}\vect{w}^m \times
            (\vect{w}^m)^\dg \del_{q_x}\mathcal{M} \vect{w}^n }{(\tilde{\Omega}_{n}-\tilde{\Omega}_{m})^2}
            +\text{c.c.}\right)+\mathcal{O}(F_x^2).
        \end{split}
    \end{align}    
    By making use of the eigenvalue equation for the Bogoliubov modes, 
    $D(\qv)\vect{w}^n(\qv) =\tilde{\Omega}_n(\qv) \vect{w}^n(\qv)$ 
   ~\eqref{eq:supp:gen-eigenvect-eq-vector}, 
    as well as the spectral representation of the symplectic
    Berry curvature $B^n_{\mu \nu}=-2\, \im[\eta^n_{\mu\nu}] $~\eqref{eq:supp:SQGT-SpectralForm}, 
    we see that the velocity $v_y$ of the Bogoliubov state in $n$-th band is given by
    \begin{align}
        \ep{v_y(\qv)}=\frac{\del \tilde{\Omega}_n(\qv)}{\del q_y}+F_x B^n_{xy}(\qv).
        \label{eq:supp:ADPT-anomal-corretion-velocity}
    \end{align}
    By symmetry, we can immediately generalize this result~\eqref{eq:supp:ADPT-anomal-corretion-velocity} to 
    general directions of the applied force $\Fv$
        \begin{align}
            \ep{\vv(\qv)}=\nabla_{\qv} \tilde{\Omega}_n(\qv)+\Bv_n(\qv)\times \Fv,
            \label{eq:supp:ADPT-anomal-corretion-velocity-general}
        \end{align}
    with $\Bv_n(\qv)\equiv\epsilon_{\mu\nu\kappa}B^n_{\mu\nu}(\qv)\vect{e}_\kappa$.
    Here, the first term corresponds to the slope of the occupied Bogoliubov energy band $\tilde{\Omega}_n(\qv)$ and the second term
    is proportional to the symplectic Berry curvature $\Bv_n(\qv)$.
    The second term describes a transverse velocity component 
    and may be viewed as a symplectic generalization of the conventional anomalous velocity term~\cite{Xiao2010,Sundaram1999,Chang1995}.
    
    The results above may easily be generalized to the case of an initial Bogoliubov wave packet consisting
    of a superposition (particle) Bogoliubov modes $\vect{\psi}^n(\qv)=e^{i\qv \cdot \rv}\vect{w}^n(\qv)$ of a
    given band $n$, 
    \begin{align}
        \vect{\phi}^n=\sum_{\qv} c(\qv) e^{i\qv \cdot \rv}\vect{w}^n(\qv),
        \label{eq:supp:Wave-packet-1}
    \end{align}
    which is normalized with respect to the Bogoliubov norm, i.e., 
    $\expval{\phi^n}{\tau_z}{\phi^n}=\sum_{\qv} \abs{c(\qv)}^2=1$~\cite{Zhang2006}.
    The velocity of such a Bogoliubov wave packet may then be related to a bulk observable, 
    such as the total current density $j_y$ in real space
    \begin{align}
        j_y=\frac{1}{L^2}\sum_{\qv} \rho(\qv) \ep{v_y(\qv)}
        =\int_{1.\, \text{BZ}} d^2 \qv \ D(\rv,\qv) \rho(\qv)  \ep{v_y(\qv)},
        \label{eq:supp:Wave-packet-2}
    \end{align}
    where $L^2$ is the area of the system and 
    $\rho(\qv)=\abs{c(\qv)}^2$ the distribution function of Bogoliubov quasiparticles in the
    respective band.
    We note that special care is required when switching to the continuum of quasimomentum states in~\cref{eq:supp:Wave-packet-2}, as a so-called modified density of states $D(\rv,\qv)$ must be included for the integration when a finite Berry curvature is present (see also ~\Ccite{Zhang2006,Price2016,Xiao2005,Xiao2010}).
    Although $j_y$~\eqref{eq:supp:Wave-packet-2} is not quantized by the (symplectic) Chern number of the respective Bogoliubov band (for the typical case of a non-uniformly populated Bogoliubov band), 
    the current density still gives rise to a finite transverse velocity contribution, 
    which is directly related to the anomalous term governed by the underlying symplectic Berry curvature of the Bogoliubov band.
    Our results~\eqref{eq:supp:ADPT-anomal-corretion-velocity-general}, which are based on first order adiabatic perturbation theory, are complementary to those of \Ccite{Zhang2006}.
    The authors of \Ccite{Zhang2006} have also studied the dynamics of the Bogoliubov wave packets using a time-dependent variational principle, however, no explicit connection to the symplectic Berry curvature has been made therein.
    Based on \Ccite{Zhang2006}, we see that the anomalous velocity term can also be verified by measuring the modified thermal depletion of a condensate for a weakly interacting BEC.
    
\section{Bogoliubov Haldane model}{\label{app:BH-model}}
    In this appendix, we provide a brief derivation of the Bogoliubov-Haldane model
    considered as an example in the main text, which mainly follows~\Ccite{Furukawa2015}.
    Adding to this, we also review the numerical algorithm described in~\Ccite{Colpa1978,Shindou2013} 
    for obtaining the Bogoliubov matrices 
   ~\eqref{eq:supp:paraunitary-matrix-bog-transf-gen} 
    with the correct paraunitarity conditions~\eqref{eq:supp:paraunitary-cond}.
    
    The starting point is a Bose-Hubbard model on a two-dimensional lattice with two sublattice sites,
     \begin{align}
        \hat{H}=\sum_{\ell \ell',ss'}h_{\ell s,\ell ' s'}\hat{a}^{\dagger}_{\ell s}\hat{a}_{\ell' s'}
        +\frac{U}{2}\sum_{\ell,s} \hat{n}_{\ell s}(\hat{n}_{\ell s}-1)
         \label{eq:supp:BH-Hamiltonian}
    \end{align}
    where the elementary unit cells are labeled by $\ell$, the sublattice sites by $s \in \{A,B\}$.
    The non-interacting (quadratic) part is governed by the matrix elements $h_{\ell s,\ell ' s'}$ 
    in position space and the interacting part by the on-site interaction strength $U$.
    
    The first summand is the non-interacting part, which may diagonalized by first moving into quasimomentum space via
    \begin{align}
        \mathcal{H}_{s s'}(\qv)& \equiv \sum_{\ell \ell '}  h_{\ell s,\ell ' s'}
        e^{-i(\rv_{\ell s}-\rv_{\ell' s'})\cdot \qv}
           =\left(h_0(\qv) \mathbbm{1} +\hv(\qv)\cdot \vect{\hat{\sigma}}\right)_{s s'},
       \label{eq:supp:Hk-Pauli-exp}
   \end{align}
   where $\vect{\hat{\sigma}} = (\sigma_x,\sigma_y,\sigma_z)^T$ is the vector of Pauli matrices.
   At each quasimomentum point, $\qv$, the Hamiltonian $\mathcal{H}_{s s'}(\qv)$ describes an operator acting on a two-dimensional 
   Hilbert space, whose elements can be represented on the Bloch sphere. 
   The north and south pole are denoted by $\qa$ and $\qb$ respectively. 
   Due to the non-interacting nature of the system, the model can be further diagonalized by rotating into the eigenbasis $\qpm$,
    \begin{align}
        \begin{split}
            \qm &=\sin(\thq/2) \qa -\cos(\thq/2)e^{i \phq} \qb, \quad
            \qp =\cos(\thq/2) \qa +\sin(\thq/2)e^{i \phq} \qb.
        \end{split}
        \label{eq:supp:kvec-plusmin-param}
    \end{align}
    which reside at $\pm \hat{\hv}(\qv)$ on the Bloch sphere, with
    $\hat{\hv}(\qv)=\hv(\qv)/|\hv(\qv)|=(\cos(\phq) \sin(\thq),\sin(\phq)\sin(\thq),\cos(\thq))^T$.
    The corresponding eigenenergies $\epsilon_{\pm}(\qv)$ of $\qpm$ are given by $\epsilon_{\pm}(\qv)=h_0(\qv)\pm \abs{\hv(\qv)}$
    which define the band structure of the lattice. 
    Now, an ideal Bose gas at zero temperature [with $U=0$ in~\eqref{eq:supp:BH-Hamiltonian}]
    forms a perfect Bose-Einstein condensate (BEC)
    characterized by all bosons condensing into the lowest energy eigenstate $\ket{\qv_0,-}$ at some quasimomentum $\qv_0$ 
    in the lower band $\epsilon_{-}(\qv)$, 
    with the ground state given by $\ket{\psi_B}= (\sqrt{N!})^{-1} (\hat{a}^{\dagger}_{\qv_0 -})^{N} \vac$.
    Here, $N$ shall denote the total number of bosons in the system and $\hat{a}^{\dagger}_{\qv,\pm}$ the creation operator
    for a boson at quasimomentum $\qv$ in the lower or upper band, respectively.

\begin{figure}[bt!]
        \begin{center}
            \includegraphics{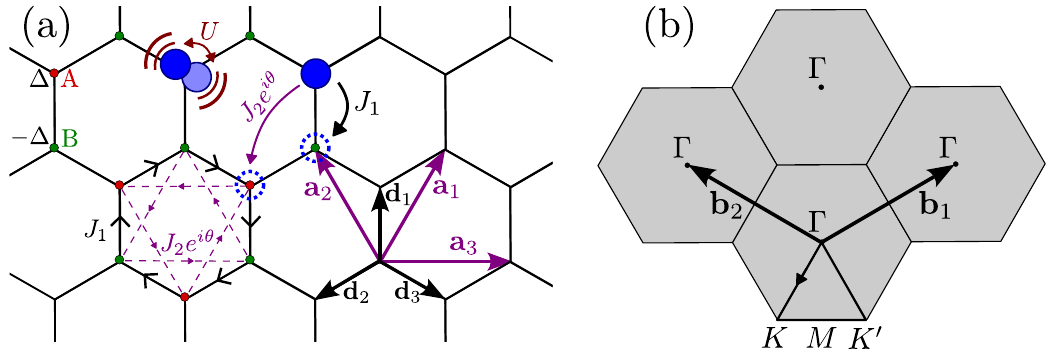}
            \caption{
            (a) Illustration of the Bogoliubov-Haldane model~\eqref{eq:supp:Bogoliubov-Hamiltonian} 
            on a hexagonal lattice with two
            sublattice sites $A$ ($B$) with potential energy offset $\Delta$ ($-\Delta$),
            real (complex-valued) tunneling matrix element $J_1$ ($J_2e^{i\theta}$) for the nearest (next-) nearest neighbors,
            and on-site interaction strength $U$.
            The translation vectors connecting the (next-) nearest neighbors ($\dv_i$) $\av_i$ 
           ~\eqref{eq:supp:translation-vectors} are shown.
            (b) Illustration of reciprocal lattice of the hexagonal lattice structure of (a) including the 
            reciprocal lattice vectors $\bv_i$~\eqref{eq:supp:reciprocal-lattice-vectors}.
            In the 1st Brillouin zone, we depict a high-symmetry path along the high-symmetry points $\Gamma$, $K$, $M$,
            and $K'$.
            }
            \label{fig:hex-latt}
        \end{center}
    \end{figure}

    In this work, we consider the Haldane model~\cite{Haldane1988} on a hexagonal lattice for
    which the non-interacting part 
   ~\eqref{eq:supp:Hk-Pauli-exp} reads, 
        \begin{align}
            \begin{split}
                    h_x(\qv) &=-J_1 \sum_i \cos (\qv \cdot \dv_i), \
                    h_y(\qv) = -J_1 \sum_i \sin (\qv \cdot \dv_i),\
                    h_0(\qv) = -2J_2 \cos(\theta) \sum_i \cos(\qv \cdot \av_i),\\
                    h_z(\qv) &= \Delta -2J_2 \sin(\theta) [\sin (\qv \cdot (-\av_1))+\sin (\qv \cdot \av_2)+\sin (\qv \cdot \av_3)], \
                \end{split}
            \label{eq:supp:Haldane-Model}
            \end{align}
    where $J_1 (J_2e^{i\theta})$ is the real nearest neighbor (complex next-nearest neighbor) hopping amplitude and 
    $\Delta$ the mass term (sublattice offset).
    The complex next-nearest neighbor hopping amplitude breaks the time reversal symmetry of the system, and
    the mass term $\Delta$ breaks the inversion symmetry of the lattice.
    On the hexagonal lattice, translation vectors connecting the nearest neighbors are denoted by $\dv_i$, 
    while the vectors $\av_i$ connect the next-nearest neighbors [Fig.~\ref{fig:hex-latt}], 
        \begin{align}
            \dv_3 &= \frac{a}{2} \begin{pmatrix} \sqrt{3} \\ -1 \end{pmatrix}, \quad
            \dv_2= \frac{a}{2} \begin{pmatrix} -\sqrt{3} \\ -1 \end{pmatrix}, \quad
            \dv_1= -(\dv_2+\dv_3) =\av \begin{pmatrix} 0 \\ 1 \end{pmatrix}, \\
            \av_3 &= \dv_3 - \dv_2,\quad \av_2 = \dv_1 - \dv_3, \quad \av_1 = \dv_1 - \dv_2,
            \label{eq:supp:translation-vectors}
        \end{align}
    where $a$ is the lattice constant.
    The reciprocal lattice (quasimomentum space) is spanned by the reciprocal lattice vectors $\bv_i$
        \begin{align}
            \bv_1 &= \frac{2\pi}{3a} \begin{pmatrix} \sqrt{3} \\ 1 \end{pmatrix}, \quad
            \bv_2= \frac{2\pi}{3a} \begin{pmatrix} -\sqrt{3} \\ 1 \end{pmatrix}, \quad
            \bv_3 = \bv_1+\bv_3,
            \label{eq:supp:reciprocal-lattice-vectors}
        \end{align}
    such that $\av_i\cdot \bv_i = 2\pi \delta_{ij}$ for $i,j \in \{1,2\}$ [see Fig.~\ref{fig:hex-latt}].
    
    Having discussed the non-interacting part, we turn to the Bogoliubov theory, which accounts for additional 
    quantum (and thermal) fluctuations on top of a mean-field solution for a weakly interacting bosonic system~\cite{Ueda2010,Pethick2008,Dalfovo1999}.
    For this, we first obtain a mean-field solution via a minimization of an associated Gross-Pitaevskii (GP) energy functional.
    The stationary solutions gained by minimizing the GP energy functional will then yield the mean-field ground state, which
    is assumed to be macroscopically occupied.
    Then, we will describe quantum fluctuations on top of the condensate, by considering
    non-condensed deviations from this mean-field solution.
    This procedure can be formally understood by replacing the operators within the Hamiltonian $\hat{H}$~\eqref{eq:supp:BH-Hamiltonian}
    via $\hat{a}_{\ell s} \to \sqrt{n_0} \zeta_{\ell s} + \hat{a}_{\ell s}$,
    where the first term denotes the complex-valued mean-field solution, $\zeta_{\ell s} \in \mathbb{C}$, and the second term
    the additional quantum fluctuations where $n_0$ denotes condensate 
    density of the system.
    Expanding the Bose-Hubbard Hamiltonian~\eqref{eq:supp:BH-Hamiltonian} in orders of the condensate 
    density $n_0$ then yields~\cite{Engelhardt2015} 
        \begin{align}
            \hat{H}= n_0 E_{\text{GP}}+n_0^{1/2} \hat{H}^{(L)}+\hat{H}^{(B)}+\mathcal{O}\left(n_0^{-1/2}\right).
            \label{eq:supp-General-Bog-Expansion-Dens}
        \end{align}
    The first term denotes the GP energy functional $ E_{\text{GP}}$ which is a function of the mean-field solutions
    $\zeta_{\ell s}$.
    The next two terms $\hat{H}^{(L)}$ and $\hat{H}^{(B)}$ are linear and quadratic in the bosonic operators
    $\hat{a}^{(\dagger)}_{\ell s}$, where the latter is known as the Bogoliubov Hamiltonian, which is sought after in the following.
    
    First, the energy functional $E_{\text{GP}}$ is found by replacing the operators $\hat{a}^{(\dagger)}_{\ell s}$ in
    $\hat{H}$ by complex-valued functions $\sqrt{n}\zeta_{\ell s}^{(*)}$, 
    where it is assumed that almost all bosons are condensed, i.e.,
    $n_0 \approx n = N/V$, with $V=2N_{\text{uc}}$ and $N$ denoting the total number of bosons in the system 
    and $N_{\text{uc}}$ the number of unit cells in the lattice~\cite{Engelhardt2015}.
    For the parameters of the bosonic Haldane model considered here, the condensate is formed at the single zero quasimomentum mode
    $\qv=\zv$ in the lower band such that a suitable ansatz for the ground state is given
    by a spatially homogeneous function $\zeta_{\ell s} = \chi_s,\ s\in\{A,B\}$.
    This leads to the following energy functional in quasimomentum space
    \begin{align}
        E_{GP}=\colvec{\chi_A^* \ \chi^*_B}(\mathcal{H}(\qv=\zv)-\mu_{\text{eff}}\mathbbm{1}_2)
        \colvec{\chi_A \\ \chi_B}+\frac{U}{2}n\left(\abs{\chi_A}^4+\abs{\chi_B}^4 \right),
        \label{eq:supp:EGP-Functional}
    \end{align}
    where we have introduced $\mu_{\text{eff}}$ as a Lagrange multiplier for satisfying the 
    conservation of the total number of bosons in the system, i.e., $\sum_{\ell s} \abs{\zeta_{\ell s}}^2 = N$.
    We minimize $E_{GP}$ by taking the variational derivative of $E_{GP}$ with respect to $\chi_{s}^{*}$ which leads to 
    the so-called GP equations, 
        \begin{align}
            \big(\mathcal{H}(\qv=\zv)-\mu_{\text{eff}}\mathbbm{1}_2\big)\colvec{\chi_A\\ \chi_B} + n U \colvec{\abs{\chi_A}^2\chi_A\\ \abs{\chi_B}^2\chi_B}=0.
            \label{eq:supp:GPE-equations}
        \end{align} 
    Since the non-interacting condensate is formed at the $\ket{\qv=\zv,-}$ mode we parametrize $\chi_s$ via the Bloch 
    sphere representation $\qm$~\eqref{eq:supp:kvec-plusmin-param}~\cite{Furukawa2015}
        \begin{align}
            \colvec{\chi_A\\ \chi_B} = \colvec{\sin(\theta/2)\\ -\cos(\theta/2)e^{i\varphi}}.
            \label{eq:supp:Mean-field-param}
        \end{align}
    Multiplying~\eqref{eq:supp:GPE-equations} by $\colvec{\chi_A^*\,\chi_B^*}$ and $\colvec{-\chi_B^*\,\chi_A^*}$ from the left
    yields three equations~\cite{Furukawa2015,Wu2017}
        \begin{align}
            &(h_0(\zv)-\mu_{\text{eff}})-h(\zv)
            [\cos(\theta_0)\cos(\theta)+\sin(\theta_0)\sin(\theta)\cos(\varphi-\varphi_0)]
            +2Un (\abs{\chi_A}^4 +\abs{\chi_B}^4)=0,\label{eq:supp:GPE-Min-eq1}\\[.2cm]
            &h(\zv)[\cos(\theta_0)\sin(\theta)\cos(\varphi)-\cos(\varphi_0)\sin(\theta_0)\cos(\theta)]
            -Un\sin(\theta)\cos(\theta)\cos(\varphi)=0,\label{eq:supp:GPE-Min-eq2}\\[.2cm]
            &(h_0(\zv)-\mu_{\text{eff}})\sin(\theta)\sin(\varphi)+h(\zv)\sin(\theta_0)\sin(\varphi_0)
            +Un\sin(\theta)\sin(\varphi)=0,\label{eq:supp:GPE-Min-eq3}
        \end{align}
    where $\hv(\zv)=h(\zv)\colvec{\cos(\varphi_0) \sin(\theta_0),\sin(\varphi_0)\sin(\theta_0),\cos(\theta_0)}^T$
    and $h_0(\zv)$ stem from~\eqref{eq:supp:Hk-Pauli-exp}.
    For a given interaction energy $U$ and density $n$, we numerically minimize the
    \crefrange{eq:supp:GPE-Min-eq1}{eq:supp:GPE-Min-eq3} and gain the stationary solutions $(\theta,\varphi,\mu_{\text{eff}})$
    specifying the mean-field ground state in~\eqref{eq:supp:Mean-field-param}.

    Since the BEC occurs in the lowest band at $\qv=\zv$, we further specify the unitary
    rotation (provided by the eigenstates of the non-interacting system~\eqref{eq:supp:kvec-plusmin-param})
    for the obtained solution~\cite{Furukawa2015}
            \begin{align}
                &\colvec{\hat{a}_{\zv, A}\\ \hat{a}_{\zv , B}}
                = U(\theta,\varphi)
                \colvec{\hat{a}_{\zv , +}\\ \hat{a}_{\zv , -}}
                = \begin{pmatrix} f_{A,+} & f_{A,-} \\
                f_{B,+} & f_{B,-} \end{pmatrix}
            \colvec{\hat{a}_{\zv , +}\\ \hat{a}_{\zv , -}},
            \\[.2cm]
            &\text{with}\
            U(\theta,\varphi)
            = \begin{pmatrix} \cos(\theta/2) & \sin(\theta/2) \\
                \sin(\theta/2)e^{i\varphi} & -\cos(\theta/2)e^{i\varphi} \end{pmatrix},
            \end{align}
    where $f_{s,\pm}(\theta,\varphi)$ with $s \in \{A,B\}$ denote the stationary solutions 
    of the GP functional~\eqref{eq:supp:EGP-Functional}.
    
    The next step is to go beyond this mean-field description and derive the corresponding Bogoliubov Hamiltonian
    $\hat{H}^{(B)}$~\eqref{eq:supp-General-Bog-Expansion-Dens}.
    The starting point is the quasimomentum space representation of the Bose-Hubbard Hamiltonian~\eqref{eq:supp:BH-Hamiltonian}
            \begin{align}
                \hat{H}=\sum_{\qv} \sum_{s,s'} \hat{a}^{\dagger}_{\qv, s}
                 \mathcal{H}_{s,s'}(\qv) \hat{a}_{\qv, s'}
                +\frac{U}{V} \sum_{\qv,\qv',\qv''}\sum_{s}
                \hat{a}^{\dagger}_{\qv+\qv'' , s}\hat{a}^{\dagger}_{\qv'-\qv'' , s}
                \hat{a}_{\qv , s}\hat{a}_{\qv' , s},
                \label{eq:supp:Ham-Full-2band-comb-momentum}
            \end{align}
    where $\hat{a}_{\qv , s} = \sum_{\ell } \braket{\qv s}{\ell s} \hat{a}_{\ell s}$ denote the bosonic operators
    transformed into quasimomentum space.
    Utilizing another rotation into the eigenbasis~\eqref{eq:supp:kvec-plusmin-param}
    the bosonic operators in quasimomentum space are decomposed into a condensed and a non-condensed part via~\cite{Furukawa2015}
        \begin{align}
            \sum_{\qv}\hat{a}_{\qv, s} &= \hat{a}_{\zv, s} +\sum_{\qv \neq \zv}\hat{a}_{\qv, s}
                = f_{s,-}\hat{a}_{\zv, -}+f_{s,+}\hat{a}_{\zv, +}+\sum_{\qv \neq \zv}\hat{a}_{\qv, s},
                \label{eq:supp:Op-Cond-decomp}
            \end{align}
    where the condensation mode is assumed to be located at $\qv=\zv$ in the lowest band and
    $f_{s,\pm} \equiv f_{s,\pm}(\varphi,\theta)$ are functions depending on the stationary solutions of the energy functional
    $E_{GP}$~\eqref{eq:supp:EGP-Functional}.

    Next, the decomposition~\eqref{eq:supp:Op-Cond-decomp} is inserted into $\hat{H}$~\eqref{eq:supp:Ham-Full-2band-comb-momentum}
    and the Hamiltonian is separated into parts of $\qv=\zv$ and $\qv\neq\zv$ while only terms up to second order
    in $\hat{a}^{\dagger}_{\qv \neq \zv, s}$ are retained, leading to an approximate Hamiltonian.
    For the actual Bogoliubov approximation, the condensation mode is then replaced via $\hat{a}_{\zv, -} \to \sqrt{N_0}$,
    where $N_0$ denotes the number of particles in the condensate mode.
    Performing all these steps will eventually lead to the following Bogoliubov Hamiltonian in quasimomentum space 
        \begin{align}
        \begin{split}
                \hat{H}^{(B)}=&
                \frac{1}{2}\sum_{\qv \neq \zv} 
                \rowvec{\hat{a}^\dg_{\qv,A} \, \hat{a}^\dg_{\qv,B}\,\hat{a}_{-\qv,A},\hat{a}_{-\qv,B}}
                \begin{pmatrix} \mathcal{H}(\qv)+\mathcal{H}_1 & 2\mathcal{H}_2 \\
                    2\mathcal{H}_{2}^{*}& (\mathcal{H}(-\qv)+\mathcal{H}_{1})^{*}
                    \end{pmatrix}
                \colvec{\hat{a}_{\qv,A} \\ 
                \hat{a}_{\qv,B} \\
                \hat{a}^\dg_{-\qv,A} \\ 
                \hat{a}^\dg_{-\qv,B}} 
                \\
                +&\frac{1}{2} 
                \rowvec{\hat{a}^\dg_{\zv,+} \, \hat{a}_{\zv,+}}
                \begin{pmatrix} h_{3}  & 2h_{4}  \\
                    2h_{4}^{*} & h_{3}^*
                \end{pmatrix}
                \colvec{\hat{a}_{\zv,+} \\ \hat{a}^\dg_{\zv,+}}
                +\text{const.}. 
            \end{split}
                \label{eq:supp:Bogoliubov-Hamiltonian}
            \end{align}
    We can also compactly write this Bogoliubov Hamiltonian~\eqref{eq:supp:Bogoliubov-Hamiltonian} 
    in terms of the Nambu spinor representation
            \begin{align}
            \hat{H}^{(B)}= \frac{1}{2}\sum_{\qv \neq \zv} \Psi_{\qv}^{\dagger}
                \mathcal{M}_{\qv} \Psi_{\qv}
                +\frac{1}{2}
                \Psi^{\dagger}_{+} 
                \mathcal{M}_{+}
                \Psi_{+}
                +\text{const.}, 
                \label{eq:supp:Bogoliubov-Hamiltonian-SpinorRep}
        \end{align}
    where $\Psi^{\dagger}_{\qv} \equiv \colvec{\hat{a}^{\dagger}_{\qv,A} \,\hat{a}^{\dagger}_{\qv,B}\,\hat{a}_{-\qv,A},\hat{a}_{-\qv,B}}$
    and $\Psi^{\dagger}_{+} \equiv \colvec{\hat{a}^{\dagger}_{\zv,+} \, \hat{a}_{\zv,+}}$.
    Here, $\mathcal{M}_+$ is the $2\times 2$ matrix and $\mathcal{M}_{\qv}$ 
    the quasimomentum dependent $4\times 4$ matrix from~\eqref{eq:supp:Bogoliubov-Hamiltonian}, 
    which are explicitly given by~\cite{Furukawa2015,Wu2017}
    \begin{align}
        \begin{split}
                \mathcal{M}_{\qv} &= \begin{pmatrix} \mathcal{H}(\qv)+\mathcal{H}_1 & 2\mathcal{H}_2 \\
                2\mathcal{H}_{2}^{*}& (\mathcal{H}(-\qv)+\mathcal{H}_{1})^{*}
                \end{pmatrix}, \quad
                \mathcal{M}_{+}= \begin{pmatrix} h_{3}  & 2h_{4}  \\
                    2h_{4}^{*} & h_{3}^*
                \end{pmatrix},\\[.3cm]
                \text{with} \ 
                \mathcal{H}_1&=4 Un \abs{F_{-}}^2
                -\mu_{\text{eff}} \mathbbm{1}_2,\quad
                \mathcal{H}_2=Un (F_{-})^2,\quad 
            \end{split}
            \label{eq:supp:block-matrix}
        \end{align}
    where $F_{-}=\diag{f_{A,-},f_{B,-}}$ and 
    where the entries of $\mathcal{H}(\qv)$ are given by~\eqref{eq:supp:Hk-Pauli-exp} and~\eqref{eq:supp:Haldane-Model}. 
    The entries of $\Mp$ are given by 
            \begin{align}
                h_3=4Un \sum_{s}\left( \abs{f_{s,-}}^2\abs{f_{s,+}}^2\right)
                +\sum_{s,s'} f_{s,+}^{*}\Hc_{s,s'}(0)f_{s',+}-\mu_{\text{eff}}, \quad
                h_4 = Un \sum_{s}(f_{s,-})^2 \,(f_{s,+}^*)^2,
            \end{align}
    where $\mu_{\text{eff}}$ follows the stationary solutions of \crefrange{eq:supp:GPE-Min-eq1}{eq:supp:GPE-Min-eq3}.
    
    The Bogoliubov Hamiltonian~\eqref{eq:supp:Bogoliubov-Hamiltonian} can now be diagonalized by applying 
    the paraunitary (Bogoliubov transformation) 
    $\Wq \,(W_+)$~\cite{Furukawa2015} 
        \begin{align}
            \colvec{\hat{a}_{\qv,A} \\ 
                \hat{a}_{\qv,B} \\
                \hat{a}^\dg_{-\qv,A} \\ 
                \hat{a}^\dg_{-\qv,B}} 
                = \Wq
                \colvec{\hat{b}_{\qv,+} \\ \hat{b}_{\qv,-} \\
                \hat{b}^\dg_{-\qv,+} \\ \hat{b}^\dg_{-\qv,-}}
                \ \Leftrightarrow \ \Psi_{\qv}=\Wq \beta_{\qv},
                \quad 
                \colvec{\hat{a}_{\zv,+} \\ \hat{a}^\dg_{\zv,+}}
                =\Wp
                \colvec{\hat{b}_{\zv,+} \\ \hat{b}^{\dagger}_{\zv,+}}
                \ \Leftrightarrow \
                \Psi_{+}=\Wp \beta_{+}
            \label{eq:supp:Bog-Transf-simple}
        \end{align}
    with 
    $\beta_{\qv}\equiv \colvec{\hat{b}_{\qv,+} \, \hat{b}_{\qv,-} \,\hat{b}^\dg_{-\qv,+} \, \hat{b}^\dg_{-\qv,-}}^T$ 
    and $\beta_{+}\equiv \colvec{\hat{b}_{\zv,+} \, \hat{b}^{\dagger}_{\zv,+}}^T$
    consisting of the new bosonic quasi-particle annihilation (creation) operators 
    $\hat{b}^{(\dg)}_{\qv,\pm}$ at quasimomentum $\qv$ in the upper ($+$) or lower Bogoliubov energy band ($-$).
    The $4\times 4$ ($2\times 2$) Bogoliubov transformation matrix $W_{\qv}$ ($W_+$) can be parametrized 
   ~\eqref{eq:supp:paraunitary-matrix-bog-transf-gen} as
    \begin{align}
        W_+ &=
        \begin{pmatrix}
            u_{\zv} & v_{\zv}^*\\
            v_{\zv} & u_{\zv}^*
        \end{pmatrix}, \quad 
            W_{\qv} = 
            \begin{pmatrix}
                U_{\qv} & V_{-\qv}^*\\
                V_{\qv} & U_{-\qv}^*
            \end{pmatrix}
            \label{eq:supp:Bog-Transf-BogHald-paraunitary-param-1}
            \\[.3cm]
             \text{with} \  
            U_{\qv} &= \begin{pmatrix}
                u_{A,+}^{\qv} & u_{A,-}^{\qv}\\
                u_{B,+}^{\qv} & u_{B,-}^{\qv}
            \end{pmatrix}, \quad 
            V_{\qv} = \begin{pmatrix}
                v_{A,+}^{\qv} & v_{A,-}^{\qv}\\
                v_{B,+}^{\qv} & v_{B,-}^{\qv}
            \end{pmatrix}.
            \label{eq:supp:Bog-Transf-BogHald-paraunitary-param-2}
        \end{align}

    Since these quasi-particle operators still have to obey the bosonic commutation relations, 
    the Bogoliubov transformation matrix is paraunitary [cf.~\cref{eq:supp:paraunitary-cond}], i.e., 
    $W^\dg \tau W=W\tau W^\dg =\tau,\  \text{with}\ \tau=\sigma_z \otimes \mathbbm{1}_2$ 
    ($\tau=\sigma_z$) for $W=W_{\qv}$ ($W=W_+$).
    These paraunitary matrices are specifically constructed~\cite{Colpa1978,Shindou2013} such that
        \begin{align}
            \Wqd \Mqc \Wq = \diag{E_{+}(\qv),E_{-}(\qv),E_{+}(-\qv),E_{-}(-\qv)}, \quad
            \Wpd \Mp \Wp = \diag{E_{+}(\zv),E_{+}(\zv)},
        \end{align}
    holds,
    by which the Bogoliubov Hamiltonian is eventually diagonalized~\cite{Furukawa2015} to
        \begin{align}
            \hat{H}^{(B)}=\sum_{\qv} E_{+}(\qv)\hat{b}^{\dagger}_{\qv,+} \hat{b}_{\qv,+}
            +\sum_{\qv \neq 0} E_{-}(\qv)\hat{b}^{\dagger}_{\qv,-} \hat{b}_{\qv,-} +\text{const.}\,.
            \label{eq:supp:Bogoliubov-Hamiltonian-Diag}
        \end{align}
    Here, $E_{\pm}(\qv)$ is the Bogoliubov energy spectrum of the upper ($+$) and lower ($-$) band.

    The Bogoliubov energy spectrum can be obtained by (numerically) solving the energy eigenvalue problem 
   ~\eqref{eq:supp:gen-eigenvect-eq-vector} of the (non-hermitian) dynamical matrix 
    $D(\qv)=\tau_z \mathcal{M}_{\qv}$, i.e., $D(\qv)\vect{w}^n(\qv) =\tilde{\Omega}_n \vect{w}^n(\qv)$.
    However, the eigenmodes obtained from this diagonalization procedure 
    do \emph{not} lead to the correct paraunitary conditions for the Bogoliubov modes 
   ~\eqref{eq:supp:paraunitary-cond}.
    
    As described in~\Ccite{Colpa1978,Shindou2013}, 
    the correct paraunitary Bogoliubov matrices can be obtained via an algorithm, which 
    includes a Cholesky decomposition of the coefficient matrices ($\Mqc,\ \Mp$).
    For completeness, we briefly review this algorithm in the next section.

    \subsection{Algorithm for constructing paraunitary Bogoliubov transformation matrix}
    \label{app:algorithm-paraunitary-Bogoliubov-transformation}
    Here, mainly following~\Ccite{Colpa1978,Shindou2013}, 
    we briefly discuss how to (numerically) obtain a paraunitary Bogoliubov transformation matrix 
    such as $W_{\qv}$ and $W_+$~\eqref{eq:supp:Bog-Transf-BogHald-paraunitary-param-1} 
    when given a bosonic BdG Hamiltonian.
    
    We discuss the algorithm for a translationally invariant bosonic BdG Hamiltonian 
    such that we have quasimomentum dependent $2 N\times 2 N$ (hermitian) coefficient matrix $\Mqc$~\eqref{eq:supp:BdG-Ham} 
    of the form
        \begin{align}
            \Mqc = \begin{pmatrix}
                K_{\qv} & G_{\qv}\\
                G^*_{-\qv} & K^*_{-\qv}
            \end{pmatrix},
            \label{eq:supp:gen-Mcoeff-BdG-Ham}
        \end{align}
    with $K_{\qv}$ and $G_{\qv}$ being quasimomentum dependent $N\times N$ matrices with $N$ denoting 
    the number internal degrees of freedom (e.g., the number sublattice sites within a unit cell).
    For each quasimomentum $\qv$ the bosonic system 
    can then be diagonalized by a paraunitary matrix $W_{\qv}$~\eqref{eq:supp:BdG-Ham-diag-cond} 
    such that
    \begin{align}
        \Wqd \Mqc \Wq = 
        \begin{pmatrix}
            \vect{\omega}_{\qv} & 0\\
            0 & \vect{\omega}_{-\qv}
        \end{pmatrix}, \  
        \text{with} \ \vect{\omega}_{\qv} = \diag{\omega_{1,\qv},\ldots,\omega_{N,\qv}},
        \label{eq:supp:gen-eigenvect-eq-vector-gen-mom}
    \end{align}
    where $\Wq$ again fulfill the paraunitary conditions~\eqref{eq:supp:paraunitary-cond},
        \begin{align}
            \Wqd \tau_z \Wq = \Wq \tau_z \Wqd = \tau_z \ \text{with} \ \tau_z = \sigma_z \otimes \mathbbm{1}_N.
            \label{eq:supp:paraunitary-cond-mom}
        \end{align}
    Now, given that $\Mqc$ is a positive-definite matrix 
    (all eigenvalues of $\Mqc$ are strictly positive), 
    the paraunitary matrix $W_{\qv}$ can then be constructed
    via the following steps~\cite{Colpa1978}.
    
    First, we apply a Cholesky decomposition to the coefficient matrix $\Mqc$ 
    (which is possible since $\Mqc$ is positive-definite)
   ~\eqref{eq:supp:gen-Mcoeff-BdG-Ham}, which decomposes 
    $\Mqc$ into a product between an upper triangular matrix $K_{\qv}$ 
    and its hermitian conjugate $K_{\qv}^{\dagger}$, i.e., $\Mqc=K_{\qv}^{\dagger}K_{\qv}$.
    We note that the positive-definiteness of $\Mqc$ also ensures the existence of 
    the inverse $K_{\qv}^{-1}$. 
   
    Second, we introduce a hermitian matrix $T_{\qv}\equiv K_{\qv} \tau_z K_{\qv}^{\dg}$
    from which we obtain the unitary matrix $U_{\qv}$ via the eigendecomposition of $T_{\qv}$,
        \begin{align}
            U^\dg_{\qv} T_{\qv} U_{\qv} = \begin{pmatrix}
                \vect{\omega}_{\qv} & 0\\
                0 & - \vect{\omega}_{-\qv}
            \end{pmatrix}.
            \label{eq:supp:gen-eigenvect-eq-matrix-T-diag-Cholesk-constr}
        \end{align}
    These two diagonal matrices 
    $\vect{\omega}_{\qv}$ and $\vect{\omega}_{-\qv}$ already provide 
    the Bogoliubov energy spectrum of~\cref{eq:supp:gen-eigenvect-eq-vector-gen-mom}~\cite{Shindou2013}.
    One can further show that both $\vect{\omega}_{\qv}$ and $\vect{\omega}_{-\qv}$ are real and strictly positive~\cite{Colpa1978}.

    Third, the paraunitary matrix $W_{\qv}$ is then constructed via 
        \begin{align}
            W_{\qv} = K_{\qv}^{-1} U_{\qv} \begin{pmatrix}
                \vect{\omega}^{1/2}_{\qv} & 0\\
                0 & \vect{\omega}^{1/2}_{-\qv}
            \end{pmatrix},
            \label{eq:supp:paraunitary-matrix-Cholesky-constr}
        \end{align}
    which will diagonalize the coefficient matrix $\Mqc$~\eqref{eq:supp:gen-Mcoeff-BdG-Ham} 
    and importantly fulfill the paraunitary conditions~\eqref{eq:supp:paraunitary-cond-mom}.
    This concludes the construction of the paraunitary matrix $W_{\qv}$ for each quasimomentum $\qv$.

    One can then show the diagonalization of~\cref{eq:supp:gen-eigenvect-eq-vector-gen-mom} by the 
    direct application of~\eqref{eq:supp:paraunitary-matrix-Cholesky-constr}
        \begin{align}
            \Wqd \Mqc \Wq = \Wqd K_{\qv}^{\dagger}K_{\qv} \Wq 
            = 
            \begin{pmatrix}
                \vect{\omega}^{1/2}_{\qv} & 0\\
                0 & \vect{\omega}^{1/2}_{-\qv}
            \end{pmatrix}  U_{\qv}^\dg (K_{\qv}^{\dg})^{-1}
            K_{\qv}^{\dagger} K_{\qv} 
            K_{\qv}^{-1} U_{\qv} \begin{pmatrix}
                \vect{\omega}^{1/2}_{\qv} & 0\\
                0 & \vect{\omega}^{1/2}_{-\qv}
            \end{pmatrix}
            = \begin{pmatrix}
                \vect{\omega}_{\qv} & 0\\
                0 & \vect{\omega}_{-\qv}
            \end{pmatrix}.
        \end{align}
    Moreover, we can see that the construction of $\Wq$~\eqref{eq:supp:paraunitary-matrix-Cholesky-constr}
    also fulfills the paraunitary conditions~\eqref{eq:supp:paraunitary-cond-mom},
        \begin{align}
            \Wqd \tau_z \Wq 
            &=\begin{pmatrix}
            \vect{\omega}^{1/2}_{\qv} & 0\\
            0 & \vect{\omega}^{1/2}_{-\qv}
        \end{pmatrix}  U_{\qv}^\dg (K_{\qv}^{\dg})^{-1} \tau_z 
        K_{\qv}^{-1} U_{\qv} \begin{pmatrix}
            \vect{\omega}^{1/2}_{\qv} & 0\\
            0 & \vect{\omega}^{1/2}_{-\qv}
        \end{pmatrix}
        =\begin{pmatrix}
            \vect{\omega}^{1/2}_{\qv} & 0\\
            0 & \vect{\omega}^{1/2}_{-\qv}
        \end{pmatrix}  U_{\qv}^\dg T_{\qv}^{-1} U_{\qv} \begin{pmatrix}
            \vect{\omega}^{1/2}_{\qv} & 0\\
            0 & \vect{\omega}^{1/2}_{-\qv}
        \end{pmatrix}
        \\
        &=\begin{pmatrix}
            \vect{\omega}^{1/2}_{\qv} & 0\\
            0 & \vect{\omega}^{1/2}_{-\qv}
        \end{pmatrix}
        \begin{pmatrix}
            \vect{\omega}^{-1}_{\qv} & 0\\
            0 & -\vect{\omega}^{-1}_{-\qv}
        \end{pmatrix}
        \begin{pmatrix}
            \vect{\omega}^{1/2}_{\qv} & 0\\
            0 & \vect{\omega}^{1/2}_{-\qv}
        \end{pmatrix}
        =\begin{pmatrix}
            \mathbbm{1}_N & 0\\
            0 & -\mathbbm{1}_N
        \end{pmatrix}=\tau_z,
        \end{align}
    where in the first step we have just inserted the definition of $\Wq$ 
   ~\eqref{eq:supp:paraunitary-matrix-Cholesky-constr}.
    In the second equality, we used that $T_{\qv}^{-1} = (K_{\qv}^\dg)^{-1} \tau_z K_{\qv}^{-1}$, 
    which itself directly follows from $T_{\qv} = K_{\qv} \tau_z K_{\qv}^{\dg}$.
    In the second equality, following from~\eqref{eq:supp:gen-eigenvect-eq-matrix-T-diag-Cholesk-constr}
    we used that $U_{\qv}^\dg T_{\qv}^{-1} U_{\qv}=\begin{pmatrix}
        \vect{\omega}^{-1}_{\qv} & 0\\
        0 & -\vect{\omega}^{-1}_{-\qv}
    \end{pmatrix}$.

    \subsection{Topological classification of the Bogoliubov-Haldane model}
    \label{app:BH-model-symm-classf}
    We briefly discuss the topological classification of the Bogoliubov-Haldane model~\eqref{eq:supp:block-matrix}.
    In BBdG systems, time-reversal symmetry (TRS) is given by $\Mc_{\qv}=(\Mc_{-\qv})^*$~\cite{Ohashi2020}, 
    where the particle-hole symmetry (PHS) is given by 
    $\tau_x(\Mc_{\qv})^*\tau_x=\Mc_{-\qv}$~\eqref{eq:supp:PHS-DynMat}.
    Here, the Bogoliubov-Haldane Hamiltonian~\eqref{eq:supp:block-matrix} 
    explicitly breaks TRS, $\Mc_{\qv}\neq (\Mc_{-\qv})^*$, due the Peierls phases $\theta$ and the chiral symmetry (CS) is also broken due to the presence of the mass term $\Delta$~\eqref{eq:supp:Haldane-Model}.

    Following \Ccite{Lein2019}, we note that only the positive energy (particle) spectrum of $D_{\qv}=\tau_z\Mc_{\qv}$~\eqref{eq:supp:block-matrix} is a genuine Hilbert space with a positive definite $\tau_z$- or Krein inner product.
    It is exactly this positive energy subspace of the Krein space $(\Hc_{\text{ph}},\tau_z,\tau_x)$ (Definition \ref{def:Real-Krein-space}) which is the relevant Bogoliubov state space for the topological classification~\cite{Lein2019}.
    Now, according to Wigner's theorem~\cite{Bracci1975} physical symmetries have to be unitarily or anti-unitarily implemented with respect to this Krein inner product structure in this space.
    This implies that the PHS in BBdG systems is actually not a physical symmetry but just a 
    \emph{constraint}, since the PHS operation of~\cref{eq:supp:PHS-DynMat} is not a valid Krein-anti-unitary on this space, as it intertwines the positive and negative energy subspaces of the Krein space 
   ~\cite{Lein2019}.
    This has direct implications on the ten-fold AZ classification scheme of topological insulators~\cite{Altland1997,Chiu2016}. 
    Falsely viewing the PHS~\eqref{eq:supp:PHS-DynMat} of the BBdG Hamiltonian~\eqref{eq:supp:BdG-Ham} as an actual symmetry would, in the absence of TRS and CS, place the Bogoliubov-Haldane model~\eqref{eq:supp:block-matrix} into class D in the AZ classification scheme (as claimed in \Ccite{Peano2016} for example).
    However, when correctly viewing the PHS as a constraint~\cite{Lein2019}, the Bogoliubov-Haldane model 
    is of class A, with an integer valued Chern invariant $\mathbb{Z}$~\cite{Lein2019,Zhou2020}.

    \subsection{SQGT from integrated Bogoliubov excitation rates}
    \label{app:SQGT-from-Bogoliubov-ExcRate}
    \begin{figure*}[bt!]
            \begin{center} 
                \includegraphics[scale = 1.0]{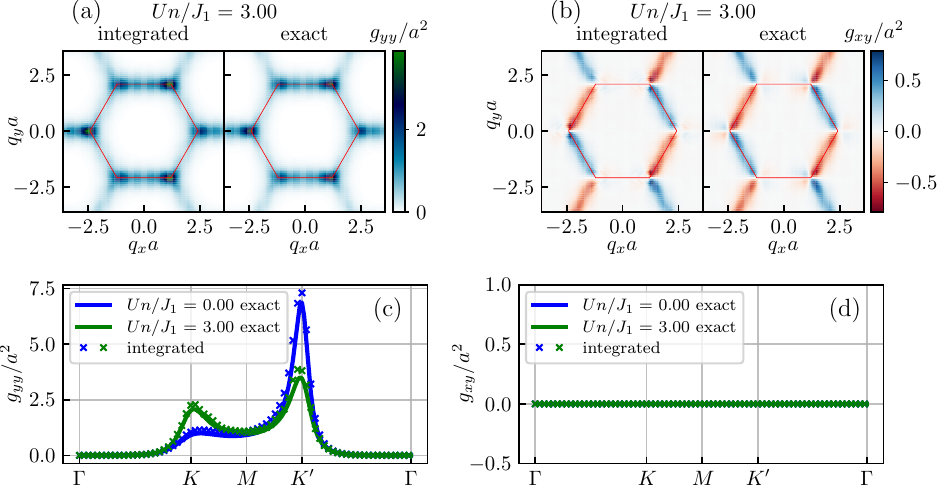}
                \caption{Integrated and exact symplectic quantum metric component $g_{yy}(\qv)$ and $g_{xy}(\qv)$
                for the upper particle band of the Bogoliubov-Haldane model~\eqref{eq:supp:Bogoliubov-Hamiltonian}.
                (a) Integrated (left) and exact (right) symplectic quantum metric component $g_{yy}(\qv)$
                for $Un/J_1=3$ shown in 2D quasimomentum space.
                (b) Same as in (a) but for the off-diagonal component $g_{xy}(\qv)$.
                (c) Exact (solid line) and integrated (crosses) symplectic quantum metric component 
                $g_{yy}(\qv)$ along a high-symmetry path in quasimomentum space 
                for two interactions strengths 
                $Un/J_1=0$ (non-interacting Haldane model) and $Un/J_1=3$.
                (d) Same as in (c) but for the off-diagonal component $g_{xy}(\qv)$.
                All the simulations have been performed for driving strengths $A/J_1=0.01$ up to a final time of
                $J_1 t/\hbar=10$ for a frequency range of $\hbar\omega/J_1\in [0.1 (0.2),5.5]$, 
                for $Un/J_1=3\,(0)$ with a spacing of $\delta \omega=0.05$ over which the excitation rates have been integrated.
                The red-colored hexagon in (a) and (b) indicates the first Brillouin zone.
                Other parameters of the underlying Haldane model~\eqref{eq:supp:Haldane-Model} are 
                the same as in~\cref{fig:Int-Rate-GxxBxy} in the main text,
                i.e., $J_2=0.1J_1$, $\theta=0.5\pi$, $\Delta/(3\sqrt{3}J_2)=0.5$.
                }
                \label{fig:Int-Rate-GyyGxy}
            \end{center}
        \end{figure*}
        Here, we briefly discuss how we numerically implement our proposed method to extract the SQGT from the integrated Bogoliubov excitation rate applied 
        to the Bogoliubov-Haldane model~\eqref{eq:supp:Bogoliubov-Hamiltonian}.
        We start with an initially populated Bogoliubov particle mode $\vect{w}^n(\qv)$ which is obtained from the 
        paraunitary Bogoliubov transformation matrix~\eqref{eq:supp:Bog-Transf-simple}.
        The time evolution of the Bogoliubov particle mode $\vect{w}^n(\qv,t)$ is performed via 
       ~\eqref{eq:supp:Effective-TDSE-Vector}
        in the translationally invariant co-moving frame.
        Specifically, when the system is shaken along the 
        $x$-direction the time-dependent dynamical matrix is given by 
       ~\cref{eq:supp:DynMat-IntPic-periodic-shaking-x}.
        The total excitation rate into other Bogoliubov in response to this perturbation is then computed via
            \begin{align}
                \Gamma^n_{\qv}(\omega) =\frac{1}{t}\sum_{m} s_m 
                \abs{(\vect{w}^m(\qv))^\dg \tau_z \vect{w}^n(\qv,t)}^2,
                \label{eq:supp:ETDPT-integrated-rate-simulation}
            \end{align}
        where $t$ is the final integration time and $\omega$ the frequency of the perturbation 
       ~\eqref{eq:supp:DynMat-IntPic-periodic-shaking-x}.
        Integrating~\eqref{eq:supp:ETDPT-integrated-rate-simulation} over a range 
        of probing frequencies $\omega$, $\Gamma^n_{\text{int},\qv}=\int d\omega\, \Gamma^n_{\qv}(\omega)$, 
        then allows us to obtain the symplectic quantum metric component $g_{xx}(\qv)$
        via~\eqref{eq:supp:ETDPT-integrated-rate-lattice}.
        This similarly applies to the $g_{yy}(\qv)$ component of the SQGT when the system is shaken along the 
        $y$-direction where the dynamical matrix in co-moving frame becomes 
        $D_y'(t)=D(\qv) + (2A/\omega)\sin(\omega t)\del_{q_y}D(\qv)+\mathcal{O}((A/\omega)^2)$.
        For the off-diagonal components, the lattice is shaken along both directions 
       ~\eqref{eq:supp:dynamical-matrix-periodic-shaking-xy}, 
        the time-dependent dynamical matrix is given by~\cref{eq:supp:DynMat-IntPic-periodic-shaking-xy}.
        The total excitation rates $\Gamma^{n,\pm}_{\qv}(\omega)$ are obtained in the same way as 
        in~\eqref{eq:supp:ETDPT-integrated-rate-simulation}.
        Taking the differential integrated rate, 
        $\Delta \Gamma ^n_{\text{int}}= \Gamma^{n,+}_{\text{int}}-\Gamma^{n,-}_{\text{int}}=\int d\omega\, (\Gamma^{n,+}_{\qv}(\omega)-\Gamma^{n,-}_{\qv}(\omega))$
        over a range of probing frequencies $\omega$ eventually 
        allows us to obtain the off-diagonal component of the quantum metric $g_{xy}(\qv)$ (for $\varphi=0$) 
        and the Berry curvature $B_{xy}(\qv)$ (for $\varphi=\pi/2$) via 
       ~\eqref{eq:supp:ETDPT-integrated-rate-offdiag-lattice}.

        The results of this simulation for $g_{xx}(\qv)$ and $B_{xy}(\qv)$ are shown in~\cref{fig:Int-Rate-GxxBxy} in the main text.
        For completeness, here we show the results for the remaining components of the SQGT, 
        $g_{yy}(\qv)$ and $g_{xy}(\qv)$ in~\cref{fig:Int-Rate-GyyGxy}.

        \subsection{Signatures for the distribution of the original particles}\label{app:Signatures-Real-Particle-Distributions}
        Our measurement protocol for extracting the SQGT 
        requires measuring the excitation fraction in specific Bogoliubov (particle) modes.
        Focusing on the Bogoliubov-Haldane model~\eqref{eq:supp:Bogoliubov-Hamiltonian}, we 
        discuss the signatures for the quasimomentum distribution of the original particles $a_{\qv,s}$ 
        given the presence of the single or multiple quasiparticles excitations $\hat{b}^{\dg}_{\qv,\pm}$ 
        on top of the Bogoliubov vacuum, which can then be measured e.g., via time-of-flight techniques 
        in optical lattice platforms~\cite{Vogels2002,Stamper-Kurn1999,Brunello2000,Bloch2008}.

        We start by noting that the ground state $\ket{\psi_0}$ of the Bogoliubov-Hamiltonian 
       ~\eqref{eq:supp:Bogoliubov-Hamiltonian-Diag} is given by the vacuum state of the Bogoliubov quasiparticles 
        fulfilling the conditions,
            \begin{align}
                \hat{b}_{\qv,+}\ket{\psi_0}=0, \ \forall \qv, 
                \quad \text{and} \quad
                \hat{b}_{\qv,-}\ket{\psi_0}=0, \ \forall \qv\neq 0,
                \label{eq:supp:Ground-State-Bogoliubov}
            \end{align}
        where in the second equality the condensation mode has to be explicitly excluded, 
        as the Bogoliubov theory is only concerned with excitations outside the condensate mode 
        at $\qv=\zv$ in the lower band.
        Even though the Bogoliubov ground state contains no quasiparticles, 
        there is still a finite amount of original particles outside the condensation mode 
        which are (virtually) excited by interparticle interactions.
        Since these excitations are only driven by quantum fluctuations, 
        this result is called quantum depletion.
        
        We now consider the more general situation where a finite number of $l$ quasiparticles are excited on top 
        of the Bogoliubov vacuum at quasimomentum $\qv$ in band $n\in\{\pm\}$, 
        i.e, $(b_{\qv,n}^{\dg})^{l}/\sqrt{l!}\,\ket{\psi_0}\equiv \ket{\phi_0}$.
        For $l=0$ this state is again the Bogoliubov ground state~\eqref{eq:supp:Ground-State-Bogoliubov}.
        Now, for the state $\ket{\phi_0}$ with $l$ quasiparticle excitations, 
        we compute the original particle contributions outside the condensation mode 
        $\expval{\phi_0}{\hat{a}^{\dg}_{\qv,s}\hat{a}_{\qv,s'}}{\phi_0}$ where $s,s'\in\{A,B\}$ 
        is again the sublattice index~\eqref{eq:supp:Bogoliubov-Hamiltonian}. 
        To this end, we explicitly use the Bogoliubov transformations for $\qv\neq \zv$
       ~\eqref{eq:supp:Bog-Transf-simple}-\eqref{eq:supp:Bog-Transf-BogHald-paraunitary-param-2} 
        and arrive at
            \begin{align}
                \ep{\hat{a}^{\dagger}_{\qv,s} \hat{a}_{\qv,s} }
                &=l\, \abs{u_{s,m}^{\qv}}^2 \delta_{m,n}+\abs{v_{s,+}^{-\qv}}^2+\abs{v_{s,-}^{-\qv}}^2,
                \quad
                \ep{\hat{a}^{\dagger}_{-\qv,s} \hat{a}_{-\qv,s} }
                =l\, \abs{v_{s,m}^{\qv}}^2 \delta_{m,n}+\abs{v_{s,+}^{\qv}}^2+\abs{v_{s,-}^{\qv}}^2,\nonumber \\
                \ep{\hat{a}^{\dagger}_{\qv,A} \hat{a}_{\qv,B} }
                &=l\, \left[(u_{A,m}^{\qv})^*u_{B,m}^{\qv}\right] \delta_{m,n}+(v_{A,+}^{-\qv}) (v_{B,+}^{-\qv})^{*}+(v_{A,-}^{-\qv}) (v_{B,-}^{-\qv})^{*}
                =\ep{\hat{a}^{\dagger}_{\qv,B} \hat{a}_{\qv,A} }^*,\nonumber \\
                \ep{\hat{a}^{\dagger}_{-\qv,A} \hat{a}_{-\qv,B} }
                &=l\, \left[v_{A,m}^{\qv}(v_{B,m}^{\qv})^*\right] \delta_{m,n}+(v_{A,+}^{\qv}) (v_{B,+}^{\qv})^{*}+(v_{A,-}^{\qv}) (v_{B,-}^{\qv})^{*}
                =\ep{\hat{a}^{\dagger}_{-\qv,B} \hat{a}_{-\qv,A} }^*.
                \label{eq:supp:Real-Particle-Excitation-general-QP}
            \end{align}
        For $l=0$ initial quasiparticle excitation present in band $n$ in 
       ~\eqref{eq:supp:Real-Particle-Excitation-general-QP}, we recover the quantum depletion result.
        As an example, let us consider $l$ initial quasiparticle excitations in the lower Bogoliubov energy band 
        at $\qv$, i.e., $\ket{\phi_0}= (b_{\qv,-}^{\dg})^{l}/\sqrt{l!}\,\ket{\psi_0}$.
        Then, for the sublattice site $s=A$, we have $l\times \abss{u_{A,-}^{\qv}}^2+\abss{v_{A,+}^{-\qv}}^2+\abss{v_{A,-}^{-\qv}}^2$ 
        original particles moving with quasimomentum $+\qv$ and 
        $(l+1)\times\abss{v_{A,-}^{\qv}}^2+\abss{v_{A,+}^{\qv}}^2$ original particles moving with quasimomentum $-\qv$.
        These results~\eqref{eq:supp:Real-Particle-Excitation-general-QP} directly 
        relate the presence of initial Bogoliubov quasiparticle excitations with the quasimomentum distribution 
        of the real bosonic particles. 

        In optical lattice platforms, the momentum distribution can be observed via time-of-flight (TOF) imaging techniques.
        The momentum distribution $N(\kv)$ of the system after TOF is given by~\cite{Bloch2008,Hauke2014}
            \begin{align}
                N(\kv)=\abs{\tilde{w}(\kv)}^2 \frac{1}{N_{\text{uc}}}\sum_{\ell \ell'}\sum_{s s'}e^{-i\kv \cdot (\rv_{\ell s}-\rv_{\ell' s'})}
                \ep{\hat{a}^{\dagger}_{\ell s}\hat{a}_{\ell' s'}}, 
                \label{eq:supp:TOF-2-band-general}
            \end{align}
        where $\abs{\tilde{w}(\kv)}^2$ is the Wannier function envelope 
        with the Fourier-transformed Wannier function $\tilde{w}(\kv)$, 
        $N_{\text{uc}}$ the number of unit cells, and 
        $\hat{a}_{\ell s}=\sum_{\qv} \braket{\ell s}{\qv s}\hat{a}_{\qv , s}=(N_{\text{uc}})^{-1/2}\sum_{\qv}e^{-i \qv \cdot \rv_{\ell s}}\hat{a}_{\qv , s}$ 
        the real bosonic operators in the original real space basis with the unit cell index $\ell$ and sublattice index $s$~\eqref{eq:supp:BH-Hamiltonian}.
        Then, for the initial state with multiple Bogoliubov quasiparticle excitations in band $n$ 
       ~\eqref{eq:supp:Real-Particle-Excitation-general-QP} 
        the TOF momentum distribution $N(\kv)$~\eqref{eq:supp:TOF-2-band-general} becomes
            \begin{align}
                N(\kv)=&\sum_{s,s'}\ep{\hat{a}^{\dg}_{\kv s}\hat{a}_{\kv s'}}=\ep{\hat{a}^{\dg}_{\kv A}\hat{a}_{\kv A}}+\ep{\hat{a}^{\dg}_{\kv B}\hat{a}_{\kv B}}
                +\left( \ep{\hat{a}^{\dg}_{\kv A}\hat{a}_{\kv B}} + \text{c.c.} \right) \nonumber \\
                =&\sum_{s\in\{A,B\}}\left\{ l\, \abs{u_{s,m}^{\kv}}^2 \delta_{m,n}+\abs{v_{s,+}^{-\kv}}^2+\abs{v_{s,-}^{-\kv}}^2\right\} \nonumber \\
                +&\left[\left( l\, \left[(u_{A,m}^{\kv})^*u_{B,m}^{\kv}\right] \delta_{m,n}+(v_{A,+}^{-\kv}) (v_{B,+}^{-\kv})^{*}+(v_{A,-}^{-\kv}) (v_{B,-}^{-\kv})^{*}\right) + \text{c.c.} \right],
                \label{eq:supp:TOF-2-band-general-final}
            \end{align}
            where we neglected the overall Wannier envelope function $\abs{\tilde{w}(\kv)}^2$ for now.
        Thus, equipped with the quasimomentum distribution results obtained from Bogoliubov theory 
       ~\eqref{eq:supp:Real-Particle-Excitation-general-QP} and the TOF images~\eqref{eq:supp:TOF-2-band-general-final}, 
        the distribution of Bogoliubov quasiparticle excitations can be inferred to extract the SQGT.
            
        We believe that the following experimental protocol 
        can be used to measure the Bogoliubov excitation fraction 
        after periodic modulation of the systems' parameters to extract the SQGT 
        as described in the main text. 
        First, one should measure the TOF momentum distribution 
        without performing the periodic modulation and without exciting any quasiparticles 
        to obtain a reference measurement. 
        Then, as a second step, we imagine exciting Bogoliubov quasiparticles at 
        a desired quasimomentum $\qv$ in band $n$ on top of the Bogoliubov vacuum 
        with the protocol described in the main text or the one of~\Ccite{Stamper-Kurn1999,Vogels2002}, 
        whose momentum distribution should then be measured via TOF.
        The difference of these first two measurements then provides 
        a measure of the excited Bogoliubov quasiparticle distribution of step two.
        As a third step, after exciting initial Bogoliubov quasiparticles, 
        one should perform the periodic parameter modulation protocol 
        to induce scattering of Bogoliubov quasiparticles into the other band.
        Then, the momentum distribution of the system should be measured again via TOF.
        Eventually, the difference of this final momentum distribution 
        and the one after step two gives us 
        the induced Bogoliubov quasiparticle excitation fraction 
        which can be used to extract the SQGT as described in the main text.
        
\end{document}